\documentclass[showpacs,nofootinbib]{revtex4}
\usepackage{amsfonts}
\usepackage{amsmath}
\usepackage{graphicx}
\usepackage{amssymb}

\input{tcilatex}
\begin{document}

\title{Covariant gravitational dynamics in 3+1+1 dimensions}
\author{Zolt\'{a}n Keresztes$^{1,2}$, L\'{a}szl\'{o} \'{A}. Gergely$^{1,2}$}
\affiliation{$^{1}$ Department of Theoretical Physics, University of Szeged, Tisza Lajos
krt 84-86, Szeged 6720, Hungary \\
$^{2}$ Department of Experimental Physics, University of Szeged, D\'{o}m T%
\'{e}r 9, Szeged 6720, Hungary}

\begin{abstract}
We develop a 3+1+1 covariant formalism with cosmological and astrophysical
applications. First we give the evolution and constraint equations both on
the brane and off-brane in terms of 3-space covariant kinematical,
gravito-electro-magnetic (Weyl) and matter variables. We discuss the
junction conditions across the brane in terms of the new variables. Then we
establish a \textit{closure condition} for the equations on the brane. We
also establish the connection of this formalism with isotropic and
anisotropic cosmological brane-worlds. Finally we derive a new brane
solution in the framework of our formalism: the \textit{tidal charged
Taub-NUT-(A)dS brane}, which obeys the closure condition.
\end{abstract}

\date{\today}
\maketitle

\section{Introduction}

In recent years the idea of exploring the possibility of the gravitational
interaction acting in more than 4-dimensions (4d) attracted a lot of
attention. In particular, one of the simplest such models arises from the
curved generalization of the one-brane Randall-Sundrum model \cite{RS2} (for
a review see \cite{MaartensLivRev}), where gravity acts in 5-dimensions
(5d), however standard model fields are confined to the 4d brane.
Nonstandard model fields can occur in the 5d space-time. A $\mathcal{Z}_{2}$%
-symmetric embedding looks natural only if the brane is envisaged as a
boundary; otherwise generic asymmetric embeddings should be allowed. The
generic dynamics was given in Refs. \cite{Gergely Friedmann} and \cite%
{VarBraneTension} in a 5d-covariant approach. Although promising at the
level of allowing new degrees of freedom, which seem to be adjustable to
explain for example dark matter \cite{HarkoRC}-\cite{HarkoClusters}, the
complexity of the brane-world dynamics also represents a major impediment in
obtaining simple exact solutions or in monitoring the evolution of
perturbations. Therefore new approaches to gravitational dynamics in
brane-worlds are worth to study.

In Ref. \cite{KOZO1} the 5d space-time was foliated first by a family of 4d
space-times and a second family of 4d space-like hypersurfaces (see Fig 1.
in Ref. \cite{KOZO1}). Such an 5d space-time is \textit{double foliable}.
This structure of the 5d space-time allowed to describe the gravitational
degrees of freedom in terms of tensorial, vectorial and scalar quantities
with respect to the 3-space emerging as the intersection of the two
hypersurfaces. They represent the gravitons, a gravi-photon, and a
gravi-scalar and are given by quantities with well-defined geometrical
meaning, namely the tensorial, vectorial and scalar projections of the
spatial 4d metric and their canonically conjugated momenta: the extrinsic
curvature (second fundamental form of the 3-spaces with respect to the
temporal normal), normal fundamental form and normal fundamental scalar. The
evolution equations for these variables were given \`{a} la ADM both on the
brane and outside it. An extension of this formalism towards a full
Hamiltonian description was advanced in Ref. \cite{KOZO2}. It has been
shown, that among all gravitational variables on the brane, only the
momentum of the gravi-vector has a jump across the brane, related to the
energy transport (heat flow) on the brane.

The formalism presented in Refs. \cite{KOZO1}-\cite{KOZO2} however is not
straightforward to be applied in brane cosmology. A formal difference would
be the definition of the time derivative. In Refs. \cite{KOZO1}-\cite{KOZO2}
this was defined in the tradition of canonical gravity as a Lie-derivative
taken along the temporal direction (which is not necessarily orthogonal to
the 3-space) projected to the 3-space, whereas in cosmology by tradition the
time derivative is defined as a covariant derivative along the normal to the
3-surfaces (this derivative happens to enter only in expressions projected
onto the 3-space). Obviously this mismatch in the definitions of the time
derivatives is not crucial, as the two definitions differ only in terms
taken on the 3-space. However the formalism developed in Refs. \cite{KOZO1}
and \cite{KOZO2} relying on the double foliability of the 5d space-time is
unable to deal with the possible \textit{vorticity of the word-lines of
observers}.

In this paper we develop a formalism overcoming this inconvenience and
derive the full set of evolution and constraint equations governing
gravitational dynamics in a 3+1+1 covariant form. Provided the space-like
normal $n$ has vanishing vorticity on a hypersurface (the time evolution
vector can still have vorticity), the formalism becomes suitable for
describing gravitational dynamics on the brane (then $n$ becomes the brane
normal). In this sense the formalism is a generalization of the brane 3+1
covariant cosmology \cite{Maartens eqs} (which in turn is a generalization
of the general relativistic 3+1 covariant cosmology \cite{Ehlers}-\cite%
{Bonometto}).

This newly developed formalism also generalizes the s+1+1 covariant
brane-world dynamics developed in Ref. \cite{KOZO1}. Both the vector field $%
n $ and the time evolution vector $u$ are allowed to have vorticity in the
present formalism. Though observational evidence suggest that the directly
detectable 3+1 part of the universe is best described by a Friedmann brane
with perfect fluid, when discussing cosmological perturbations, the
vorticity of $u$ should be allowed, similarly as in existing formalisms of
covariant cosmology in both general relativity and brane-worlds. We can also
argue for the need of keeping the vorticity of $n$. One reason would be,
that if it is vanishing at some initial instant, it should stay zero; and
the formalism should be able to handle its "evolution". Secondly, and more
important, the vorticity of $n$ should not necessarily vanish at other
locations, than the brane, a gauge freedom worth to explore.

We note that a lower-dimensional, 2+1+1 formalism was developed in Refs. 
\cite{Clarkson Barrett 211}, \cite{Clarkson 211} and applied in the general
relativistic covariant perturbations of Schwarzschild black holes and
rotationally symmetric space-time; then for investigating spherically
symmetric static solutions in $f(R)$ gravity theories in Ref. \cite{NCGD 211}%
.

In general relativity the important topic of cosmological perturbations,
related to both the Cosmic Microwave Background and structure formation, has
a rich literature, from which (without claiming completeness) we mention
Refs. \cite{Hawking1966}-\cite{Gebbie Dunsby Ellis}, all based on the 3+1
covariant approach.

Brane-world cosmological perturbations are equally important, however
additional difficulties emerge due to the complexity of brane-world theory
and the impediment to predict and perform observations on the brane. At a
technical level, the latter is obstructed by the lack of closure of the
perturbation equations on the brane. Despite impressive developments \cite%
{MukohyamaMaster}-\cite{PalStructure2}, many questions remain unanswered.
Although we cannot overcome well-known difficulties, we expect our new
formalism will provide the most convenient and complete tool-chest for
approaching the problems. We also foresee the possibility of important
applications for brane-world exact solutions.

We establish the generic 3+1+1 covariant formalism in Sec. \ref{decomp_sec},
by defining the kinematical quantities and the decomposition of the Weyl and
energy-momentum tensors. We also relate the curvature and Ricci tensors and
the 3-dimensional (3d) scalar curvature to kinematic, non-local (Weyl) and
matter-defined variables. In Appendix \ref{comm_app} the commutation
relations among all types of derivatives emerging in the formalism are given.

Sec. \ref{dynamics_sec} contain the full 3+1+1 decomposed covariant
gravitational dynamics and constraints, together with the available matter
field evolutions. In Appendix \ref{frame_app} we discuss the gauge freedom
in the frame choice and give the transformations of all relevant quantities
under infinitesimal frame transformations.

Taking\ into account that the brane is a 4d time-like hypersurface, in Sec. %
\ref{brane_sec} we discuss the decomposition of the Lanczos equation and of
the sources of the effective Einstein equation. Then we specify the generic
evolution and constraint equations on the brane, expressed in terms of
quantities defined on the brane. These equations arise from combinations of
the equations given in Section \ref{dynamics_sec}, evaluated on the brane.
Appendix \ref{asym_app} contains the brane equations for an asymmetric
embedding. We continue Section \ref{brane_sec} by specializing to a
symmetrically embedded brane, by taking into account the Lanczos equation.
Then we conclude the section by deriving a \textit{closure condition} for
the equations on the brane.

Sec. \ref{cosmological} contains the derivation of the most important
cosmological equations to a hypersurface (a generic brane), then the
discussion of the particular case of cosmological symmetries and perfect
fluid source on the brane. As a test of our formalism we recover the
Friedmann, Raychaudhuri and energy balance equations and compare them with
the relevant results of Ref. \cite{VarBraneTension}. In subsection \ref{ani}
and Appendix \ref{BianchiBraneWorld} we also relate our formalism to
anisotropic brane-world cosmologies.

Sec. \ref{astrophysical} contains an astrophysical application of our
formalism devoted to locally rotationally symmetric (LRS) space-times, at
the end of which we recover a new brane solution with NUT charge. This
solution obeys the previously derived closure condition.

Sec. \ref{concl_sec} contains the Concluding Remarks.

\textit{Notations.} Quantities defined on the 3-space orthogonal to both $%
u^{a}$ and $n^{a}$ carry no distinguishing mark and the 3d metric is denoted 
$h_{ab}$. Quantities defined on the brane carry the pre-index $^{\left(
4\right) }$, the only exception being the 4d metric $g_{ab}$. Quantities
defined on the full 5-space carry a distinguishing $\widetilde{}$ mark.
Exceptionally, other quantities also carry the distinguishing $\widetilde{}$
mark. These are (a) the 3+1+1 decomposed components of the 5d
energy-momentum tensor, which are defined on the 3-space orthogonal to both $%
u^{a}$ and $n^{a}$, and (b) the kinematic and extrinsic curvature type
quantities related to another singled out spatial vector $e^{a}$ in Sec. \ref%
{astrophysical}. Calligraphic symbols denote 3+1+1 decomposed components of
the 5d Weyl tensor. Whenever possible, identical symbols are used for
quantities related to the temporal vector field $u^{a}$ and the brane normal 
$n^{a}$, the latter distinguished by a $\widehat{}$ mark. We denote the
average of a quantity $f$ taken over the two sides of the brane by $%
\left\langle f\right\rangle $, and its jump by $\Delta f$. \ Angular
brackets $\langle ~\rangle $ on abstract indices denote tensors which are
projected in all indices with the metric $h_{ab}$, symmetrized and
trace-free. Round brackets $(~)$ and square brackets $[~]$ on indices denote
the symmetric and antisymmetric parts, respectively.

\section{The 3+1+1 covariant formalism\label{decomp_sec}}

\subsection{Decomposition of the metric}

Let $u^{a}=dx^{a}/d\tau $ and $n^{a}=dx^{a}/dy$ be a time-like and a
space-like vector field in the 5d space-time, with $\tau $ and $y$ the
affine parameters of the respective non-null integral curves. They obey the
normalization conditions $-u^{a}u_{a}=1=n^{a}n_{a}$ and the perpendicularity
condition $u^{a}n_{a}=0$. The 5d metric is decomposed as%
\begin{equation}
\widetilde{g}_{ab}=n_{a}n_{b}+g_{ab}\ ,
\end{equation}%
with 
\begin{equation}
g_{ab}=-u_{a}u_{b}+h_{ab}\ ,
\end{equation}%
the metric on the 4d temporal leaves $y=$const (with the brane at $y=0$) and
the spatial part $h_{ab}$ of this metric obeying $u^{a}h_{ab}=n^{a}h_{ab}=0$%
. We denote by $\varepsilon _{abc}$ the volume element associated with $%
h_{ab}$. The 4-vector $u^{a}$ represents the time-like velocity field of
brane observers (see Fig 1).

\begin{figure}[tbp]
\includegraphics[height=8cm, angle=360]{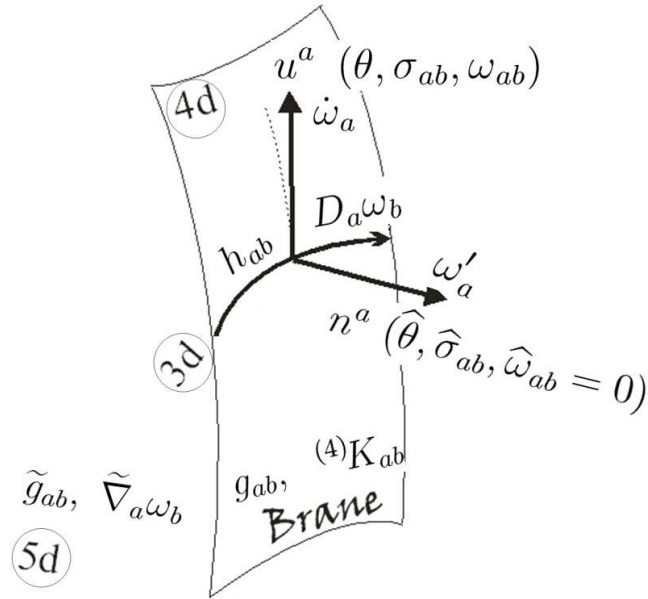}
\caption{Elements of the 3+1+1 decomposition of the 5d space-time geometry
with metric $\widetilde{g}_{ab}$ and compatible connection $\widetilde{%
\protect\nabla }$. The 4d brane with normal $n^{a}$ has induced metric $%
g_{ab}=\widetilde{g}_{ab}-n_{a}n_{b}$ and extrinsic curvature $^{\left(
4\right) }\!K_{ab}=g_{a}^{c}g_{b}^{d}\widetilde{\protect\nabla }_{c}n_{d}$.
The 4-velocity field $u^{a}$ of observers in the brane defines local 3d
orthogonal spatial patches, hypersurface-forming only when the vorticity $%
\protect\omega _{ab}=0$. The other kinematical characteristics of $u^{a}$
are the expansion $\Theta $ and shear $\protect\sigma _{ab}$. The expansion,
shear and vorticity of $n^{a}$ are $\widehat{\Theta },~\widehat{\protect%
\sigma }_{ab}$ and $\widehat{\protect\omega }_{ab}$, the latter vanishing on
the brane. The temporal, off-brane and 3d covariant derivative are shown for
the vorticity 1-form $\protect\omega _{a}.$}
\end{figure}

We employ three type of derivatives, all associated with projections of the
5d connection $\widetilde{\nabla }_{i}$. A dot and a prime denote covariant
derivatives along the integral curves of $u^{a}$ and $n^{a}$, respectively,
while $D$ is the 3d covariant derivative compatible with the metric $h$:%
\begin{eqnarray}
\dot{T}_{b..c} &=&u^{a}\widetilde{\nabla }_{a}T_{b..c}~, \\
T_{b..c}^{\prime } &=&n^{a}\widetilde{\nabla }_{a}T_{b..c}~, \\
D_{a}T_{b..c} &=&h_{a}^{\,\,\,d}h_{b}^{\,\,\,i}..h_{c}^{\,\,\,\,j}\widetilde{%
\nabla }_{d}T_{i..j}\ .
\end{eqnarray}%
Note that the $D$-derivative is the same as introduced in general relativity
employing the corresponding projection of the $\nabla $-derivative ($\nabla $
being the connection compatible with $g_{ab}$). This is, because both
generate the covariant derivatives formed with the connection compatible
with $h_{ab}$. Concerning the time-derivative defined above, except for
scalars, it differs from the corresponding general relativistic time
derivative employed in the 3+1 covariant formalism (which is defined with $%
\nabla $) in $n^{a}$ and $u^{a}$ terms. Nevertheless, when projected to the
3-manifold with $h_{b}^{\,\,\,a}$, the two definitions agree.

\subsection{Kinematic quantities}

We introduce the kinematic quantities through the decomposition of the 5d
covariant derivative of $u^{a}$, $n^{a}$ as 
\begin{subequations}
\begin{eqnarray}
\widetilde{\nabla }_{a}u_{b} &=&-u_{a}A_{b}+\widehat{K}%
u_{a}n_{b}+Kn_{a}n_{b}+n_{a}K_{b}+L_{a}n_{b}+K_{ab}\ ,  \label{nablau} \\
\widetilde{\nabla }_{a}n_{b} &=&n_{a}\widehat{A}_{b}+Kn_{a}u_{b}+\ \widehat{K%
}u_{a}u_{b}-u_{a}\widehat{K}_{b}+L_{a}u_{b}+\widehat{K}_{ab}\ ,
\label{nablan}
\end{eqnarray}%
where 
\end{subequations}
\begin{eqnarray}
A_{a} &=&\dot{u}_{\langle a\rangle }\ ,\ \quad \widehat{A}_{a}=n_{\langle
a\rangle }^{\prime }\ ,  \notag \\
K_{a} &=&u_{\langle a\rangle }^{\prime }\ ,\ \quad \widehat{K}_{a}=\dot{n}%
_{\langle a\rangle }\ ,  \notag \\
K &=&n^{b}u_{b}^{\prime }\ ,\quad \widehat{K}=u^{b}\dot{n}_{b}\ ,  \notag \\
L_{a} &=&h_{a}^{\,\,\,c}n^{d}\widetilde{\nabla }_{c}u_{d}\ ,  \notag \\
K_{ab} &=&D_{a}u_{b}\ ,\quad \widehat{K}_{ab}=D_{a}n_{b}\ .
\end{eqnarray}%
As a rule, an overhat is used for kinematical quantities related to the
vector field $n^{a}$ in order to distinguish them from the similar
kinematical quantities related to the vector field $u^{a}$. Here $A_{a}$ is
the acceleration. All scalars, vectors and tensors in the above
decomposition are defined on the 3d manifold orthogonal both to $n^{a}$ and $%
u^{a}$. The tensorial expressions $K_{ab}$ and $\widehat{K}_{ab}$ can be
further decomposed into (trace-, trace-free symmetric and antisymmetric)
irreducible parts as follows:%
\begin{eqnarray}
K_{ab} &=&\frac{\Theta }{3}h_{ab}+\sigma _{ab}+\omega _{ab}\ ,  \label{Kab}
\\
\widehat{K}_{ab} &=&\frac{\widehat{\Theta }}{3}h_{ab}+\widehat{\sigma }_{ab}+%
\widehat{\omega }_{ab}\ ,  \label{Kabhat}
\end{eqnarray}%
where we have defined the expansion, vorticity and shear of the vector
fields $u^{a}$ and $n^{a}$ as%
\begin{eqnarray}
\Theta &=&D^{a}u_{a}\ ,\quad \omega _{ab}=D_{[a}u_{b]}\ ,\quad \sigma
_{ab}=D_{\langle a}u_{b\rangle }\ ,  \notag \\
\ \widehat{\Theta } &=&D^{a}n_{a}\ ,\quad \widehat{\omega }%
_{ab}=D_{[a}n_{b]}\ ,\quad \widehat{\sigma }_{ab}=D_{\langle a}n_{b\rangle
}\ .
\end{eqnarray}%
The antisymmetric 3d tensors $\omega _{ab}$ and $\widehat{\omega }_{ab}$ can
be also encoded in the vorticity vectors $\omega ^{a}=\varepsilon
^{abc}\omega _{bc}/2$ and$\ \widehat{\omega }^{a}=\varepsilon ^{abc}\widehat{%
\omega }_{bc}/2$. When the vorticities of both vector field $n^{a}$ and $%
u^{a}$ vanish $\omega _{ab}=\widehat{\omega }_{ab}=0$, the tensorial
expressions $K_{ab}$ and $\widehat{K}_{ab}$ are symmetric and they represent
the two extrinsic curvatures of a 3d hypersurface. The condition $\widehat{%
\omega }_{ab}=0$ is also necessary in order to have a brane at $y=0$, but
not sufficient. Indeed the brane is a 3+1 dimensional hypersurface which can
exist only if the higher-dimensional vorticity of its normal $\widehat{%
\omega }_{ab}=\nabla _{\lbrack a}n_{b]}$ vanishes. This condition translates
into $0=g_{[a}^{\,\,\,\,\,\,c}g_{b]}^{\,\,\,\,d}\widetilde{\nabla }%
_{c}n_{d}=-u_{[a}\left( \widehat{K}_{b]}+L_{b]}\right) +\widehat{\omega }%
_{ab}$, therefore (due to Frobenius' theorem) the necessary and sufficient
conditions for the existence of the 3+1 brane are%
\begin{eqnarray}
\left. \widehat{\omega }_{ab}\right\vert _{y=0} &=&0~,  \notag \\
\left. L_{a}\right\vert _{y=0} &=&-\left. \widehat{K}_{a}\right\vert _{y=0}~.
\label{brane}
\end{eqnarray}

In summary, the independent components of $\widetilde{\nabla }_{a}u_{b}$ are
expressed by three scalars ($K,~\widehat{K},~\Theta $), four $3$-vectors ($%
A_{a},~K_{a},~L_{a},~\omega _{a}$) and a symmetric trace-free $3$-tensor ($%
\sigma _{ab}$). The corresponding decomposition of $\widetilde{\nabla }%
_{a}n_{b}$ consists of the three scalars ($\widehat{K},~K,$ $\widehat{\Theta 
}$), four $3$-vectors ($\widehat{A}_{a},~\widehat{K}_{a},~L_{a},~\widehat{%
\omega }_{a}$) and a symmetric trace-free $3$-tensor ($\widehat{\sigma }%
_{ab} $). The irreducible decompositions of the covariant derivatives $%
\widetilde{\nabla }_{a}u_{b}$ and $\widetilde{\nabla }_{a}n_{b}$ have
therefore $20$ independent components each (due to the constraints $u^{b}%
\widetilde{\nabla }_{a}u_{b}=0$ and $n^{b}\widetilde{\nabla }_{a}n_{b}=0$).

\subsection{Gravito-electro-magnetic quantities\label%
{gravito-electric-magnetic}}

The non-local gravitational properties of the 5d space-time are carried by
the 5d Weyl tensor, the principal directions of which lead to a
classification scheme generalizing the general relativistic Petrov
classification \cite{Coley}. Here we are interested in the 3+1+1
decomposition\footnote{%
The case of a generic $n+1$ decomposition of the Weyl tensor was discussed
in Ref. \cite{Senovilla}.} of $\widetilde{C}_{abcd}$, which can be given in
terms of the quantities (with a total of 35 independent components):%
\begin{eqnarray}
\mathcal{E} &=&\widetilde{C}_{abcd}n^{a}u^{b}n^{c}u^{d}\ ,\   \notag \\
\mathcal{H}_{k} &=&\frac{1}{2}\varepsilon _{k}^{\ \ ab}\widetilde{C}%
_{abcd}u^{c}n^{d}\ ,~\quad \mathcal{F}_{kl}=\widetilde{C}_{abcd}h_{(k}^{\,\,%
\,\,a}u^{b}h_{l)}^{\,\,\,\,c}n^{d}\ ,  \notag \\
\widehat{\mathcal{E}}_{k} &=&\widetilde{C}_{abcd}h_{k}^{\,\,\,%
\,a}n^{b}u^{c}n^{d}\ ,\quad \mathcal{E}_{k}=\widetilde{C}_{abcd}h_{k}^{\,\,%
\,\,a}u^{b}n^{c}u^{d}\ ,  \notag \\
\widehat{\mathcal{E}}_{kl} &=&\widetilde{C}_{abcd}h_{\langle
k}^{\,\,\,\,\,\,a}n^{b}h_{l\rangle }^{\,\,\,\,c}n^{d}\ ,\ \quad \mathcal{E}%
_{kl}=\widetilde{C}_{abcd}h_{\langle k}^{\,\,\,\,\,a}u^{b}h_{l\rangle
}^{\,\,\,\,c}u^{d}\ ,  \notag \\
\widehat{\mathcal{H}}_{kl} &=&\frac{1}{2}\varepsilon _{(k}^{\ \ \ \
ab}h_{l)}^{\,\,\,\,c}\widetilde{C}_{abcd}n^{d}\ ,\ \quad \mathcal{H}_{kl}=%
\frac{1}{2}\varepsilon _{(k}^{\ \ \ \ ab}h_{l)}^{\,\,\,\,c}\widetilde{C}%
_{abcd}u^{d}\ .  \label{WeylS3}
\end{eqnarray}%
We note that all tensorial quantities defined above are trace-free (from the
properties of the Weyl tensor), and further in the tensors $\mathcal{F}_{kl}$%
, $\mathcal{H}_{kl}$ and $\widehat{\mathcal{H}}_{kl}$ the brackets $\langle
~\rangle $ are equivalent with the round brackets $(~)$. The Weyl tensor in
terms of the quantities defined in (\ref{WeylS3}) is%
\begin{eqnarray}
\widetilde{C}_{abcd} &=&-2\left( \mathcal{E}_{d[a}h_{b]c}-\mathcal{E}%
_{c[a}h_{b]d}\right) +2\left( \widehat{\mathcal{E}}_{d[a}h_{b]c}-\widehat{%
\mathcal{E}}_{c[a}h_{b]d}\right) -\frac{2}{3}\mathcal{E}h_{c[a}h_{b]d} 
\notag \\
&&+2\left( \varepsilon _{cd}^{\ \ \,\,\,\,i}\widehat{\mathcal{H}}%
_{i[a}n_{b]}+\varepsilon _{ab}^{\ \ \,\,\,\,i}\widehat{\mathcal{H}}%
_{i[c}n_{d]}\right) -2\left( \varepsilon _{cd}^{\ \ \,\,\,\,i}\mathcal{H}%
_{i[a}u_{b]}+\varepsilon _{ab}^{\ \ \,\,\,\,i}\mathcal{H}_{i[c}u_{d]}\right)
\notag \\
&&+2\left( n_{c}n_{[a}\widehat{\mathcal{E}}_{b]d}-n_{d}n_{[a}\widehat{%
\mathcal{E}}_{b]c}\right) +2\left( u_{c}u_{[a}\mathcal{E}_{b]d}-u_{d}u_{[a}%
\mathcal{E}_{b]c}\right)  \notag \\
&&-2\left( u_{c}n_{[a}\mathcal{F}_{b]d}-u_{d}n_{[a}\mathcal{F}%
_{b]c}+n_{c}u_{[a}\mathcal{F}_{b]d}-n_{d}u_{[a}\mathcal{F}_{b]c}\right) 
\notag \\
&&-\left( n_{[c}u_{a]}\varepsilon _{bdk}+u_{[c}n_{b]}\varepsilon
_{adk}\right) \mathcal{H}^{k}-\left( u_{[d}n_{a]}\varepsilon
_{bck}+n_{[d}u_{b]}\varepsilon _{ack}\right) \mathcal{H}^{k}  \notag \\
&&-2\left( u_{[c}n_{d]}\varepsilon _{abk}+u_{[a}n_{b]}\allowbreak
\varepsilon _{cdk}\right) \mathcal{H}^{k}-2\left( \mathcal{E}%
_{[a}h_{b][c}n_{d]}+\mathcal{E}_{[c}h_{d][a}n_{b]}\right)  \notag \\
&&-2\left( \widehat{\mathcal{E}}_{[a}h_{b][c}u_{d]}+\widehat{\mathcal{E}}%
_{[c}h_{d][a}u_{b]}\right) +4\mathcal{E}u_{[a}n_{b]}u_{[c}n_{d]}  \notag \\
&&-2\left( n_{c}\allowbreak u_{d}n_{[a}\widehat{\mathcal{E}}%
_{b]}-u_{c}n_{d}n_{[a}\widehat{\mathcal{E}}_{b]}+n_{a}u_{b}n_{[c}\widehat{%
\mathcal{E}}_{d]}-u_{a}n_{b}n_{[c}\widehat{\mathcal{E}}_{d]}\right)  \notag
\\
&&+2\left( u_{c}\allowbreak n_{d}u_{[a}\mathcal{E}_{b]}-n_{c}u_{d}u_{[a}%
\mathcal{E}_{b]}+u_{a}n_{b}u_{[c}\mathcal{E}_{d]}-n_{a}u_{b}u_{[c}\mathcal{E}%
_{d]}\right)  \notag \\
&&-\frac{2}{3}\mathcal{E}\left( n_{d}n_{[a}h_{b]c}-n_{c}n_{[a}h_{b]d}\right)
+\frac{2}{3}\mathcal{E}\left( u_{d}u_{[a}h_{b]c}-u_{c}u_{[a}h_{b]d}\right) \
.  \label{Weyldec}
\end{eqnarray}

This relation generalizes the general relativistic 3+1 covariant
decomposition of the 4d Weyl tensor $C_{abcd}$ (which has only 10
independent components), where only two tensors $E_{kl}=C_{abcd}h_{\langle
k}^{\,\,\,\,\,\,a}u^{b}h_{l\rangle }^{\,\,\,\,c}u^{d}$ and $H_{kl}=\frac{1}{2%
}\varepsilon _{(k}^{\ \ \ ab}h_{l)}^{\,\,\,\,c}C_{abcd}u^{d}$ appear, the
electric and magnetic parts of the Weyl curvature. On the brane the relation
between the set of variables $\mathcal{E}_{kl}$ and $\mathcal{H}_{kl}$ and
the variables $E_{kl}$ and $H_{kl}$ is given by 
\begin{eqnarray}
\mathcal{E}_{ab} &=&E_{ab}+\frac{1}{2}\widehat{\mathcal{E}}_{ab}-\frac{1}{2}%
\left( \widehat{K}+\frac{\widehat{\Theta }}{3}\right) \widehat{\sigma }_{ab}+%
\frac{1}{2}\widehat{\sigma }_{c\langle a}\widehat{\sigma }_{b\rangle
}^{\,\,\,\,c}+\frac{1}{2}\widehat{K}_{\langle a}\widehat{K}_{b\rangle }\ ,
\label{Eab} \\
\mathcal{H}_{ab} &=&H_{ab}-\varepsilon _{\langle a}^{\ \ \ \ cd}\widehat{%
\sigma }_{b\rangle c}\widehat{K}_{d}\ .  \label{Hab}
\end{eqnarray}

\subsection{Decomposition of the energy-momentum tensor}

The 5d gravitational dynamics is governed by the Einstein equation%
\begin{equation}
\widetilde{G}_{ab}=-\widetilde{\Lambda }\widetilde{g}_{ab}+\widetilde{\kappa 
}^{2}\left[ \widetilde{T}_{ab}+\tau _{ab}\delta \left( y\right) \right] ~,
\end{equation}%
where $\widetilde{\kappa }^{2}$ denotes the 5d coupling constant. The
sources of gravity are the 5d cosmological constant $\widetilde{\Lambda }$,
the regular part of the 5d energy-momentum tensor $\widetilde{T}_{ab}$ and a
distributional energy-momentum tensor with support on the brane:%
\begin{equation}
\tau _{ab}=-\lambda g_{ab}+T_{ab}~.
\end{equation}%
Here $\lambda $ is the constant brane tension and $T_{ab}$ describes
standard model matter fields on the brane\footnote{%
For DGP / induced gravity type models \cite{Induced}, $T_{ab}$ should be
replaced by $T_{ab}-\left( \gamma /\kappa ^{2}\right) G_{ab}$, with $G_{ab}$
the Einstein tensor constructed from the metric $g_{ab}$ and $\gamma $ the
dimensionless induced gravity parameter. Randall-Sundrum type brane-worlds
are recovered for $\gamma \rightarrow 0$, on the $\varepsilon =-1$ branch.},
decomposed with respect to a brane observer with $4$-velocity $u^{a}$ as%
\begin{equation}
T_{ab}=\rho u_{a}u_{b}+q_{(a}u_{b)}+ph_{ab}+\pi _{ab}\ .  \label{Tab}
\end{equation}%
The quantities $\rho $, $q_{a}$, $p$ and $\pi _{ab}$ are the energy density,
the energy current vector, the isotropic pressure and the symmetric
trace-free anisotropic pressure tensor of the matter on the brane.

We decompose the regular part of the 5d energy-momentum tensor relative to $%
u^{a}$ and $n^{a}$ as:%
\begin{equation}
\widetilde{T}_{ab}=\widetilde{\rho }u_{a}u_{b}+2\widetilde{q}_{(a}u_{b)}+2%
\widetilde{q}u_{(a}n_{b)}+\widetilde{p}h_{ab}+\widetilde{\pi }n_{a}n_{b}+2%
\widetilde{\pi }_{(a}n_{b)}+\widetilde{\pi }_{ab}\ ,
\end{equation}%
where $\widetilde{\rho }$ is the relativistic energy density relative to $%
u^{a}$, $\widetilde{p}$ the isotropic pressure, $\widetilde{q}_{a}$ and $%
\widetilde{q}$ the relativistic momentum densities on the 3-space and in the
direction $n^{a}$. The quantities $\widetilde{\pi },~\widetilde{\pi }_{a}$
and the symmetric and trace-free$~$tensor $\widetilde{\pi }_{ab}$ are
related to the scalar, vectorial and tensorial components (with respect to
the 3d space) of the 5d anisotropic pressure tensor, which is%
\begin{equation}
\frac{3\left( \widetilde{\pi }-\widetilde{p}\right) }{4}n_{a}n_{b}+2%
\widetilde{\pi }_{(a}n_{b)}+\frac{\widetilde{p}-\widetilde{\pi }}{4}h_{ab}+%
\widetilde{\pi }_{ab}~.
\end{equation}

By employing the definitions given in this Section we give the commutation
relations among the temporal, off-brane and brane covariant derivatives in
Appendix A.

\subsection{The Gauss equation and its contractions\label{curvature_sec}}

We define the local 3d curvature tensor $\mathcal{R}_{abcd}$ of the space
orthogonal to both $u^{a}$ and $n^{a}$ as%
\begin{equation}
D_{[a}D_{b]}V_{c}-\omega _{ab}\dot{V}_{\langle c\rangle }+\widehat{\omega }%
_{ab}V_{\langle c\rangle }^{\prime }=\frac{1}{2}\mathcal{R}_{abcd}V^{d}\ ,
\end{equation}%
resulting in%
\begin{eqnarray}
\mathcal{R}_{abcd}
&=&h_{a}^{\,\,\,\,i}h_{b}^{\,\,\,\,j}h_{c}^{\,\,\,k}h_{d}^{\,\,\,\,l}%
\widetilde{R}_{ijkl}-\left( D_{a}u_{c}\right) \left( D_{b}u_{d}\right)
+\left( D_{a}u_{d}\right) \left( D_{b}u_{c}\right)  \notag \\
&&+\left( D_{a}n_{c}\right) \left( D_{b}n_{d}\right) -\left(
D_{a}n_{d}\right) \left( D_{b}n_{c}\right) \ .  \label{localRiemann}
\end{eqnarray}%
This is a natural generalization of the definition used in general
relativity \cite{EBH}, \cite{Barrow Maartens Tsagas}, \cite{Tsagas Challinor
Maartens}. By the definitions of the kinematical, gravito-electro-magnetic
and matter variables, we have%
\begin{eqnarray}
\mathcal{R}_{abcd} &=&\left[ -\frac{2}{3}\left( \Theta ^{2}-\widehat{\Theta }%
^{2}\right) -\mathcal{E}+\widetilde{\Lambda }+\widetilde{\kappa }^{2}\left( 
\widetilde{\rho }+\widetilde{p}-\widetilde{\pi }\right) \right] \frac{%
h_{c[a}h_{b]d}}{3}  \notag \\
&&-2\left( \mathcal{E}_{d[a}h_{b]c}-\mathcal{E}_{c[a}h_{b]d}\right) +2\left( 
\widehat{\mathcal{E}}_{d[a}h_{b]c}-\widehat{\mathcal{E}}_{c[a}h_{b]d}\right)
-\frac{2\widetilde{\kappa }^{2}}{3}\left( h_{d[a}\widetilde{\pi }%
_{b]c}-h_{c[a}\widetilde{\pi }_{b]d}\right)  \notag \\
&&-\frac{2\Theta }{3}\left[ \left( \sigma _{a[c}+\omega _{a[c}\right)
h_{d]b}+\left( \sigma _{b[d}+\omega _{b[d}\right) h_{c]a}\right] -2\left(
\sigma _{a[c}+\omega _{a[c}\right) \left( \sigma _{d]b}-\omega _{d]b}\right)
\notag \\
&&+\frac{2\widehat{\Theta }}{3}\left[ \left( \widehat{\sigma }_{a[c}+%
\widehat{\omega }_{a[c}\right) h_{d]b}+\left( \widehat{\sigma }_{b[d}+%
\widehat{\omega }_{b[d}\right) h_{c]a}\right] +2\left( \widehat{\sigma }%
_{a[c}+\widehat{\omega }_{a[c}\right) \left( \widehat{\sigma }_{d]b}-%
\widehat{\omega }_{d]b}\right) \ ,
\end{eqnarray}%
and the local 3d Ricci tensor and Ricci scalar become%
\begin{eqnarray}
\mathcal{R}_{ac} &=&h^{bd}\mathcal{R}_{abcd}=\left[ -\frac{2}{3}\left(
\Theta ^{2}-\widehat{\Theta }^{2}\right) -2\mathcal{E}+\widetilde{\Lambda }+%
\widetilde{\kappa }^{2}\left( \widetilde{\rho }+\widetilde{p}-\widetilde{\pi 
}\right) \right] \frac{h_{ac}}{3}  \notag \\
&&+\mathcal{E}_{ac}-\widehat{\mathcal{E}}_{ac}+\frac{\widetilde{\kappa }^{2}%
}{3}\widetilde{\pi }_{ac}-\frac{2\Theta }{3}\left( \sigma _{ac}+\omega
_{ac}\right) +\frac{2\widehat{\Theta }}{3}\left( \widehat{\sigma }_{ac}+%
\widehat{\omega }_{ac}\right)  \notag \\
&&+\sigma _{ab}\sigma _{c}^{\,\,\,\,b}-\omega _{b}\omega ^{b}h_{ac}+\omega
_{a}\omega _{c}-2\sigma _{\lbrack a}^{\,\,\,\,\,b}\omega _{c]b}  \notag \\
&&-\widehat{\sigma }_{ab}\widehat{\sigma }_{c}^{\,\,\,\,b}+\widehat{\omega }%
_{b}\widehat{\omega }^{b}h_{ac}-\widehat{\omega }_{a}\widehat{\omega }_{c}+2%
\widehat{\sigma }_{[a}^{\,\,\,\,\,\,b}\widehat{\omega }_{c]b}\ , \\
\mathcal{R} &=&h^{ac}\mathcal{R}_{ac}=-2\mathcal{E}+\widetilde{\Lambda }+%
\widetilde{\kappa }^{2}\left( \widetilde{\rho }+\widetilde{p}-\widetilde{\pi 
}\right) -\frac{2}{3}\left( \Theta ^{2}-\widehat{\Theta }^{2}\right)  \notag
\\
&&-2\omega _{a}\omega ^{a}+\sigma _{ab}\sigma ^{ab}+2\widehat{\omega }_{a}%
\widehat{\omega }^{a}-\widehat{\sigma }_{ab}\widehat{\sigma }^{ab}\ .
\label{localRicciscalar}
\end{eqnarray}%
Eq. (\ref{localRicciscalar}) can be referred as a generalized Friedmann
equation in the 5d space-time, as will become evident in Section \ref%
{cosmological}. The general relativistic counterpart is presented in Refs. 
\cite{Barrow Maartens Tsagas}, \cite{Tsagas Challinor Maartens}.

\section{3+1+1 covariant dynamics\label{dynamics_sec}}

The \textit{full set} of the 3+1+1 covariant evolution and constraint
equations for the kinematic, gravito-electro-magnetic and matter variables
are given by the projections of the Bianchi and Ricci identities for $u^{a}$
and $n^{a}$.\textbf{\ }These equations hold as the main result of the paper,
and they are presented in the following three subsections. In particular,
the first subsection contains all Ricci identities; the second subsection
contains twice contracted Bianchi identities, which by virtue of the
Einstein equations describe evolutions for the energy density and currents;
while the third subsection contains the rest of independent Bianchi
identities. A related Appendix \ref{frame_app} gives the transformation
rules under a frame change for the totality of the kinematic,
gravito-electro-magnetic and matter variables, to linear order accuracy.

\subsection{Ricci identities\label{Ricci_sec}}

The Ricci identity for $u^{a}$ gives the following independent equations:%
\begin{eqnarray}
0 &=&\dot{K}+\widehat{K}^{\prime }+\widehat{A}_{a}A^{a}+K^{2}-\widehat{K}%
^{2}+L_{a}K^{a}-\widehat{K}_{a}K^{a}-L_{a}\widehat{K}^{a}  \notag \\
&&+\mathcal{E}-\frac{\widetilde{\Lambda }}{6}+\frac{\widetilde{\kappa }^{2}}{%
6}\left( \widetilde{\rho }-\widetilde{\pi }+3\widetilde{p}\right) \ , \\
0 &=&\dot{\Theta}-D^{a}A_{a}+\frac{\Theta ^{2}}{3}+\widehat{\Theta }\widehat{%
K}-A^{a}A_{a}-2\omega _{a}\omega ^{a}+\sigma _{ab}\sigma ^{ab}-\mathcal{E} 
\notag \\
&&+\left( \widehat{K}^{a}K_{a}+K^{a}L_{a}+L^{a}\widehat{K}_{a}\right) -\frac{%
\widetilde{\Lambda }}{2}+\frac{\widetilde{\kappa }^{2}}{2}\left( \widetilde{%
\rho }+\widetilde{\pi }+\widetilde{p}\right) \ ,  \label{Theta_dot} \\
0 &=&\dot{K}_{\langle a\rangle }-A_{\langle a\rangle }^{\prime }+\left( K+%
\frac{\Theta }{3}\right) K_{a}+\left( K-\frac{\Theta }{3}\right) \widehat{K}%
_{a}+\mathcal{E}_{a}  \notag \\
&&+\widehat{K}\left( \widehat{A}_{a}+A_{a}\right) +\left( \omega
_{ba}+\sigma _{ba}\right) \left( K^{b}-\widehat{K}^{b}\right) -\frac{%
\widetilde{\kappa }^{2}}{3}\widetilde{\pi }_{a}\ ,  \label{Ka_dot} \\
0 &=&\dot{L}_{\langle a\rangle }+D_{a}\widehat{K}+\left( K+\frac{\Theta }{3}%
\right) L_{a}+\left( \widehat{K}+\frac{\widehat{\Theta }}{3}\right)
A_{a}+\left( K-\frac{\Theta }{3}\right) \widehat{K}_{a}  \notag \\
&&+\left( \widehat{\omega }_{ab}+\widehat{\sigma }_{ab}\right) A^{b}-\left(
\omega _{ab}+\sigma _{ab}\right) \left( \widehat{K}^{b}-L^{b}\right) +%
\mathcal{E}_{a}-\frac{\widetilde{\kappa }^{2}}{3}\widetilde{\pi }_{a}\ ,
\label{L_dot} \\
0 &=&\dot{\omega}_{\langle a\rangle }-\frac{1}{2}\varepsilon _{a}^{\ \
cd}D_{c}A_{d}+\widehat{K}\widehat{\omega }_{a}+\frac{2\Theta }{3}\omega
_{a}-\sigma _{ab}\omega ^{b}+\frac{1}{2}\varepsilon _{a}^{\ \ cd}\left( K_{c}%
\widehat{K}_{d}+K_{c}L_{d}+\widehat{K}_{c}L_{d}\right) \ ,  \label{omega_dot}
\end{eqnarray}%
\begin{eqnarray}
0 &=&\dot{\sigma}_{\langle ab\rangle }-D_{\langle a}A_{b\rangle }+\frac{%
2\Theta }{3}\sigma _{ab}+\widehat{K}\widehat{\sigma }_{ab}-A_{\langle
a}A_{b\rangle }+K_{\langle a}L_{b\rangle }  \notag \\
&&+\widehat{K}_{\langle a}K_{b\rangle }+L_{\langle a}\widehat{K}_{b\rangle
}+\omega _{\langle a}\omega _{b\rangle }+\sigma _{c\langle a}\sigma
_{b\rangle }^{\,\,\,\,c}+\mathcal{E}_{ab}-\frac{\widetilde{\kappa }^{2}}{3}%
\widetilde{\pi }_{ab}\ ,  \label{sigma_dot} \\
0 &=&\Theta ^{\prime }-D^{a}K_{a}-\left( K-\frac{\Theta }{3}\right) \widehat{%
\Theta }+\left( K^{a}+L^{a}\right) \widehat{A}_{a}-A^{a}\left(
K_{a}-L_{a}\right) -2\widehat{\omega }_{a}\omega ^{a}+\widehat{\sigma }%
_{ab}\sigma ^{ab}-\widetilde{\kappa }^{2}\widetilde{q}\ ,
\end{eqnarray}%
\begin{eqnarray}
0 &=&L_{\langle a\rangle }^{\prime }-D_{a}K+\left( K-\frac{\Theta }{3}%
\right) \widehat{A}_{a}-\left( \widehat{K}-\frac{\widehat{\Theta }}{3}%
\right) L_{a}+\left( \widehat{K}+\frac{\widehat{\Theta }}{3}\right) K_{a} 
\notag \\
&&+\left( \widehat{\sigma }_{ab}+\widehat{\omega }_{ab}\right) \left(
K^{b}+L^{b}\right) -\left( \omega _{ab}+\sigma _{ab}\right) \widehat{A}^{b}-%
\widehat{\mathcal{E}}_{a}+\frac{\widetilde{\kappa }^{2}}{3}\widetilde{q}%
_{a}\ ,  \label{callLa_column} \\
0 &=&\omega _{\langle k\rangle }^{\prime }-\frac{1}{2}\varepsilon _{k}^{\ \
ab}D_{a}K_{b}-\frac{1}{2}\varepsilon _{k}^{\ \ ab}\left( A_{b}+\widehat{A}%
_{b}\right) \left( K_{a}-L_{a}\right) -\left( K-\frac{\Theta }{3}\right) 
\widehat{\omega }_{k} \\
&&+\frac{\widehat{\Theta }}{3}\omega _{k}+\frac{1}{2}\varepsilon _{k}^{\ \
ab}\omega _{a}\widehat{\omega }_{b}-\frac{1}{2}\widehat{\sigma }_{ka}\omega
^{a}-\frac{1}{2}\sigma _{ka}\widehat{\omega }^{a}+\frac{1}{2}\varepsilon
_{k}^{\ \ ab}\sigma _{cb}\widehat{\sigma }_{a}^{\,\,\,\,c}-\frac{1}{2}%
\mathcal{H}_{k}\ ,  \label{Omega_column} \\
0 &=&\sigma _{\langle ab\rangle }^{\prime }-D_{\langle a}K_{b\rangle
}-A_{\langle b}\left( K_{a\rangle }-L_{a\rangle }\right) +\widehat{A}%
_{\langle b}\left( K_{a\rangle }+L_{a\rangle }\right) +\frac{\widehat{\Theta 
}}{3}\sigma _{ab}  \notag \\
&&-\left( K-\frac{\Theta }{3}\right) \widehat{\sigma }_{ab}+\omega _{\langle
a}\widehat{\omega }_{b\rangle }+\omega _{c\langle a}\widehat{\sigma }%
_{b\rangle }^{\,\,\,\,\,\,c}+\sigma _{\langle a}^{\,\,\,\,\,\,d}\widehat{%
\omega }_{b\rangle d}+\sigma _{c\langle a}\widehat{\sigma }_{b\rangle
}^{\,\,\,\,c}+\mathcal{F}_{ab}\ , \\
0 &=&\varepsilon _{k}^{\ \ ab}D_{a}L_{b}+2\left( \widehat{K}+\frac{\widehat{%
\Theta }}{3}\right) \omega _{k}+2\left( K-\frac{\Theta }{3}\right) \widehat{%
\omega }_{k}+\varepsilon _{k}^{\ \ ab}\omega _{a}\widehat{\omega }_{b}-%
\widehat{\sigma }_{ka}\omega ^{a}+\sigma _{ka}\widehat{\omega }%
^{a}-\varepsilon _{k}^{\ \ ab}\sigma _{cb}\widehat{\sigma }_{a}^{\,\,\,c}+%
\mathcal{H}_{k}\ ,  \label{sigmak} \\
0 &=&D^{a}\omega _{a}-A_{a}\omega ^{a}+\widehat{\omega }^{a}\left(
K_{a}-L_{a}\right) \ ,  \label{Omega_const} \\
0 &=&D_{\langle c}\omega _{k\rangle }+\varepsilon _{ab\langle k}D^{b}\sigma
_{c\rangle }^{\,\,\,\,a}+2A_{\langle c}\omega _{k\rangle }-2K_{\langle c}%
\widehat{\omega }_{k\rangle }-L_{\langle c}\widehat{\omega }_{k\rangle
}+\varepsilon _{\langle k}^{\ \ \,ab}\widehat{\sigma }_{c\rangle b}L_{a}+%
\mathcal{H}_{kc}\ ,  \label{Hab_tild_const} \\
0 &=&D^{b}\sigma _{ab}+\varepsilon _{a}^{\ \ ck}D_{c}\omega _{k}-\frac{2}{3}%
D_{a}\Theta +\frac{2\widehat{\Theta }}{3}L_{a}-\varepsilon _{a}^{\ \ ck}L_{c}%
\widehat{\omega }_{k}  \notag \\
&&+2\varepsilon _{a}^{\ \ ck}\left( A_{c}\omega _{k}-K_{c}\widehat{\omega }%
_{k}\right) -\widehat{\sigma }_{ab}L^{b}+\widehat{\mathcal{E}}_{a}+\frac{2%
\widetilde{\kappa }^{2}}{3}\widetilde{q}_{b}\ .  \label{Divsigmaab_const}
\end{eqnarray}

The Ricci identity for $n^{a}$ gives the following independent equations:%
\begin{eqnarray}
0 &=&\dot{\widehat{\Theta }}-D^{a}\widehat{K}_{a}+\left( \widehat{K}+\frac{%
\widehat{\Theta }}{3}\right) \Theta -\left( \widehat{K}^{a}-L^{a}\right)
A_{a}+\widehat{A}^{a}\left( \widehat{K}_{a}+L_{a}\right) -2\omega _{a}%
\widehat{\omega }^{a}+\widehat{\sigma }_{ab}\sigma ^{ab}-\widetilde{\kappa }%
^{2}\widetilde{q}\ ,  \label{theta_tild_dot} \\
0 &=&\dot{\widehat{\omega }}_{\langle k\rangle }-\frac{1}{2}\varepsilon
_{k}^{\ \ ab}D_{a}\widehat{K}_{b}+\frac{1}{2}\varepsilon _{k}^{\ \ ab}\left( 
\widehat{A}_{b}+A_{b}\right) \left( \widehat{K}_{a}+L_{a}\right) +\left( 
\widehat{K}+\frac{\widehat{\Theta }}{3}\right) \omega _{k}  \notag \\
&&+\frac{\Theta }{3}\widehat{\omega }_{k}+\frac{1}{2}\varepsilon _{k}^{\ \
ab}\widehat{\omega }_{a}\omega _{b}-\frac{1}{2}\widehat{\sigma }_{ka}\omega
^{a}-\frac{1}{2}\sigma _{ka}\widehat{\omega }^{a}+\frac{1}{2}\varepsilon
_{k}^{\ \ ab}\widehat{\sigma }_{cb}\sigma _{a}^{\,\,\,\,c}+\frac{1}{2}%
\mathcal{H}_{k}\ , \\
0 &=&\dot{\widehat{\sigma }}_{\langle ab\rangle }-D_{\langle a}\widehat{K}%
_{b\rangle }+\widehat{A}_{\langle b}\left( \widehat{K}_{a\rangle
}+L_{a\rangle }\right) -A_{\langle b}\left( \widehat{K}_{a\rangle
}-L_{a\rangle }\right) +\frac{\Theta }{3}\widehat{\sigma }_{ab}  \notag \\
&&+\left( \widehat{K}+\frac{\widehat{\Theta }}{3}\right) \sigma _{ab}+\omega
_{\langle a}\widehat{\omega }_{b\rangle }-\omega _{c\langle a}\widehat{%
\sigma }_{b\rangle }^{\,\,\,\,c}-\sigma _{\langle a}^{\,\,\,\,\,\,d}\widehat{%
\omega }_{b\rangle d}+\sigma _{c\langle a}\widehat{\sigma }_{b\rangle
}^{\,\,\,\,c}+\mathcal{F}_{ab}\ ,  \label{sigma_tild_dot} \\
0 &=&\widehat{K}_{\langle a\rangle }^{\prime }-\dot{\widehat{A}}_{\langle
a\rangle }-\left( \widehat{K}-\frac{\widehat{\Theta }}{3}\right) \widehat{K}%
_{a}-\left( \widehat{K}+\frac{\widehat{\Theta }}{3}\right) K_{a}-K\left(
A_{a}+\widehat{A}_{a}\right)  \notag \\
&&+\left( \widehat{\omega }_{ba}+\widehat{\sigma }_{ba}\right) \left( 
\widehat{K}^{b}-K^{b}\right) +\widehat{\mathcal{E}}_{a}-\frac{\widetilde{%
\kappa }^{2}}{3}\widetilde{q}_{a}\ ,  \label{Bdot} \\
0 &=&\widehat{\Theta }^{\prime }-D^{a}\widehat{A}_{a}+\frac{\widehat{\Theta }%
^{2}}{3}-\Theta K+\widehat{A}^{a}\widehat{A}_{a}-2\widehat{\omega }_{a}%
\widehat{\omega }^{a}+\widehat{\sigma }_{ab}\widehat{\sigma }^{ab}+\mathcal{E%
}  \notag \\
&&-\left( \widehat{K}^{a}K_{a}-\widehat{K}^{a}L_{a}-L^{a}K_{a}\right) +\frac{%
\widetilde{\Lambda }}{2}+\frac{\widetilde{\kappa }^{2}}{2}\left( \widetilde{%
\pi }+\widetilde{\rho }-\widetilde{p}\right) \ ,
\end{eqnarray}%
\begin{eqnarray}
0 &=&\widehat{\omega }_{\langle a\rangle }^{\prime }-\frac{1}{2}\varepsilon
_{a}^{\ \ cd}D_{c}\widehat{A}_{d}-K\omega _{a}+\frac{2\widehat{\Theta }}{3}%
\widehat{\omega }_{a}-\widehat{\sigma }_{ab}\widehat{\omega }^{b}-\frac{1}{2}%
\varepsilon _{a}^{\ \ cd}\left( K_{c}\widehat{K}_{d}+\widehat{K}%
_{c}L_{d}+K_{c}L_{d}\right) \ , \\
0 &=&\widehat{\sigma }_{\langle ab\rangle }^{\prime }-D_{\langle a}\widehat{A%
}_{b\rangle }+\frac{2\widehat{\Theta }}{3}\widehat{\sigma }_{ab}-K\sigma
_{ab}+\widehat{A}_{\langle a}\widehat{A}_{b\rangle }+\widehat{K}_{\langle
a}L_{b\rangle }  \notag \\
&&-\widehat{K}_{\langle a}K_{b\rangle }+L_{\langle a}K_{b\rangle }+\widehat{%
\omega }_{\langle a}\widehat{\omega }_{b\rangle }+\widehat{\sigma }%
_{c\langle a}\widehat{\sigma }_{b\rangle }^{\,\,\,\,c}+\widehat{\mathcal{E}}%
_{ab}+\frac{\widetilde{\kappa }^{2}}{3}\widetilde{\pi }_{ab}\ ,
\label{sigma_tild_column} \\
0 &=&D^{a}\widehat{\omega }_{a}+\widehat{A}_{a}\widehat{\omega }^{a}-\omega
^{a}\left( \widehat{K}_{a}+L_{a}\right) \ ,
\end{eqnarray}%
\begin{eqnarray}
0 &=&D_{\langle c}\widehat{\omega }_{k\rangle }+\varepsilon _{ab\langle
k}D^{b}\widehat{\sigma }_{c\rangle }^{\,\,\,\,a}-2\widehat{A}_{\langle c}%
\widehat{\omega }_{k\rangle }+2\widehat{K}_{\langle c}\omega _{k\rangle
}-L_{\langle c}\omega _{k\rangle }+\varepsilon _{\langle k}^{\ \
\,\,ab}\sigma _{c\rangle b}L_{a}+\widehat{\mathcal{H}}_{kc}\ ,
\label{call_Hab} \\
0 &=&D^{b}\widehat{\sigma }_{ab}+\varepsilon _{a}^{\ \ ck}D_{c}\widehat{%
\omega }_{k}-\frac{2}{3}D_{a}\widehat{\Theta }+\frac{2\Theta }{3}%
L_{a}-\varepsilon _{a}^{\ \ ck}L_{c}\omega _{k}  \notag \\
&&-2\varepsilon _{a}^{\ \ ck}\left( \widehat{A}_{c}\widehat{\omega }_{k}-%
\widehat{K}_{c}\omega _{k}\right) -\sigma _{ab}L^{b}-\mathcal{E}_{a}-\frac{2%
\widetilde{\kappa }^{2}}{3}\widetilde{\pi }_{a}\ .  \label{Div_sigma_tild}
\end{eqnarray}

\subsection{Conservations laws}

The twice-contracted 5d Bianchi identities imply $\widetilde{\nabla }^{a}%
\widetilde{T}_{ab}=0$, which can be decomposed into the projections taken
with $u,~n\,$\ and$~h$, respectively:%
\begin{eqnarray}
0 &=&\dot{\widetilde{\rho }}+\widetilde{q}^{\prime }+D^{a}\widetilde{q}_{a}+%
\widetilde{\rho }\left( K+\Theta \right) +K\widetilde{\pi }+\Theta 
\widetilde{p}+\widetilde{\pi }^{ab}\sigma _{ab}  \notag \\
&&-\widetilde{q}\left( 2\widehat{K}-\widehat{\Theta }\right) +\widetilde{q}%
^{a}\left( 2A_{a}-\widehat{A}_{a}\right) +\widetilde{\pi }^{a}\left(
L_{a}+K_{a}\right) \ ,  \label{rhodot} \\
0 &=&\dot{\widetilde{q}}+\widetilde{\pi }^{\prime }+D^{a}\widetilde{\pi }%
_{a}+\widetilde{\pi }\left( \widehat{\Theta }-\widehat{K}\right) -\widehat{K}%
\widetilde{\rho }-\widehat{\Theta }\widetilde{p}-\widetilde{\pi }^{ab}%
\widehat{\sigma }_{ab}  \notag \\
&&+\widetilde{q}\left( 2K+\Theta \right) -\widetilde{\pi }^{a}\left( 2%
\widehat{A}_{a}-A_{a}\right) -\widetilde{q}^{a}\left( \widehat{K}%
_{a}-L_{a}\right) \ ,  \label{qdot} \\
0 &=&\dot{\widetilde{q}}_{\langle k\rangle }+\widetilde{\pi }_{\langle
k\rangle }^{\prime }+D_{k}\widetilde{p}+D^{a}\widetilde{\pi }_{ak}+\frac{4%
\widehat{\Theta }}{3}\widetilde{\pi }_{k}+\frac{4\Theta }{3}\widetilde{q}%
_{k}-\widehat{K}\widetilde{\pi }_{k}+K\widetilde{q}_{k}+\widetilde{\rho }%
A_{k}+\widetilde{\pi }\widehat{A}_{k}  \notag \\
&&-\widetilde{p}\left( \widehat{A}_{k}-A_{k}\right) +\widetilde{q}^{a}\omega
_{ak}+\widetilde{q}^{a}\sigma _{ak}+\widetilde{\pi }^{a}\widehat{\omega }%
_{ak}+\widetilde{\pi }^{a}\widehat{\sigma }_{ak}+\widetilde{q}\left( 
\widehat{K}_{k}+K_{k}\right) -\widetilde{\pi }_{ak}\left( \widehat{A}%
^{a}-A^{a}\right) \ .  \label{qadot}
\end{eqnarray}%
The first of these equations is the continuity equation, as can be easily
verified in the homogeneous, isotropic case ($\widetilde{q}=\widetilde{q}%
^{a}=\widetilde{\pi }^{a}=\widetilde{\pi }^{ab}=0$ and $\Theta =3H$) and for 
$K=0$. The ensemble of the equations represent an incomplete set of
evolution equations for the 5d matter (there are no evolution equations for $%
\widetilde{p}$, $\widetilde{\pi }$, $\widetilde{\pi }_{a}$, $\widetilde{\pi }%
_{ab}$).

\subsection{Bianchi identities\label{Bianchi_sec}}

The equations independent from Eqs. (\ref{rhodot})-(\ref{qadot}) arising
from the 5d Bianchi identities are:%
\begin{eqnarray}
0 &=&\dot{\mathcal{E}}-D^{a}\widehat{\mathcal{E}}_{a}+\frac{4}{3}\Theta 
\mathcal{E}+\widehat{\mathcal{E}}_{ab}\sigma ^{ab}-\mathcal{F}_{ab}\widehat{%
\sigma }^{ab}+3\mathcal{H}_{a}\widehat{\omega }^{a}-\mathcal{E}_{a}\left( 2%
\widehat{K}^{a}+L^{a}\right) -2\widehat{\mathcal{E}}_{a}A^{a}  \notag \\
&&-\frac{\widetilde{\kappa }^{2}}{2}\left( \widetilde{\rho }-\widetilde{\pi }%
+\widetilde{p}\right) ^{\cdot }-\frac{2\widetilde{\kappa }^{2}}{3}D^{a}%
\widetilde{q}_{a}-\frac{2\widetilde{\kappa }^{2}}{3}\Theta \left( \widetilde{%
\rho }+\widetilde{p}\right)  \notag \\
&&-\frac{2\widetilde{\kappa }^{2}}{3}\widehat{\Theta }\widetilde{q}-\frac{4%
\widetilde{\kappa }^{2}}{3}\widetilde{q}_{a}A^{a}-\frac{2\widetilde{\kappa }%
^{2}}{3}\widetilde{\pi }_{a}\left( 2\widehat{K}^{a}+L^{a}\right) -\frac{2%
\widetilde{\kappa }^{2}}{3}\widetilde{\pi }_{ab}\sigma ^{ab}\ ,
\label{E_dot}
\end{eqnarray}%
\begin{eqnarray}
0 &=&\dot{\mathcal{E}}_{\langle k\rangle }+D^{a}\mathcal{F}_{ka}+\frac{1}{2}%
\varepsilon _{k}^{\ \ ab}D_{a}\mathcal{H}_{b}-\mathcal{E}_{ka}\left( 
\widehat{K}^{a}+L^{a}\right) -\widehat{\mathcal{E}}_{ka}L^{a}-\left( 
\widehat{K}+\frac{\widehat{\Theta }}{3}\right) \widehat{\mathcal{E}}_{k}+%
\frac{4}{3}\Theta \mathcal{E}_{k}  \notag \\
&&+\frac{4\mathcal{E}}{3}\widehat{K}_{k}+\mathcal{F}_{ka}A^{a}-\frac{1}{2}%
\left( \sigma _{ka}+\omega _{ka}\right) \mathcal{E}^{a}-\left( \widehat{%
\sigma }_{ka}-2\widehat{\omega }_{ka}\right) \widehat{\mathcal{E}}^{a}-2%
\mathcal{H}_{ka}\widehat{\omega }^{a}-\widehat{\mathcal{H}}_{ka}\omega ^{a} 
\notag \\
&&+\varepsilon _{k}^{\ \ ab}\widehat{\mathcal{H}}_{ca}\sigma
_{b}^{\,\,\,\,\,c}-\frac{3}{2}\varepsilon _{k}^{\ \ ab}\mathcal{H}_{a}A_{b}+%
\frac{2\widetilde{\kappa }^{2}}{3}\dot{\widetilde{P}}_{\langle k\rangle }+%
\frac{2\widetilde{\kappa }^{2}}{3}D_{k}\widetilde{q}+\frac{2\widetilde{%
\kappa }^{2}}{3}\left( \widetilde{\pi }-\widetilde{p}\right) \widehat{K}_{k}
\notag \\
&&+\frac{2\widetilde{\kappa }^{2}}{3}\left( \widetilde{\rho }+\widetilde{\pi 
}\right) L_{k}-\frac{2\widetilde{\kappa }^{2}}{3}\left( \widehat{K}+\frac{%
\widehat{\Theta }}{3}\right) \widetilde{q}_{k}-\frac{2\widetilde{\kappa }^{2}%
}{3}\left( \widehat{\omega }_{ka}+\widehat{\sigma }_{ka}\right) \widetilde{q}%
^{a}  \notag \\
&&-\frac{2\widetilde{\kappa }^{2}}{3}\widetilde{\pi }_{ka}\widehat{K}^{a}+%
\frac{2\widetilde{\kappa }^{2}}{3}\left( \omega _{ka}+\sigma _{ka}\right) 
\widetilde{\pi }^{a}+\frac{2\widetilde{\kappa }^{2}\Theta }{9}\widetilde{\pi 
}_{k}+\frac{2\widetilde{\kappa }^{2}\widetilde{q}}{3}A_{k}\ ,  \label{Ea_dot}
\end{eqnarray}%
\begin{eqnarray}
0 &=&\mathcal{\dot{H}}_{\langle k\rangle }-\varepsilon _{k}^{\ \
ab}D_{a}\allowbreak \mathcal{E}_{b}-\widehat{\mathcal{H}}_{ka}A^{a}+\mathcal{%
H}_{ka}\widehat{K}^{a}-\mathcal{E}_{ka}\widehat{\omega }^{a}-\frac{8}{3}%
\mathcal{E}\widehat{\omega }_{k}+\mathcal{F}_{ka}\omega ^{a}+\varepsilon
_{k}^{\ \ ab}\mathcal{F}_{ac}\sigma _{b}^{\,\,\,c}+\Theta \mathcal{H}_{k} 
\notag \\
&&+\frac{3}{2}\left( \omega _{ka}-\sigma _{ka}\right) \mathcal{H}%
^{a}+\varepsilon _{k}^{\ \ ab}\left[ \frac{3}{2}\left( \widehat{\mathcal{E}}%
_{a}\widehat{K}_{b}-A_{a}\mathcal{E}_{b}\right) -L_{a}\widehat{\mathcal{E}}%
_{b}\right] -\varepsilon _{k}^{\ \ ab}\mathcal{E}_{ac}\widehat{\sigma }%
_{b}^{\,\,\,c}+\frac{\widetilde{\kappa }^{2}}{3}\varepsilon _{k}^{\ \
ab}D_{a}\widetilde{\pi }_{b}  \notag \\
&&-\frac{\widetilde{\kappa }^{2}}{3}\varepsilon _{k}^{\ \ ab}\widetilde{q}%
_{a}L_{b}+\frac{2\widetilde{\kappa }^{2}}{3}\widetilde{q}\omega _{k}+\frac{%
\widetilde{\kappa }^{2}}{3}\varepsilon _{k}^{\ \ ab}\widetilde{\pi }_{ac}%
\widehat{\sigma }_{b}^{\,\,\,c}+\frac{2\widetilde{\kappa }^{2}}{3}\left( 
\widetilde{\pi }-\widetilde{p}\right) \widehat{\omega }_{k}+\frac{\widetilde{%
\kappa }^{2}}{3}\widetilde{\pi }_{ka}\widehat{\omega }^{a}\ ,
\label{CallHkdot} \\
0 &=&\dot{\widehat{\mathcal{E}}}_{\langle k\rangle }+\allowbreak \mathcal{E}%
_{\langle k\rangle }^{\prime }-D_{k}\mathcal{E}-\mathcal{E}_{ka}\widehat{A}%
^{a}-\widehat{\mathcal{E}}_{ka}A^{a}+K\widehat{\mathcal{E}}_{k}-\widehat{K}%
\mathcal{E}_{k}+\frac{2}{3}\left( \Theta \widehat{\mathcal{E}}_{k}+\widehat{%
\Theta }\mathcal{E}_{k}\right)  \notag \\
&&+2\left( \omega _{ka}+\sigma _{ka}\right) \widehat{\mathcal{E}}%
^{a}+2\left( \widehat{\omega }_{ka}+\widehat{\sigma }_{ka}\right) \mathcal{E}%
^{a}+\mathcal{F}_{ka}\left( K^{a}+\widehat{K}^{a}\right) +\frac{3}{2}%
\varepsilon _{kab}\mathcal{H}^{a}\left( \widehat{K}^{b}-K^{b}\right)  \notag
\\
&&-\frac{4}{3}\mathcal{E}\left( A_{k}-\widehat{A}_{k}\right) -\frac{%
\widetilde{\kappa }^{2}}{6}D_{k}\left( \widetilde{\rho }-\widetilde{\pi }+%
\widetilde{p}\right) +\frac{\widetilde{\kappa }^{2}}{3}D^{a}\widetilde{\pi }%
_{ak}  \notag \\
&&+\frac{2\widetilde{\kappa }^{2}}{9}\left( \widehat{\Theta }\widetilde{\pi }%
_{k}+\Theta \widetilde{q}_{k}\right) -\frac{\widetilde{\kappa }^{2}}{3}\left[
\left( 3\widehat{\omega }_{ka}+\widehat{\sigma }_{ka}\right) \widetilde{\pi }%
^{a}+\left( 3\omega _{ka}+\sigma _{ka}\right) \widetilde{q}^{a}\right] \ ,
\label{Epsilona_dot}
\end{eqnarray}%
\begin{eqnarray}
0 &=&\dot{\mathcal{E}}_{\langle kj\rangle }-\mathcal{F}_{\langle kj\rangle
}^{\prime }-\varepsilon _{ab\langle k}D^{a}\mathcal{H}_{j\rangle }^{\,\,\,b}-%
\frac{1}{2}D_{\langle k}\widehat{\mathcal{E}}_{j\rangle }+\left( K-\frac{%
\Theta }{3}\right) \widehat{\mathcal{E}}_{kj}+\left( K+\Theta \right) 
\mathcal{E}_{kj}  \notag \\
&&+\frac{2\mathcal{E}}{3}\sigma _{kj}+2\left( \widehat{K}-\frac{\widehat{%
\Theta }}{3}\right) \mathcal{F}_{kj}-\widehat{\mathcal{E}}_{\langle k}\left(
A_{j\rangle }-\frac{3}{2}\widehat{A}_{j\rangle }\right) +\frac{1}{2}\mathcal{%
E}_{\langle k}\left( 4\widehat{K}_{j\rangle }-L_{j\rangle }-3K_{k\rangle
}\right) \allowbreak  \notag \\
&&+\frac{3}{2}\mathcal{H}_{\langle k}\widehat{\omega }_{j\rangle }+2\widehat{%
\mathcal{E}}_{a\langle k}\sigma _{j\rangle }^{\,\,\,a}+\frac{3}{2}%
\varepsilon _{\langle j}^{\ \ \,\,\,ab}\widehat{\sigma }_{k\rangle a}%
\mathcal{H}_{b}+\mathcal{E}_{\langle k}^{\,\,\,\,\,\,a}\left( \omega
_{j\rangle a}-3\sigma _{j\rangle a}\right) -\mathcal{F}_{\langle
j}^{\,\,\,\,a}\left( \widehat{\omega }_{k\rangle a}-\widehat{\sigma }%
_{k\rangle a}\right)  \notag \\
&&+\varepsilon _{\langle k}^{\ \ \ \ ab}\mathcal{H}_{j\rangle a}\left(
2A_{b}-\widehat{A}_{b}\right) -\varepsilon _{\langle k}^{\ \ \ \ ab}\widehat{%
\mathcal{H}}_{j\rangle a}\left( K_{b}+L_{b}\right) \allowbreak +\frac{2%
\widetilde{\kappa }^{2}}{3}\dot{\widetilde{\pi }}_{\langle kj\rangle }+\frac{%
2\widetilde{\kappa }^{2}}{3}D_{\langle k}\widetilde{q}_{j\rangle }+\frac{2%
\widetilde{\kappa }^{2}}{9}\Theta \widetilde{\pi }_{jk}  \notag \\
&&+\frac{2\widetilde{\kappa }^{2}}{3}\widetilde{\pi }_{\langle k}\left( 3%
\widehat{K}_{j\rangle }+L_{j\rangle }\right) +2\widetilde{\kappa }^{2}%
\widetilde{q}_{\langle k}A_{j\rangle }+\frac{2\widetilde{\kappa }^{2}}{3}%
\widetilde{q}\widehat{\sigma }_{jk}+\frac{2\widetilde{\kappa }^{2}}{3}\left( 
\widetilde{\rho }+\widetilde{p}\right) \sigma _{jk}+\frac{2\widetilde{\kappa 
}^{2}}{3}\widetilde{\pi }_{\langle j}^{\,\,\,\,\,a}\left( \omega _{k\rangle
a}+\sigma _{k\rangle a}\right) \ ,  \label{Etildab_dot}
\end{eqnarray}%
\begin{eqnarray}
0 &=&\mathcal{\dot{F}}_{\langle kj\rangle }-\mathcal{E}_{\langle kj\rangle
}^{\prime }+D_{\langle k}\mathcal{E}_{j\rangle }+\left( \widehat{K}-\frac{%
\widehat{\Theta }}{3}\right) \mathcal{E}_{kj}+\widehat{K}\widehat{\mathcal{E}%
}_{kj}+\frac{4\mathcal{E}}{3}\widehat{\sigma }_{kj}+\left( 2K+\frac{\Theta }{%
3}\right) \mathcal{F}_{kj}  \notag \\
&&-\frac{3}{2}\mathcal{H}_{\langle k}\omega _{j\rangle }-\widehat{\mathcal{E}%
}_{\langle j}\left( \frac{3}{2}\widehat{K}_{k\rangle }-L_{k\rangle
}-K_{k\rangle }\right) \allowbreak +\mathcal{F}_{\langle
j}^{\,\,\,\,\,\,a}\left( \omega _{k\rangle a}+\sigma _{k\rangle a}\right) -2%
\mathcal{E}_{\langle k}\left( \widehat{A}_{j\rangle }-\frac{3}{4}A_{j\rangle
}\right)  \notag \\
&&-\mathcal{E}_{\,\langle j}^{\,\,\,\,\,\,\,a}\left( \widehat{\omega }%
_{k\rangle a}+\widehat{\sigma }_{k\rangle a}\right) +\varepsilon _{\langle
k}^{\ \ \ ab}\mathcal{H}_{j\rangle a}\left( \widehat{K}_{b}-2K_{b}\right)
+\varepsilon _{\langle j}^{\ \ \ ab}\widehat{\mathcal{H}}_{k\rangle a}A_{b}+%
\frac{3}{2}\varepsilon _{\langle j}^{\ \ ab}\sigma _{k\rangle a}\mathcal{H}%
_{b}+\frac{\widetilde{\kappa }^{2}}{3}\widetilde{\pi }_{\langle kj\rangle
}^{\prime }  \notag \\
&&-\frac{\widetilde{\kappa }^{2}}{3}D_{\langle k}\widetilde{\pi }_{j\rangle
}+\frac{\widetilde{\kappa }^{2}}{9}\widehat{\Theta }\widetilde{\pi }_{kj}-%
\frac{\widetilde{\kappa }^{2}}{3}\left( \widetilde{\pi }-\widetilde{p}%
\right) \widehat{\sigma }_{kj}-\frac{\widetilde{\kappa }^{2}}{3}\widetilde{q}%
\sigma _{kj}  \notag \\
&&+\frac{\widetilde{\kappa }^{2}}{3}\widetilde{\pi }_{\langle
j}^{\,\,\,\,\,a}\left( \widehat{\omega }_{k\rangle a}+\widehat{\sigma }%
_{k\rangle a}\right) +\frac{2\widetilde{\kappa }^{2}}{3}\widetilde{\pi }%
_{\langle j}\widehat{A}_{k\rangle }+\frac{\widetilde{\kappa }^{2}}{3}%
\widetilde{q}_{j}\left( 2K_{k\rangle }-L_{k\rangle }\right) \ ,
\end{eqnarray}%
\begin{eqnarray}
0 &=&\dot{\mathcal{H}}_{\langle kj\rangle }+\varepsilon _{ab\langle k}D^{a}%
\mathcal{E}_{j\rangle }^{\,\,\,b}+\widehat{K}\widehat{\mathcal{H}}%
_{kj}+\Theta \mathcal{H}_{kj}-\frac{3}{2}\mathcal{H}_{\langle k}\widehat{K}%
_{j\rangle }-\frac{3}{2}\widehat{\mathcal{E}}_{\langle j}\omega _{k\rangle
}-\left( \omega _{a\langle k}+3\sigma _{a\langle k}\right) \mathcal{H}%
_{j\rangle }^{\,\,\,a}  \notag \\
&&+\varepsilon _{\langle k}^{\ \ \ ab}\mathcal{F}_{j\rangle a}\left( 
\widehat{K}_{b}+2L_{b}\right) +\varepsilon _{\langle k}^{\ \ \ ab}\left( 
\widehat{\mathcal{E}}_{j\rangle a}-2\mathcal{E}_{j\rangle a}\right) A_{b}-%
\frac{1}{2}\varepsilon _{\langle k}^{\ \ \ ab}\sigma _{j\rangle a}\widehat{%
\mathcal{E}}_{b}+\varepsilon _{\langle k}^{\ \ \ ab}\widehat{\sigma }%
_{j\rangle a}\mathcal{E}_{b}  \notag \\
&&+3\mathcal{E}_{\langle j}\widehat{\omega }_{k\rangle }-\frac{\widetilde{%
\kappa }^{2}}{3}\varepsilon _{ab\langle k}D^{a}\widetilde{\pi }_{j\rangle
}^{\,\,\,b}-\widetilde{\kappa }^{2}\left( \widetilde{q}_{\langle j}\omega
_{k\rangle }+\widetilde{\pi }_{\langle j}\widehat{\omega }_{k\rangle
}\right) -\frac{\widetilde{\kappa }^{2}}{3}\varepsilon _{\langle k}^{\ \ \ \
ab}\left( \sigma _{j\rangle a}\widetilde{q}_{b}+\widehat{\sigma }_{j\rangle
a}\widetilde{\pi }_{b}\right) \ ,  \label{Hab_dot}
\end{eqnarray}%
\begin{eqnarray}
0 &=&\dot{\widehat{\mathcal{E}}}_{\langle kj\rangle }-\mathcal{F}_{\langle
kj\rangle }^{\prime }-D_{\langle k}\widehat{\mathcal{E}}_{j\rangle }+\left(
K+\frac{\Theta }{3}\right) \widehat{\mathcal{E}}_{kj}+K\mathcal{E}_{kj}+%
\frac{4\mathcal{E}}{3}\sigma _{kj}+\left( 2\widehat{K}-\frac{\widehat{\Theta 
}}{3}\right) \mathcal{F}_{kj}-\frac{3}{2}\mathcal{H}_{\langle k}\widehat{%
\omega }_{j\rangle }  \notag \\
&&-\mathcal{F}_{\langle j}^{\,\,\,\,\,a}\left( \widehat{\omega }_{k\rangle
a}+\widehat{\sigma }_{k\rangle a}\right) -\mathcal{E}_{\langle j}\left( 
\frac{3}{2}K_{k\rangle }+L_{k\rangle }-\widehat{K}_{k\rangle }\right)
\allowbreak -2\widehat{\mathcal{E}}_{\langle k}\left( A_{j\rangle }-\frac{3}{%
4}\widehat{A}_{j\rangle }\right) +\widehat{\mathcal{E}}_{\langle
j}^{\,\,\,\,\,a}\left( \omega _{k\rangle a}+\sigma _{k\rangle a}\right) 
\notag \\
&&-\varepsilon _{\langle k}^{\ \ \ ab}\widehat{\mathcal{H}}_{j\rangle
a}\left( K_{b}-2\widehat{K}_{b}\right) -\varepsilon _{\langle j}^{\ \ \ ab}%
\mathcal{H}_{k\rangle a}\widehat{A}_{b}+\frac{3}{2}\varepsilon _{\langle
j}^{\ \ \,\,ab}\widehat{\sigma }_{k\rangle a}\mathcal{H}_{b}+\frac{%
\widetilde{\kappa }^{2}}{3}\dot{\widetilde{\pi }}_{\langle kj\rangle }+\frac{%
\widetilde{\kappa }^{2}}{3}D_{\langle k}\widetilde{q}_{j\rangle }+\frac{%
\widetilde{\kappa }^{2}}{3}\left( \widetilde{\rho }+\widetilde{p}\right)
\sigma _{kj}  \notag \\
&&+\frac{\widetilde{\kappa }^{2}\widetilde{q}}{3}\widehat{\sigma }_{kj}+%
\frac{\widetilde{\kappa }^{2}\Theta }{9}\widetilde{\pi }_{kj}+\frac{2%
\widetilde{\kappa }^{2}}{3}\widetilde{q}_{\langle j}A_{k\rangle }+\frac{%
\widetilde{\kappa }^{2}}{3}\widetilde{\pi }_{\langle j}\left( 2\widehat{K}%
_{k\rangle }+L_{k\rangle }\right) +\frac{\widetilde{\kappa }^{2}}{3}%
\widetilde{\pi }_{\langle j}^{\,\,\,\,\,a}\left( \omega _{k\rangle a}+\sigma
_{k\rangle a}\right) \ ,  \label{Epsilonab_dot} \\
0 &=&\dot{\widehat{\mathcal{H}}}_{\langle kj\rangle }+\varepsilon
_{ab\langle k}D^{a}\mathcal{F}_{j\rangle }^{b}-\frac{1}{2}D_{\langle j}%
\mathcal{H}_{k\rangle }+\left( \widehat{K}+\frac{\widehat{\Theta }}{3}%
\right) \mathcal{H}_{kj}+\frac{2\Theta }{3}\widehat{\mathcal{H}}_{kj}-2%
\mathcal{H}_{a\langle k}\widehat{\sigma }_{j\rangle }^{\,\,\,a}  \notag \\
&&+\widehat{\mathcal{H}}_{\langle j}^{\,\,\,\,\,a}\left( \omega _{k\rangle
a}-\sigma _{k\rangle a}\right) +\frac{3}{2}\mathcal{E}_{\langle j}\omega
_{k\rangle }-3\widehat{\omega }_{\langle k}\widehat{\mathcal{E}}_{j\rangle
}-\varepsilon _{\langle k}^{\ \ \ ab}\mathcal{E}_{j\rangle a}\left( \widehat{%
K}_{b}-L_{b}\right)  \notag \\
&&-\frac{3}{2}\mathcal{H}_{\langle k}A_{j\rangle }+\varepsilon _{\langle
k}^{\ \ \ ab}\widehat{\mathcal{E}}_{j\rangle a}\left( 2\widehat{K}%
_{b}+L_{b}\right) +\frac{3}{2}\varepsilon _{\langle k}^{\ \ \ ab}\sigma
_{j\rangle a}\mathcal{E}_{b}-\varepsilon _{\langle k}^{\ \ \ ab}\mathcal{F}%
_{j\rangle a}A_{b}\ ,  \label{CallHabhatdot} \\
0 &=&\mathcal{E}^{\prime }+D^{a}\mathcal{E}_{a}+\frac{4}{3}\widehat{\Theta }%
\mathcal{E}-2\mathcal{E}_{a}\widehat{A}^{a}-\widehat{\mathcal{E}}_{a}\left(
2K^{a}-L^{a}\right) +3\omega _{a}\mathcal{H}^{a}+\mathcal{F}_{ab}\sigma ^{ab}
\notag \\
&&-\mathcal{E}_{ab}\widehat{\sigma }^{ab}-\frac{\widetilde{\kappa }^{2}}{2}%
\left( \widetilde{\rho }-\widetilde{\pi }+\widetilde{p}\right) ^{\prime }+%
\frac{2\widetilde{\kappa }^{2}}{3}D^{a}\widetilde{\pi }_{a}+\frac{2%
\widetilde{\kappa }^{2}\Theta }{3}\widetilde{q}  \notag \\
&&-\frac{2\widetilde{\kappa }^{2}\widehat{\Theta }}{3}\left( \widetilde{p}-%
\widetilde{\pi }\right) -\frac{4\widetilde{\kappa }^{2}}{3}\widetilde{\pi }%
_{a}\widehat{A}^{a}-\frac{2\widetilde{\kappa }^{2}}{3}\widetilde{q}%
_{a}\left( 2K^{a}-L^{a}\right) -\frac{2\widetilde{\kappa }^{2}}{3}\widetilde{%
\pi }_{ab}\widehat{\sigma }^{ab}\ ,
\end{eqnarray}%
\begin{eqnarray}
0 &=&\widehat{\mathcal{E}}_{\langle k\rangle }^{\prime }-D^{a}\mathcal{F}%
_{ka}+\frac{1}{2}\varepsilon _{k}^{\ \ \ ab}D_{a}\mathcal{H}_{b}-\widehat{%
\mathcal{E}}_{ka}\left( K^{a}-L^{a}\right) +\mathcal{E}_{ka}L^{a}-\frac{1}{2}%
\left( \widehat{\sigma }_{ka}+\widehat{\omega }_{ka}\right) \widehat{%
\mathcal{E}}^{a}  \notag \\
&&-\left( \sigma _{ka}-2\omega _{ka}\right) \mathcal{E}^{a}+2\widehat{%
\mathcal{H}}_{ka}\omega ^{a}+\mathcal{H}_{ka}\widehat{\omega }^{a}+\mathcal{F%
}_{ka}\widehat{A}^{a}+\left( K-\frac{\Theta }{3}\right) \mathcal{E}_{k}-%
\frac{4\mathcal{E}}{3}K_{k}  \notag \\
&&+\frac{4\widehat{\Theta }}{3}\widehat{\mathcal{E}}_{k}-\varepsilon _{k}^{\
\ \ ab}\mathcal{H}_{ca}\widehat{\sigma }_{b}^{\,\,\,c}+\frac{3}{2}%
\varepsilon _{k}^{\ \ \ ab}\mathcal{H}_{a}\widehat{A}_{b}+\frac{2\widetilde{%
\kappa }^{2}}{3}\widetilde{q}_{\langle k\rangle }^{\prime }-\frac{2%
\widetilde{\kappa }^{2}}{3}D_{k}\widetilde{q}  \notag \\
&&+\frac{2\widetilde{\kappa }^{2}}{3}\left( \widetilde{\rho }+\widetilde{p}%
\right) K_{k}-\frac{2\widetilde{\kappa }^{2}}{3}\left( \widetilde{\pi }+%
\widetilde{\rho }\right) L_{k}+\frac{2\widetilde{\kappa }^{2}}{3}\left( K-%
\frac{\Theta }{3}\right) \widetilde{\pi }_{k}+\frac{2\widetilde{\kappa }^{2}%
\widehat{\Theta }}{9}\widetilde{q}_{k}  \notag \\
&&-\frac{2\widetilde{\kappa }^{2}}{3}\left( \omega _{ka}+\sigma _{ka}\right) 
\widetilde{\pi }^{a}+\frac{2\widetilde{\kappa }^{2}}{3}\left( \widehat{%
\omega }_{ka}+\widehat{\sigma }_{ka}\right) \widetilde{q}^{a}+\frac{2%
\widetilde{\kappa }^{2}\widetilde{q}}{3}\widehat{A}_{k}+\frac{2\widetilde{%
\kappa }^{2}}{3}\widetilde{\pi }_{ka}K^{a}\ ,
\end{eqnarray}%
\begin{eqnarray}
0 &=&\mathcal{H}_{\langle k\rangle }^{\prime }+\varepsilon _{k}^{\ \
ab}D_{a}\allowbreak \widehat{\mathcal{E}}_{b}+\mathcal{H}_{ka}\widehat{A}%
^{a}-\widehat{\mathcal{H}}_{ka}K^{a}+\widehat{\mathcal{E}}_{ka}\omega ^{a}-%
\mathcal{F}_{ka}\widehat{\omega }^{a}+\varepsilon _{k}^{\ \ ab}\widehat{%
\mathcal{E}}_{ac}\sigma _{b}^{\,\,\,c}-\varepsilon _{k}^{\ \ ab}\mathcal{F}%
_{ac}\widehat{\sigma }_{b}^{\,\,\,c}+\frac{3}{2}\left( \widehat{\omega }%
_{ka}-\widehat{\sigma }_{ka}\right) \mathcal{H}^{a}  \notag \\
&&+\widehat{\Theta }\mathcal{H}_{k}-\frac{8}{3}\mathcal{E}\omega
_{k}-\varepsilon _{k}^{\ \ ab}\left[ \frac{3}{2}\widehat{A}_{a}\widehat{%
\mathcal{E}}_{b}-\frac{3}{2}\mathcal{E}_{a}K_{b}-L_{a}\mathcal{E}_{b}\right]
-\frac{\widetilde{\kappa }^{2}}{3}\varepsilon _{k}^{\ \ ab}D_{a}\widetilde{q}%
_{b}+\frac{\widetilde{\kappa }^{2}}{3}\varepsilon _{k}^{\ \ ab}\widetilde{%
\pi }_{a}L_{b}  \notag \\
&&-\frac{2\widetilde{\kappa }^{2}\widetilde{q}}{3}\widehat{\omega }_{k}+%
\frac{\widetilde{\kappa }^{2}}{3}\varepsilon _{k}^{\ \ ab}\widetilde{\pi }%
_{ac}\sigma _{b}^{\,\,\,c}-\frac{2\widetilde{\kappa }^{2}}{3}\left( 
\widetilde{\rho }+\widetilde{p}\right) \omega _{k}+\frac{\widetilde{\kappa }%
^{2}}{3}\widetilde{\pi }_{ka}\omega ^{a}\ ,
\end{eqnarray}%
\begin{eqnarray}
0 &=&\mathcal{E}_{\langle k\rangle }^{\prime }+D^{a}\widehat{\mathcal{E}}%
_{ka}-\frac{2}{3}D_{k}\mathcal{E}+\widehat{\Theta }\mathcal{E}_{k}+\frac{4%
\mathcal{E}}{3}\widehat{A}_{k}-\mathcal{E}_{ka}\widehat{A}^{a}-3\widehat{%
\mathcal{H}}_{ka}\widehat{\omega }^{a}+\left( K-\frac{2}{3}\Theta \right) 
\widehat{\mathcal{E}}_{k}+\mathcal{F}_{ka}\left( K^{a}-2L^{a}\right)  \notag
\\
&&+\frac{1}{2}\left( 3\widehat{\sigma }_{ka}+\widehat{\omega }_{ka}\right) 
\mathcal{E}^{a}+\left( \sigma _{ka}+3\omega _{ka}\right) \widehat{\mathcal{E}%
}^{a}-\frac{3}{2}\varepsilon _{k}^{\ \ ab}\mathcal{H}_{a}K_{b}+\varepsilon
_{k}^{\ \ ab}\widehat{\mathcal{H}}_{ca}\widehat{\sigma }_{b}^{\,\,\,c}+\frac{%
2\widetilde{\kappa }^{2}}{3}\widetilde{\pi }_{\langle k\rangle }^{\prime }+%
\frac{\widetilde{\kappa }^{2}}{3}D^{a}\widetilde{\pi }_{ka}  \notag \\
&&-\frac{\widetilde{\kappa }^{2}}{3}D_{k}\left( \widetilde{\rho }+\widetilde{%
\pi }-\widetilde{p}\right) +\frac{2\widetilde{\kappa }^{2}\widetilde{q}}{3}%
\left( K_{k}-2L_{k}\right) +\frac{2\widetilde{\kappa }^{2}}{3}\left( K+\frac{%
\Theta }{3}\right) \widetilde{q}_{k}-\frac{2\widetilde{\kappa }^{2}}{3}%
\widetilde{\pi }_{ka}\widehat{A}^{a}  \notag \\
&&+\frac{2\widetilde{\kappa }^{2}}{3}\left( \widetilde{\pi }-\widetilde{p}%
\right) \widehat{A}_{k}+\frac{2\widetilde{\kappa }^{2}}{3}\widehat{\Theta }%
\widetilde{\pi }_{k}+\frac{\widetilde{\kappa }^{2}}{3}\left[ \left( \widehat{%
\omega }_{ka}+3\widehat{\sigma }_{ka}\right) \widetilde{\pi }^{a}-\left(
3\omega _{ka}+\sigma _{ka}\right) \widetilde{q}^{a}\right] \ ,
\label{Ea_column}
\end{eqnarray}%
\begin{eqnarray}
0 &=&\widehat{\mathcal{E}}_{\langle kj\rangle }^{\prime }-\mathcal{E}%
_{\langle kj\rangle }^{\prime }+\varepsilon _{ab\langle k}D^{a}\widehat{%
\mathcal{H}}_{j\rangle }^{\,\,\,b}+\frac{1}{2}D_{\langle k}\mathcal{E}%
_{j\rangle }+\frac{2\mathcal{E}}{3}\widehat{\sigma }_{jk}-\frac{2\widehat{%
\Theta }}{3}\mathcal{E}_{kj}\allowbreak +\widehat{\Theta }\widehat{\mathcal{E%
}}_{kj}-\mathcal{E}_{\langle k}\widehat{A}_{j\rangle }  \notag \\
&&-3\mathcal{H}_{\langle k}\omega _{j\rangle }-\frac{1}{2}\widehat{\mathcal{E%
}}_{\langle k}\left( 2K_{j\rangle }-L_{j\rangle }\right) +2\mathcal{F}%
_{a\langle k}\sigma _{j\rangle }^{\,\,\,a}-\mathcal{E}_{\langle
k}^{\,\,\,\,\,\,a}\left( \widehat{\omega }_{j\rangle a}-\widehat{\sigma }%
_{j\rangle a}\right) +\widehat{\mathcal{E}}_{\langle
k}^{\,\,\,\,\,\,a}\left( \widehat{\omega }_{j\rangle a}-3\widehat{\sigma }%
_{j\rangle a}\right)  \notag \\
&&-\frac{\Theta }{3}\mathcal{F}_{kj}+2\varepsilon _{\langle k}^{\ \ \ \ ab}%
\widehat{\mathcal{H}}_{j\rangle a}\widehat{A}_{b}-\varepsilon _{\langle
k}^{\ \ \ \ ab}\mathcal{H}_{j\rangle a}\left( 2K_{b}-L_{b}\right) -\frac{%
\widetilde{\kappa }^{2}}{3}\widetilde{\pi }_{\langle kj\rangle }^{\prime }-%
\frac{\widetilde{\kappa }^{2}}{3}\widetilde{q}_{\langle k}\left(
4K_{j\rangle }-L_{j\rangle }\right)  \notag \\
&&+\frac{\widetilde{\kappa }^{2}}{3}D_{\langle k}\widetilde{\pi }_{j\rangle
}-\frac{4\widetilde{\kappa }^{2}}{3}\widetilde{\pi }_{\langle k}\widehat{A}%
_{j\rangle }+\frac{\widetilde{\kappa }^{2}}{3}\widetilde{q}\sigma _{jk}+%
\frac{\widetilde{\kappa }^{2}}{3}\left( \widetilde{\pi }-\widetilde{p}%
\right) \widehat{\sigma }_{jk}-\frac{\widetilde{\kappa }^{2}\widehat{\Theta }%
}{9}\widetilde{\pi }_{jk}-\frac{\widetilde{\kappa }^{2}}{3}\widetilde{\pi }%
_{\langle k}^{\,\,\,\,\,\,a}\left( \widehat{\omega }_{j\rangle a}+\widehat{%
\sigma }_{j\rangle a}\right) \ , \\
0 &=&\mathcal{H}_{\langle kj\rangle }^{\prime }+\varepsilon _{ab\langle
k}D^{a}\mathcal{F}_{j\rangle }^{\,\,\,\,b}+\frac{1}{2}D_{\langle j}\mathcal{H%
}_{k\rangle }-\left( K-\frac{\Theta }{3}\right) \widehat{\mathcal{H}}_{kj}+%
\frac{2\widehat{\Theta }}{3}\mathcal{H}_{kj}-\frac{3}{2}\mathcal{H}_{\langle
k}\widehat{A}_{j\rangle }  \notag \\
&&-\frac{3}{2}\widehat{\mathcal{E}}_{\langle j}\widehat{\omega }_{k\rangle
}+3\omega _{\langle k}\mathcal{E}_{j\rangle }-2\widehat{\mathcal{H}}%
_{a\langle k}\sigma _{j\rangle }^{\,\,\,a}+\mathcal{H}_{\langle
j}^{\,\,\,\,\,a}\left( \widehat{\omega }_{k\rangle a}-\widehat{\sigma }%
_{k\rangle a}\right) +\varepsilon _{\langle k}^{\ \ \ ab}\widehat{\mathcal{E}%
}_{j\rangle a}\left( K_{b}+L_{b}\right)  \notag \\
&&-\varepsilon _{\langle k}^{\ \ \ ab}\mathcal{E}_{j\rangle a}\left(
2K_{b}-L_{b}\right) -\frac{3}{2}\varepsilon _{\langle k}^{\ \ \ ab}\widehat{%
\sigma }_{j\rangle a}\widehat{\mathcal{E}}_{b}+\varepsilon _{\langle k}^{\ \
\ ab}\mathcal{F}_{j\rangle a}\widehat{A}_{b}\ , \\
0 &=&\widehat{\mathcal{H}}_{\langle kj\rangle }^{\prime }+\varepsilon
_{ab\langle k}D^{a}\widehat{\mathcal{E}}_{j\rangle }^{\,\,\,\,b}-\frac{3}{2}%
\mathcal{H}_{\langle k}K_{j\rangle }-K\mathcal{H}_{kj}+\widehat{\Theta }%
\widehat{\mathcal{H}}_{kj}-\left( \widehat{\omega }_{a\langle k}+3\widehat{%
\sigma }_{a\langle k}\right) \widehat{\mathcal{H}}_{j\rangle }^{\,\,\,a}+%
\frac{3}{2}\mathcal{E}_{\langle j}\widehat{\omega }_{k\rangle }  \notag \\
&&-\varepsilon _{\langle k}^{\ \ \ ab}\left( \mathcal{E}_{j\rangle a}-2%
\widehat{\mathcal{E}}_{j\rangle a}\right) \widehat{A}_{b}+\frac{1}{2}%
\varepsilon _{\langle k}^{\ \ \ ab}\widehat{\sigma }_{j\rangle a}\mathcal{E}%
_{b}-\varepsilon _{\langle k}^{\ \ \ ab}\sigma _{j\rangle a}\widehat{%
\mathcal{E}}_{b}-\varepsilon _{\langle k}^{\ \ \ ab}\mathcal{F}_{j\rangle
a}\left( K_{b}-2L_{b}\right)  \notag \\
&&-3\widehat{\mathcal{E}}_{\langle j}\omega _{k\rangle }+\frac{\widetilde{%
\kappa }^{2}}{3}\varepsilon _{ab\langle k}D^{a}\widetilde{\pi }_{j\rangle
}^{\,\,\,b}+\widetilde{\kappa }^{2}\left( \widetilde{q}_{\langle j}\omega
_{k\rangle }+\widetilde{\pi }_{\langle j}\widehat{\omega }_{k\rangle
}\right) +\frac{\widetilde{\kappa }^{2}}{3}\varepsilon _{\langle k}^{\ \ \ \
ab}\left( \sigma _{j\rangle a}\widetilde{q}_{b}+\widehat{\sigma }_{j\rangle
a}\widetilde{\pi }_{b}\right) \ , \\
0 &=&D^{k}\mathcal{H}_{k}-\widehat{\mathcal{H}}_{ab}\sigma ^{ab}+\mathcal{H}%
_{ab}\widehat{\sigma }^{ab}-3\mathcal{E}_{a}\omega ^{a}-3\widehat{\mathcal{E}%
}_{a}\widehat{\omega }^{a}\ ,  \label{sigma_a_const}
\end{eqnarray}%
\begin{eqnarray}
0 &=&D^{a}\widehat{\mathcal{E}}_{ak}-D^{a}\mathcal{E}_{ak}-\frac{1}{3}D_{k}%
\mathcal{E}-\frac{\widehat{\Theta }}{3}\mathcal{E}_{k}-\frac{\Theta }{3}%
\widehat{\mathcal{E}}_{k}+3\mathcal{H}_{ka}\omega ^{a}-\varepsilon _{k}^{\ \
ab}\mathcal{H}_{ac}\sigma _{b}^{\,\,\,\,c}  \notag \\
&&+\frac{1}{2}\left[ \left( 3\widehat{\omega }_{ka}+\widehat{\sigma }%
_{ka}\right) \mathcal{E}^{a}+\left( 3\omega _{ka}+\sigma _{ka}\right) 
\widehat{\mathcal{E}}^{a}\right] -3\widehat{\mathcal{H}}_{ka}\widehat{\omega 
}^{a}+\varepsilon _{k}^{\ \ ab}\widehat{\mathcal{H}}_{ac}\widehat{\sigma }%
_{b}^{\,\,\,\,c}-\frac{\widetilde{\kappa }^{2}}{3}D^{a}\widetilde{\pi }_{ak}
\notag \\
&&+\frac{\widetilde{\kappa }^{2}}{6}D_{k}\left( \widetilde{\rho }-\widetilde{%
\pi }+\widetilde{p}\right) -\frac{2\widetilde{\kappa }^{2}\Theta }{9}%
\widetilde{q}_{k}-\frac{2\widetilde{\kappa }^{2}\widehat{\Theta }}{9}%
\widetilde{\pi }_{k}+\widetilde{\kappa }^{2}\widetilde{q}^{a}\omega _{ka}+%
\frac{\widetilde{\kappa }^{2}}{3}\widetilde{q}^{a}\sigma _{ka}+\widetilde{%
\kappa }^{2}\widetilde{\pi }^{a}\widehat{\omega }_{ka}+\frac{\widetilde{%
\kappa }^{2}}{3}\widetilde{\pi }^{a}\widehat{\sigma }_{ka}\ ,
\label{Div_Eab} \\
0 &=&D^{a}\mathcal{H}_{ak}+\frac{1}{2}\varepsilon _{k}^{\ \ ab}D_{a}\widehat{%
\mathcal{E}}_{b}-\frac{\widehat{\Theta }}{3}\mathcal{H}_{k}-\frac{4\mathcal{E%
}}{3}\omega _{k}-\widehat{\mathcal{H}}_{ka}L^{a}-2\mathcal{F}_{ka}\widehat{%
\omega }^{a}-\left( \widehat{\sigma }_{ka}-2\widehat{\omega }_{ka}\right) 
\mathcal{H}^{a}  \notag \\
&&-\left( \widehat{\mathcal{E}}_{ka}-3\mathcal{E}_{ka}\right) \omega
^{a}-\varepsilon _{k}^{\ \ ab}\left( \mathcal{E}_{ac}-\widehat{\mathcal{E}}%
_{ac}\right) \sigma _{b}^{\,\,\,\,c}-\frac{1}{2}\varepsilon _{k}^{\ \ ab}%
\mathcal{E}_{a}L_{b}+\frac{\widetilde{\kappa }^{2}}{3}\varepsilon _{k}^{\ \
ab}D_{a}\widetilde{q}_{b}  \notag \\
&&-\frac{\widetilde{\kappa }^{2}}{3}\varepsilon _{k}^{\ \ ab}\widetilde{\pi }%
_{a}L_{b}+\frac{2\widetilde{\kappa }^{2}}{3}\widetilde{q}\widehat{\omega }%
_{k}+\frac{2\widetilde{\kappa }^{2}}{3}\left( \widetilde{\rho }+\widetilde{p}%
\right) \omega _{k}-\frac{\widetilde{\kappa }^{2}}{3}\widetilde{\pi }%
_{ka}\omega ^{a}-\frac{\widetilde{\kappa }^{2}}{3}\varepsilon _{k}^{\ \ ab}%
\widetilde{\pi }_{a}^{\,\,\,\,c}\sigma _{bc}\ ,  \label{Div_Hab} \\
0 &=&D^{a}\widehat{\mathcal{H}}_{ak}-\frac{1}{2}\varepsilon _{k}^{\ \
ab}D_{a}\mathcal{E}_{b}-\frac{\Theta }{3}\mathcal{H}_{k}-\mathcal{H}%
_{ka}L^{a}+2\mathcal{F}_{ka}\omega ^{a}+\left( \mathcal{E}_{ka}-3\widehat{%
\mathcal{E}}_{ka}\right) \widehat{\omega }^{a}-\frac{4\mathcal{E}}{3}%
\widehat{\omega }_{k}  \notag \\
&&-\left( \sigma _{ka}-2\omega _{ka}\right) \mathcal{H}^{a}-\frac{1}{2}%
\varepsilon _{k}^{\ \ ab}L_{a}\widehat{\mathcal{E}}_{b}+\varepsilon _{k}^{\
\ ab}\widehat{\sigma }_{b}^{\,\,\,\,c}\left( \widehat{\mathcal{E}}_{ca}-%
\mathcal{E}_{ca}\right) -\frac{\widetilde{\kappa }^{2}}{3}\varepsilon
_{k}^{\ \ ab}D_{a}\widetilde{\pi }_{b}  \notag \\
&&-\frac{\widetilde{\kappa }^{2}}{3}\varepsilon _{k}^{\ \ ab}L_{a}\widetilde{%
q}_{b}-\frac{2\widetilde{\kappa }^{2}\widetilde{q}}{3}\omega _{k}-\frac{2%
\widetilde{\kappa }^{2}}{3}\left( \widetilde{\pi }-\widetilde{p}\right) 
\widehat{\omega }_{k}-\frac{\widetilde{\kappa }^{2}}{3}\widetilde{\pi }_{ka}%
\widehat{\omega }^{a}+\frac{\widetilde{\kappa }^{2}}{3}\varepsilon _{k}^{\ \
ab}\widetilde{\pi }_{cb}\widehat{\sigma }_{a}^{\,\,\,\,c}\ .
\label{DivCallHabhat}
\end{eqnarray}

\section{3+1 gravitational dynamics on the brane\label{brane_sec}}

In this section we consider distributional energy-momentum tensor sources on
the brane, in addition to the regular energy-momentum tensor $\widetilde{T}%
_{ab}$. Such a distributional source comes together with a discontinuity in
the extrinsic curvature, as related by the Lanczos equation.

\subsection{The Lanczos equation\label{lanczos_sec}}

The extrinsic curvature of the brane is $^{\left( 4\right) }\!K_{ab}=\nabla
_{(c}n_{d)}=g_{(c}^{a}g_{d)}^{b}\widetilde{\nabla }_{a}n_{b}$, equal to the
symmetrized version of the last four terms of Eq. (\ref{nablan}). Replacing $%
\widehat{K}_{ab}$ by the expression (\ref{Kabhat}), and specializing to the
brane cf. Eqs. (\ref{brane}), the extrinsic curvature is expressed as 
\begin{equation}
^{\left( 4\right) }\!K_{ab}=\ \widehat{K}u_{a}u_{b}-2u_{(a}\widehat{K}_{b)}+%
\frac{\widehat{\Theta }}{3}h_{ab}+\widehat{\sigma }_{ab}~.  \label{Kab4}
\end{equation}%
As we approach the brane from left or right, the limiting values of the
extrinsic curvature could be different, according to the embedding and 5d
metric in the two regions. Therefore we introduce averages and differences
of the extrinsic curvature.

The Lanczos equation \cite{Israel}, \cite{Lanczos} relates the jump of the
extrinsic curvature across the brane to the distributional matter layer: 
\begin{equation}
\Delta ^{\left( 4\right) }\!K_{ab}=-\widetilde{\kappa }^{2}\left( \tau _{ab}-%
\frac{\tau }{3}g_{ab}\right) \ .  \label{Lanczos}
\end{equation}%
The $u^{a}u^{b},~u^{a}h_{c}^{b},$ trace and trace-free parts of the $%
h_{c}^{a}h_{d}^{b}$ projections give: 
\begin{eqnarray}
\Delta \widehat{K} &=&\frac{\widetilde{\kappa }^{2}}{3}\left( \lambda -2\rho
-3p\right) \ ,  \label{Bound1} \\
\Delta \widehat{K}_{a} &=&\widetilde{\kappa }^{2}q_{a}\ ,  \label{Bound2} \\
\Delta \widehat{\Theta } &=&-\widetilde{\kappa }^{2}\left( \lambda +\rho
\right) \ ,  \label{Bound3} \\
\Delta \widehat{\sigma }_{ab} &=&-\widetilde{\kappa }^{2}\pi _{ab}\ .
\label{Bound4}
\end{eqnarray}

The Lanczos equation is necessary in order to derive the gravitational
dynamics on the brane, given by a scalar (the twice-contracted Gauss), a 4d
vectorial (the Codazzi) and a 4d tensorial (the effective Einstein)
equations \cite{Gergely Friedmann}. The latter has been first derived in\ 
\cite{SMS}, later generalized to include bulk matter and asymmetric
embedding contributions (Eq. (1) in \cite{Gergely Friedmann}).

\subsection{The 3+1 decomposition of the source terms of the effective
Einstein equation\label{sources_sec}}

We give the 3+1 covariant decomposition of the source terms of the effective
Einstein equation. This equation is \cite{Gergely Friedmann}:%
\begin{equation}
G_{ab}=-\left( \Lambda -\frac{\widetilde{\kappa }^{2}\left\langle \widetilde{%
\pi }\right\rangle }{2}\right) g_{ab}+\kappa ^{2}T_{ab}+\widetilde{\kappa }%
^{4}S_{ab}-\left\langle ^{\left( 4\right) }\!\mathcal{E}_{ab}\right\rangle
+\left\langle L_{ab}\right\rangle +\left\langle \mathcal{P}%
_{ab}\right\rangle \ .
\end{equation}%
The sources are: the stress-energy tensor $T_{ab}$ representing standard
model matter [decomposed in Eq. (\ref{Tab})]; the source term $S_{ab}$
quadratic in $T_{ab}$ (dominant at high energies), $\left\langle \mathcal{P}%
_{ab}\right\rangle $ the pull-back to the brane of the bulk matter; $%
\left\langle L_{ab}\right\rangle $ a source term originating in the
asymmetry of the embedding and $^{\left( 4\right) }\!\mathcal{E}_{ab}$ the
contribution of the electric part (relative to the vector $n^{a}$) of the 5d
Weyl tensor. We have defined the 4d coupling constant $\kappa ^{2}$ and the
brane cosmological constant $\Lambda $ as 
\begin{eqnarray}
6\kappa ^{2} &=&\widetilde{\kappa }^{4}\lambda ~,  \label{kappa4d} \\
2\Lambda &=&\kappa ^{2}\lambda +\left\langle \widetilde{\Lambda }%
\right\rangle ~.  \label{Lambdabrane}
\end{eqnarray}

The quadratic source term is decomposed as%
\begin{eqnarray}
S_{ab} &=&\frac{1}{24}\left( 2\rho ^{2}-3\pi _{cd}\pi ^{cd}\right)
u_{a}u_{b}+\frac{1}{24}\left( 2\rho ^{2}+4p\rho -4q_{c}q^{c}+\pi _{cd}\pi
^{cd}\right) h_{ab}  \notag \\
&&+\frac{1}{4}q_{\langle a}q_{b\rangle }+\frac{\rho }{3}q_{(a}u_{b)}-\frac{1%
}{2}q^{c}\pi _{c(a}u_{b)}-\frac{\rho +3p}{12}\pi _{ab}-\frac{1}{4}\pi
_{c\langle a}\pi _{b\rangle }^{\,\,\,\,\,c}~.
\end{eqnarray}%
Some of the numerical coefficients here are corrected with respect to the
corresponding expression (7) given in Ref. \cite{Maartens eqs}.

The electric part of the 5d Weyl tensor expressed in terms of
gravito-electro-magnetic quantities defined in Section \ref%
{gravito-electric-magnetic} is 
\begin{equation}
\left\langle ^{\left( 4\right) }\!\mathcal{E}_{ab}\right\rangle
=\left\langle \mathcal{E}\right\rangle \left( u_{a}u_{b}+\frac{1}{3}%
h_{ab}\right) -2\left\langle \widehat{\mathcal{E}}_{(a}\right\rangle
u_{b)}+\left\langle \widehat{\mathcal{E}}_{ab}\right\rangle \ .
\label{epdecomp}
\end{equation}

The asymmetry source term is decomposed as 
\begin{eqnarray}
\left\langle L_{ab}\right\rangle &=&\frac{1}{3}\left[ \left\langle \widehat{%
\Theta }\right\rangle ^{2}-\frac{3}{2}\left\langle \widehat{\sigma }%
_{cd}\right\rangle \left\langle \widehat{\sigma }^{cd}\right\rangle \right]
u_{a}u_{b}  \notag \\
&&-u_{(a}\left[ \frac{4}{3}\left\langle \widehat{\Theta }\right\rangle
h_{b)c}-2\left\langle \widehat{\sigma }_{b)c}\right\rangle \right]
\left\langle \widehat{K}^{c}\right\rangle  \notag \\
&&+\frac{1}{9}\left[ 6\left\langle \widehat{\Theta }\right\rangle
\left\langle \widehat{K}\right\rangle -9\left\langle \widehat{K}%
^{c}\right\rangle \left\langle \widehat{K}_{c}\right\rangle -\left\langle 
\widehat{\Theta }\right\rangle ^{2}+\frac{9}{2}\left\langle \widehat{\sigma }%
_{cd}\right\rangle \left\langle \widehat{\sigma }^{cd}\right\rangle \right]
h_{ab}  \notag \\
&&+\left\langle \widehat{K}_{a}\right\rangle \left\langle \widehat{K}%
_{b}\right\rangle +\left\langle \frac{\widehat{\Theta }}{3}-\widehat{K}%
\right\rangle \left\langle \widehat{\sigma }_{ab}\right\rangle -\left\langle 
\widehat{\sigma }_{a}^{\,\,\,\,c}\right\rangle \left\langle \widehat{\sigma }%
_{bc}\right\rangle ~.
\end{eqnarray}%
As the induced metric is continuous, the average of the trace $%
L=g^{ab}L_{ab} $ is the trace of the average:%
\begin{equation}
\left\langle L\right\rangle =\left\langle \widehat{\sigma }%
_{cd}\right\rangle \left\langle \widehat{\sigma }^{cd}\right\rangle
-2\left\langle \widehat{K}_{b}\right\rangle \left\langle \widehat{K}%
^{b}\right\rangle -\frac{2}{3}\left\langle \widehat{\Theta }\right\rangle
^{2}+2\left\langle \widehat{\Theta }\right\rangle \left\langle \widehat{K}%
\right\rangle ~.  \label{Lave}
\end{equation}%
For a symmetric embedding $\left\langle ^{\left( 4\right)
}\!K_{ab}\right\rangle =0$, therefore cf. Eq. (\ref{Kab4}) $\left\langle 
\widehat{\Theta }\right\rangle =\left\langle \widehat{K}\right\rangle
=0=\left\langle \widehat{K}_{c}\right\rangle =\left\langle \widehat{\sigma }%
_{cd}\right\rangle $, so that $\left\langle L_{ab}\right\rangle =0$.

Finally, 
\begin{equation}
\frac{6\mathcal{P}_{ab}}{\widetilde{\kappa }^{2}}=3\left( \widetilde{\rho }+%
\widetilde{p}\right) u_{a}u_{b}+8\widetilde{q}_{(a}u_{b)}+\left( \widetilde{%
\rho }+\widetilde{p}\right) h_{ab}+4\widetilde{\pi }_{ab}\ .
\end{equation}

\subsection{Gravitational dynamics on the brane: generic embedding\label%
{branedynamics_sec}}

In order to obtain the evolution and constraint equations on the brane, we
select a subset of the Ricci and Bianchi equations given in subsections \ref%
{Ricci_sec} and \ref{Bianchi_sec}, by combining them in such a way, that the
off-brane derivatives of the kinematical and gravito-electro-magnetic
quantities drop out. Additionally, the equations of this subset contain only
quantities appearing in the effective Einstein equation and in the 4d theory.

First we express $\mathcal{H}_{a}$, $\mathcal{E}_{a}$, $\mathcal{F}_{ab}$
and $\widehat{\mathcal{H}}_{ab}$ from (\ref{sigmak}), (\ref{Div_sigma_tild}%
), (\ref{sigma_tild_dot}) and (\ref{call_Hab}) respectively and we employ
the definitions (\ref{Eab}), (\ref{Hab}) in order to introduce $E_{ab}$ and $%
H_{ab}$ in place of $\mathcal{E}_{ab}$ and $\mathcal{H}_{ab}$. Inserting
these into the system of equations given by the following equations: (\ref%
{theta_tild_dot}), (\ref{L_dot}), (\ref{E_dot}), (\ref{Epsilona_dot})-(\ref%
{Ea_column}), (\ref{Theta_dot}), (\ref{omega_dot}), (\ref{sigma_dot}), (\ref%
{Omega_const}), (\ref{Hab_tild_const}), (\ref{Divsigmaab_const}), (\ref%
{Etildab_dot})-(\ref{Epsilonab_dot}), (\ref{Hab_dot}), (\ref{Div_Eab}) and (%
\ref{Div_Hab}), evaluated at the brane, we obtain a system of equations to
be referred as the brane Eqs. These equations are either evolution or
constraint equations on the brane and for a generic asymmetric embedding are
presented in Appendix \ref{asym_app}. The evolutions refer to the quantities 
$\widehat{\Theta },~\widehat{K}_{a},~\Theta ,~\omega _{a},~\sigma _{ab},~%
\mathcal{E},~\widehat{\mathcal{E}}_{a},~E_{ab},~H_{ab}$.

\subsection{Gravitational evolution and constraint equations on a
symmetrically embedded brane\label{sym_sec}}

In this subsection we restrict ourselves to \textit{symmetrically embedded}
branes. The $Z_{2}$-symmetric embedding arises when there is a perfect
symmetry between the 5d space-time regions on the two sides of the brane. In
this case the extrinsic curvatures on the two sides of the brane are
opposite. This is due to the fact, that the normal vectors to the brane on
its two sides are $n^{a}$ and $-n^{a}$, respectively. Therefore $\Delta
^{\left( 4\right) }\!K_{ab}=2^{\left( 4\right) }\!K_{ab}$ and $\overline{K}%
_{ab}=0$.

We present a system of evolution and constrain equations, which \textit{hold
on the brane and contain no off-brane derivatives}$.$We obtain these
equations by specifying Eqs. (\ref{Thetahat_brane_dot})-(\ref%
{DivHab_brane_dot}) for a symmetrically embedded brane; then replacing $%
\widehat{\Theta },~\widehat{K}_{a},~\widehat{K}$ and $\widehat{\sigma }%
_{ab}~ $with the corresponding matter variables as given by the projections (%
\ref{Bound1})-(\ref{Bound4}) of the Lanczos equation, again specified for a
symmetrically embedded brane. Finally, we employ the definitions (\ref%
{kappa4d}) and (\ref{Lambdabrane}), whenever possible.

For improved clarity we group all general relativistic contributions on the
left hand side of the equations, keeping the brane-world contributions on
the right hand side:%
\begin{eqnarray}
\dot{\rho}+D^{a}q_{a}+\left( \rho +p\right) \Theta +2q^{a}A_{a}+\pi
_{ab}\sigma ^{ab} &=&-2\widetilde{q}\ ,  \label{rho_brane_dot} \\
\dot{q}_{\langle a\rangle }+D_{a}p+D^{b}\pi _{ab}+\frac{4\Theta }{3}%
q_{a}+\left( \rho +p\right) A_{a}+\pi _{ab}A^{b}-\omega _{ab}q^{b}+\sigma
_{ab}q^{b} &=&-2\widetilde{\pi }_{a}\ ,  \label{qa_brane_dot}
\end{eqnarray}%
\begin{eqnarray}
0 &=&\dot{\mathcal{E}}-D^{a}\widehat{\mathcal{E}}_{a}+\frac{4}{3}\Theta 
\mathcal{E}+\widehat{\mathcal{E}}_{ab}\sigma ^{ab}-2\widehat{\mathcal{E}}%
_{a}A^{a}+\frac{\widetilde{\kappa }^{4}}{4}\Biggl[\pi ^{ab}\pi _{\langle
ab\rangle }^{\cdot }+\pi ^{ab}D_{a}q_{b}+q^{a}D^{b}\pi _{ab}-\frac{2}{3}%
q^{a}D_{a}\rho  \notag \\
&&+2A_{b}q_{a}\pi ^{ab}+\frac{\Theta }{3}\pi _{ab}\pi ^{ab}+\left( \rho
+p\right) \sigma _{ab}\pi ^{ab}+\sigma _{ca}\pi _{b}^{\,\,\,\,c}\pi ^{ab}+%
\frac{2\Theta }{3}q_{a}q^{a}-\sigma _{ab}q^{a}q^{b}\Biggr]  \notag \\
&&-\frac{\widetilde{\kappa }^{2}}{2}\left( \widetilde{\rho }-\widetilde{\pi }%
+\widetilde{p}\right) ^{\cdot }-\frac{2\widetilde{\kappa }^{2}}{3}D^{a}%
\widetilde{q}_{a}-\frac{2\widetilde{\kappa }^{2}}{3}\Theta \left( \widetilde{%
\rho }+\widetilde{p}\right) +2\kappa ^{2}\widetilde{q}+\frac{\widetilde{%
\kappa }^{4}}{3}\rho \widetilde{q}-\frac{4\widetilde{\kappa }^{2}}{3}%
\widetilde{q}_{a}A^{a}-\frac{2\widetilde{\kappa }^{2}}{3}\widetilde{\pi }%
_{ab}\sigma ^{ab}\ ,  \label{E_brane_dot} \\
0 &=&\dot{\widehat{\mathcal{E}}}_{\langle k\rangle }+\frac{4}{3}\Theta 
\widehat{\mathcal{E}}_{k}-\frac{1}{3}D_{k}\mathcal{E}-\frac{4\mathcal{E}}{3}%
A_{k}-D^{a}\widehat{\mathcal{E}}_{ka}-\widehat{\mathcal{E}}_{ka}A^{a}-\left(
\omega _{ka}-\sigma _{ka}\right) \widehat{\mathcal{E}}^{a}  \notag \\
&&+\frac{\widetilde{\kappa }^{4}}{4}\Biggl[-\dot{\pi}_{\langle ka\rangle
}q^{a}-\left( \rho +p\right) D^{b}\pi _{kb}-q^{a}\sigma _{ck}\pi
_{a}^{\,\,\,\,c}+\frac{2}{3}\left( \rho +p\right) D_{k}\rho -\frac{2\Theta }{%
3}\left( \rho +p\right) q_{k}  \notag \\
&&-2q^{a}D_{k}q_{a}+q^{a}D_{a}q_{k}+\frac{1}{3}q_{k}D^{a}q_{a}-2q^{a}A_{%
\langle k}q_{a\rangle }+\sigma _{ba}\pi _{k}^{\,\,\,\,b}q^{a}-\frac{2}{3}%
q_{k}\sigma _{ab}\pi ^{ab}  \notag \\
&&+\pi ^{ab}D_{k}\pi _{ab}-\pi _{b}^{\,\,\,\,a}D^{b}\pi _{ka}-\frac{1}{3}\pi
_{k}^{\,\,\,\,a}D_{a}\rho -\varepsilon _{cab}q^{a}\omega ^{b}\pi
_{k}^{\,\,\,\,c}-\varepsilon _{k}^{\ \ ab}q_{c}\omega _{a}\pi
_{b}^{\,\,\,\,c}\Biggr]  \notag \\
&&+\frac{\widetilde{\kappa }^{2}}{3}\Biggl[-2\widetilde{\pi }_{\langle
k\rangle }^{\prime }+\frac{1}{2}D_{k}\left( \widetilde{\rho }+3\widetilde{%
\pi }-3\widetilde{p}\right) +\frac{\widetilde{\kappa }^{2}}{3}\left(
2\lambda -\rho -3p\right) \widetilde{\pi }_{k}-2K\widetilde{q}_{k}  \notag \\
&&-\frac{2}{3}\widetilde{q}\left( 2\widehat{K}_{k}+K_{k}\right) +2\widetilde{%
\pi }_{ka}\widehat{A}^{a}-2\left( \widetilde{\pi }-\widetilde{p}\right) 
\widehat{A}_{k}+\frac{5\widetilde{\kappa }^{2}}{2}\pi _{ka}\widetilde{\pi }%
^{a}\Biggr]\ ,  \label{CallEa_brane_dot}
\end{eqnarray}%
\begin{eqnarray}
&&\dot{\Theta}-D^{a}A_{a}+\frac{\Theta ^{2}}{3}-A^{a}A_{a}-2\omega
_{a}\omega ^{a}+\sigma _{ab}\sigma ^{ab}+\frac{\kappa ^{2}}{2}\left( \rho
+3p\right) -\Lambda  \notag \\
&=&\mathcal{E}+\frac{\widetilde{\kappa }^{4}}{4}q^{a}q_{a}-\frac{\widetilde{%
\kappa }^{4}}{12}\rho \left( 2\rho +3p\right) -\frac{\widetilde{\kappa }^{2}%
}{2}\left( \widetilde{\rho }+\widetilde{\pi }+\widetilde{p}\right) \ ,
\label{thetadot_brane} \\
&&\dot{\sigma}_{\langle ab\rangle }-D_{\langle a}A_{b\rangle }+\frac{2\Theta 
}{3}\sigma _{ab}-A_{\langle a}A_{b\rangle }+\omega _{\langle a}\omega
_{b\rangle }+\sigma _{c\langle a}\sigma _{b\rangle }^{\,\,\,\,c}+E_{ab}-%
\frac{\kappa ^{2}}{2}\pi _{ab}  \notag \\
&=&\frac{\widetilde{\kappa }^{4}}{8}q_{\langle a}q_{b\rangle }-\frac{%
\widetilde{\kappa }^{4}}{8}\pi _{c\langle a}\pi _{b\rangle }^{\,\,\,\,c}-%
\frac{\widetilde{\kappa }^{4}}{24}\left( \rho +3p\right) \pi _{ab}-\frac{1}{2%
}\widehat{\mathcal{E}}_{ab}+\frac{\widetilde{\kappa }^{2}}{3}\widetilde{\pi }%
_{ab}\ ,  \label{sigmadot_brane}
\end{eqnarray}%
\begin{equation}
\dot{\omega}_{\langle a\rangle }-\frac{1}{2}\varepsilon _{a}^{\ \
cd}D_{c}A_{d}+\frac{2\Theta }{3}\omega _{a}-\sigma _{ab}\omega ^{b}=0\ ,
\end{equation}%
\begin{equation}
D^{b}\sigma _{ab}-\frac{2}{3}D_{a}\Theta +\varepsilon _{a}^{\ \
ck}D_{c}\omega _{k}+2\varepsilon _{a}^{\ \ ck}A_{c}\omega _{k}+\kappa
^{2}q_{a}=-\frac{\widetilde{\kappa }^{4}}{6}\rho q_{a}+\frac{\widetilde{%
\kappa }^{4}}{4}\pi _{ab}q^{b}-\widehat{\mathcal{E}}_{a}-\frac{2\widetilde{%
\kappa }^{2}}{3}\widetilde{q}_{a}\ ,  \label{Dsigma_brane}
\end{equation}%
\begin{eqnarray}
D_{\langle c}\omega _{k\rangle }+\varepsilon _{ab\langle k}D^{b}\sigma
_{c\rangle }^{\,\,\,\,a}+2A_{\langle c}\omega _{k\rangle }+H_{ab} &=&0\ ,
\label{Hab_brane} \\
D^{a}\omega _{a}-A_{a}\omega ^{a} &=&0\ ,  \label{Domega_brane}
\end{eqnarray}%
\begin{eqnarray}
&&\dot{E}_{\langle kj\rangle }-\varepsilon _{ab\langle k}D^{a}H_{j\rangle
}^{\,\,\,\,b}+\Theta E_{kj}+E_{\,\,\,\langle k}^{a}\left( \omega _{j\rangle
a}-3\sigma _{j\rangle a}\right) +2\varepsilon _{\langle k}^{\ \ \ \
ab}H_{j\rangle a}A_{b}  \notag \\
&&+\frac{\kappa ^{2}}{2}\Biggl[\dot{\pi}_{\langle kj\rangle }+D_{\langle
k}q_{j\rangle }+\left( \rho +p\right) \sigma _{kj}+2q_{\langle k}A_{j\rangle
}+\frac{\Theta }{3}\pi _{kj}+\pi _{\langle k}^{\,\,\,\,\,\,a}\left( \omega
_{j\rangle a}+\sigma _{j\rangle a}\right) \Biggr]  \notag \\
&=&\frac{1}{2}\dot{\widehat{\mathcal{E}}}_{\langle kj\rangle }-\frac{1}{2}%
D_{\langle k}\widehat{\mathcal{E}}_{j\rangle }+\frac{1}{2}\widehat{\mathcal{E%
}}_{\langle j}^{\,\,\,\,\,a}\left( \omega _{k\rangle a}+\sigma _{k\rangle
a}\right) +\frac{\Theta }{6}\widehat{\mathcal{E}}_{kj}-\widehat{\mathcal{E}}%
_{\langle k}A_{j\rangle }+\frac{2\mathcal{E}}{3}\sigma _{kj}  \notag \\
&&+\frac{\widetilde{\kappa }^{4}}{24}\Biggl[\left( \rho +3p\right) \dot{\pi}%
_{\langle kj\rangle }+6\pi _{\langle j}^{\,\,\,\,\,a}\dot{\pi}_{k\rangle
a}+\left( \rho +3p\right) ^{\cdot }\pi _{kj}-2q_{\langle k}D_{j\rangle
}\left( \rho -3p\right) -2\rho D_{\langle k}q_{j\rangle }+3\pi _{\langle
j}^{\,\,\,\,\,a}D_{k\rangle }q_{a}  \notag \\
&&+6q_{\langle k}D^{b}\pi _{j\rangle b}+3q_{a}D_{\langle k}\pi _{j\rangle
}^{\,\,\,\,\,a}-2\rho \left( \rho +p\right) \sigma _{kj}+7\Theta q_{\langle
k}q_{j\rangle }+2\left( \rho +3p\right) q_{\langle k}A_{j\rangle }+\Theta
\pi _{\langle j}^{\,\,\,\,\,a}\pi _{k\rangle a}  \notag \\
&&+\frac{\Theta }{3}\left( \rho +3p\right) \pi _{kj}+\left( \rho +3p\right)
\pi _{\langle k}^{\,\,\,\,\,\,a}\left( \omega _{j\rangle a}+\sigma
_{j\rangle a}\right) +3q_{\langle k}\sigma _{j\rangle b}q^{b}+3\pi _{\langle
j}^{\,\,\,\,\,a}\omega _{k\rangle c}\pi _{a}^{\,\,\,\,c}+3\sigma
_{jk}q^{a}q_{a}  \notag \\
&&-9q_{\langle k}\omega _{j\rangle a}q^{a}+6q_{\langle k}\pi _{j\rangle
a}A^{a}+6\pi _{a\langle j}A_{k\rangle }q^{a}+3\pi _{c}^{\,\,\,\,a}\pi
_{\langle k}^{\,\,\,\,\,\,c}\sigma _{j\rangle a}\Biggr]  \notag \\
&&+\frac{\widetilde{\kappa }^{2}}{3}\Biggl[\dot{\widetilde{\pi }}_{\langle
kj\rangle }+D_{\langle k}\widetilde{q}_{j\rangle }-\frac{\widetilde{\kappa }%
^{2}}{2}\widetilde{\pi }_{\langle k}q_{j\rangle }+4\widetilde{q}_{\langle
k}A_{j\rangle }+\left( \widetilde{\rho }+\widetilde{p}\right) \sigma _{jk}+%
\frac{\Theta }{3}\widetilde{\pi }_{jk}+\widetilde{\pi }_{\langle
j}^{\,\,\,\,\,a}\left( \omega _{k\rangle a}+\sigma _{k\rangle a}\right) %
\Biggr]\ ,
\end{eqnarray}%
\begin{eqnarray}
&&\dot{H}_{\langle kj\rangle }+\varepsilon _{ab\langle k}D^{a}E_{j\rangle
}^{\,\,\,\,b}+\Theta H_{kj}-3\sigma _{a\langle k}H_{j\rangle
}^{\,\,\,\,a}-\omega _{a\langle k}H_{j\rangle }^{\,\,\,\,a}-2\varepsilon
_{\langle k}^{\ \ \ ab}E_{j\rangle a}A_{b}-\frac{\kappa ^{2}}{2}\left(
\varepsilon _{ab\langle k}D^{a}\pi _{j\rangle }^{\,\,\,\,b}+3\omega
_{\langle k}q_{j\rangle }+\varepsilon _{\langle k}^{\ \ \ ab}\sigma
_{j\rangle a}q_{b}\right)  \notag \\
&=&-\frac{1}{2}\varepsilon _{ab\langle k}D^{a}\widehat{\mathcal{E}}%
_{j\rangle }^{\,\,\,\,b}+\frac{1}{2}\varepsilon _{\langle k}^{\ \ \
ab}\sigma _{j\rangle a}\widehat{\mathcal{E}}_{b}+\frac{3}{2}\widehat{%
\mathcal{E}}_{\langle j}\omega _{k\rangle }+\frac{\widetilde{\kappa }^{4}}{24%
}\Biggl[\varepsilon _{\langle k}^{\ \ \ \ cd}\pi _{j\rangle c}D_{d}\left(
\rho +3p\right) -\left( \rho +3p\right) \varepsilon _{ab\langle k}D^{a}\pi
_{j\rangle }^{\,\,\,\,b}  \notag \\
&&-3\varepsilon _{ab\langle k}D^{a}\pi _{j\rangle c}\pi ^{cb}-3\varepsilon
_{ab\langle k}\pi _{j\rangle c}D^{a}\pi ^{cb}+6\rho \omega _{\langle
k}q_{j\rangle }+2\rho \varepsilon _{\langle k}^{\ \ \ ab}\sigma _{j\rangle
a}q_{b}+3\varepsilon _{ab\langle k}q_{j\rangle }D^{a}q^{b}+3\varepsilon
_{ab\langle k}q^{b}D^{a}q_{j\rangle }  \notag \\
&&+3\varepsilon _{\langle k}^{\ \ \ ab}\sigma _{j\rangle b}\pi
_{a}^{\,\,\,\,c}q_{c}-9\pi _{\langle j}^{\,\,\,\,\,\,a}\omega _{k\rangle
}q_{a}\Biggr]+\frac{\widetilde{\kappa }^{2}}{3}\left( \varepsilon
_{ab\langle k}D^{a}\widetilde{\pi }_{j\rangle }^{\,\,\,\,b}+3\widetilde{q}%
_{\langle j}\omega _{k\rangle }+\varepsilon _{\langle k}^{\ \ \ \ ab}\sigma
_{j\rangle a}\widetilde{q}_{b}\right) \ ,
\end{eqnarray}%
\begin{eqnarray}
&&D^{a}E_{ak}-3H_{ka}\omega ^{a}+\varepsilon _{k}^{\ \ ab}H_{ac}\sigma
_{b}^{\,\,\,\,c}+\frac{\kappa ^{2}}{2}\Biggl[-\frac{2}{3}D_{k}\rho +D^{a}\pi
_{ak}+\frac{2}{3}\Theta q_{k}-\sigma _{kb}q^{b}-3\varepsilon _{k}^{\ \
cd}q_{c}\omega _{d}\Biggr]  \notag \\
&=&\frac{1}{2}D^{a}\widehat{\mathcal{E}}_{ak}-\frac{1}{3}D_{k}\mathcal{E}-%
\frac{\Theta }{3}\widehat{\mathcal{E}}_{k}+\frac{1}{2}\left( 3\omega
_{ka}+\sigma _{ka}\right) \widehat{\mathcal{E}}^{a}+\frac{\widetilde{\kappa }%
^{4}}{24}\Biggl[\frac{4}{3}\rho D_{k}\rho +\pi _{ak}D^{a}\left( \rho
+3p\right)  \notag \\
&&+\left( \rho +3p\right) D^{a}\pi _{ak}+3\pi _{a}^{\,\,\,\,c}D^{a}\pi
_{ck}-4\pi ^{ab}D_{k}\pi _{ab}+3\pi _{kb}D^{a}\pi
_{a}^{\,\,\,\,\,b}+2q^{a}D_{k}q_{a}-3q_{a}D^{a}q_{k}  \notag \\
&&-3q_{k}D^{a}q_{a}+9\varepsilon _{k}^{\ \ ad}q_{c}\omega _{a}\pi
_{d}^{\,\,\,\,c}-3\pi _{a}^{\,\,\,\,b}q^{a}\sigma _{kb}-\frac{4}{3}\rho
\Theta q_{k}+2\rho \sigma _{kb}q^{b}+6\rho \varepsilon _{k}^{\ \
cd}q_{c}\omega _{d}+2\Theta \pi _{k}^{\,\,\,\,a}q_{a}\Biggr]  \notag \\
&&-\frac{\widetilde{\kappa }^{2}}{3}\Biggl[D^{a}\widetilde{\pi }_{ak}-\frac{1%
}{2}D_{k}\left( \widetilde{\rho }-\widetilde{\pi }+\widetilde{p}\right) +%
\frac{2}{3}\Theta \widetilde{q}_{k}-\widetilde{q}^{a}\left( 3\omega
_{ka}+\sigma _{ka}\right) \Biggr]\ ,  \label{DElectric_brane}
\end{eqnarray}%
\begin{eqnarray}
&&D^{a}H_{ak}+3E_{ka}\omega ^{a}-\varepsilon _{k}^{\ \ ab}E_{ac}\sigma
_{b}^{\,\,\,\,c}+\frac{\kappa ^{2}}{2}\left[ \varepsilon _{k}^{\ \
ab}D_{a}q_{b}+2\left( \rho +p\right) \omega _{k}-\pi _{k}^{\,\,\,\,c}\omega
_{c}-\varepsilon _{k}^{\ \ ab}\pi _{ac}\sigma _{b}^{\,\,\,\,c}\right]  \notag
\\
&=&-\frac{1}{2}\varepsilon _{k}^{\ \ ab}D_{a}\widehat{\mathcal{E}}_{b}+\frac{%
4\mathcal{E}}{3}\omega _{k}-\frac{1}{2}\widehat{\mathcal{E}}_{ak}\omega ^{a}-%
\frac{1}{2}\varepsilon _{k}^{\ \ ab}\widehat{\mathcal{E}}_{ac}\sigma
_{b}^{\,\,\,\,c}+\frac{\widetilde{\kappa }^{4}}{24}\Biggl[3q^{c}\omega
_{c}q_{k}-2\varepsilon _{k}^{\ \ ab}q_{b}D_{a}\rho  \notag \\
&&-2\rho \varepsilon _{k}^{\ \ ab}D_{a}q_{b}-3\varepsilon
_{abk}q^{c}D^{b}\pi _{c}^{\,\,\,\,a}-3\varepsilon _{k}^{\ \ \ ab}\pi
_{a}^{\,\,\,\,c}D_{b}q_{c}-4\rho \left( \rho +p\right) \omega
_{k}+3q_{a}q^{a}\omega _{k}-\left( \rho +3p\right) \pi
_{k}^{\,\,\,\,c}\omega _{c}  \notag \\
&&-3\varepsilon _{k}^{\ \ ac}\sigma _{ab}q^{b}q_{c}+3\pi _{ab}\pi
^{ab}\omega _{k}-3\pi _{ca}\pi _{k}^{\,\,\,\,c}\omega ^{a}-\left( \rho
+3p\right) \varepsilon _{k}^{\ \ ab}\pi _{ac}\sigma
_{b}^{\,\,\,c}-3\varepsilon _{k}^{\ \ ab}\pi _{da}\pi _{c}^{\,\,\,\,d}\sigma
_{b}^{\,\,\,\,c}\Biggr]  \notag \\
&&+\frac{\widetilde{\kappa }^{2}}{3}\Biggl[\widetilde{\pi }_{ka}\omega
^{a}-\varepsilon _{k}^{\ \ ab}D_{a}\widetilde{q}_{b}-2\left( \widetilde{\rho 
}+\widetilde{p}\right) \omega _{k}+\varepsilon _{k}^{\ \ ab}\widetilde{\pi }%
_{a}^{\,\,\,\,c}\sigma _{bc}\Biggr]\ .  \label{DMagnetic_brane}
\end{eqnarray}%
For vanishing 5d matter these equations reduce to the corrected form of the
Eqs. (26)-(29) and Appendix A of Ref. \cite{Maartens eqs}.

Eqs. (\ref{rho_brane_dot}) and (\ref{qa_brane_dot}) express the interchange
of energy density and energy current between the brane and the 5d space-time
(due to the nonvanishing right hand sides). In the absence of the 5d
sources, these equations become evolution equations for the brane matter
alone. Similar relations for the effective nonlocal energy density and
effective nonlocal energy current are given by Eqs. (\ref{E_brane_dot}) and (%
\ref{CallEa_brane_dot}).

Even for the chosen $\mathcal{Z}_{2}$-symmetric embedding and with the
simplifying assumption $\widetilde{T}_{ab}=0$ (thus $0=\widetilde{\rho }=%
\widetilde{q}=\widetilde{q}_{a}=\widetilde{\pi }=\widetilde{\pi }_{a}=%
\widetilde{\pi }_{ab}=\widetilde{p}$) the above equations do not close on
the brane, due to the absence of evolution equations for $\widehat{\mathcal{E%
}}_{ab}$, $\pi _{ab}$, and $p$. Assumptions fixing $\pi _{ab}$ (typically by
kinetic considerations employing the Boltzmann equation) and $p$ (by choice
of a continuity equation) are required already in general relativity for
closing the system, however on the brane (even with empty 5d space-time) an
additional assumption for $\widehat{\mathcal{E}}_{ab}$ is equally required 
\cite{Maartens eqs}. From these considerations it is immediate to establish
the closure in the special case $\widehat{\mathcal{E}}_{ab}=0=\pi _{ab}$,
provided the equation of state is known.

As the closure is difficult to achieve even in the simple 5d vacuum case, in
order to tackle realistic problems we need to consider the ensemble of all
dynamical and constraint equations given in the preceding section.

\subsection{Closure conditions}

The general relativistic 3+1 covariant formalism contains 10
gravito-electro-magnetic variables with 10 evolution equations (besides
there are also 12 kinematic variables with 9 evolution equations and 15
constraints altogether). The 3+1+1 brane-world formalism developed in this
paper contains 35 gravito-electro-magnetic variables with 35 evolution
equations (beside 35 kinematical variables with 28 evolution equations and
77 constraints altogether).

The subset of equations on the brane, given in subsection \ref{sym_sec} has
10+9 gravito-electro-magnetic variables with only 10+4 evolution equations
(there are also 12 kinematical variables with 9 evolution equations and 15
constraints altogether.) The 9 new gravito-electro-magnetic variables are
the quantities appearing on the right hand side of Eq. (\ref{epdecomp});
among them the last term $\widehat{\mathcal{E}}_{ab}$, representing 5
independent variables, has no evolution equation. It has been known that the
system of equation is closed by the condition $\widehat{\mathcal{E}}_{ab}=0$ 
\cite{Maartens eqs}.

In this subsection we want to explore the extra information we have in the
complete system of evolution equations derived. In particular, Eq. (\ref%
{Epsilonab_dot}) contains the desired temporal evolution, however it also
contains terms not appearing in subsection \ref{sym_sec}. Remembering, that
on the brane $\widehat{\omega }_{ab}=0$ we could impose%
\begin{eqnarray}
\mathcal{F}_{\langle kj\rangle }^{\prime } &=&\left( 2\widehat{K}-\frac{%
\widehat{\Theta }}{3}\right) \mathcal{F}_{kj}-\mathcal{F}_{\langle
j}^{\,\,\,\,\,a}\widehat{\sigma }_{k\rangle a}+K\widehat{\mathcal{E}}_{kj}+K%
\mathcal{E}_{kj}-\mathcal{E}_{\langle j}\left( \frac{3}{2}K_{k\rangle
}+L_{k\rangle }-\widehat{K}_{k\rangle }\right)  \notag \\
&&+\frac{3}{2}\widehat{\mathcal{E}}_{\langle k}\widehat{A}_{j\rangle
}-\varepsilon _{\langle k}^{\ \ \ ab}\widehat{\mathcal{H}}_{j\rangle
a}\left( K_{b}-2\widehat{K}_{b}\right) -\varepsilon _{\langle j}^{\ \ \ ab}%
\mathcal{H}_{k\rangle a}\widehat{A}_{b}+\frac{3}{2}\varepsilon _{\langle
j}^{\ \ \,\,ab}\widehat{\sigma }_{k\rangle a}\mathcal{H}_{b}~,
\label{closurecondgen}
\end{eqnarray}%
such that Eq. (\ref{Epsilonab_dot}) becomes 
\begin{equation}
0=\dot{\widehat{\mathcal{E}}}_{\langle kj\rangle }-D_{\langle k}\widehat{%
\mathcal{E}}_{j\rangle }+\frac{\Theta }{3}\widehat{\mathcal{E}}_{kj}+\frac{4%
\mathcal{E}}{3}\sigma _{kj}-2\widehat{\mathcal{E}}_{\langle k}A_{j\rangle }+%
\widehat{\mathcal{E}}_{\langle j}^{\,\,\,\,\,a}\left( \omega _{k\rangle
a}+\sigma _{k\rangle a}\right) +\mathcal{M}_{kj}~,  \label{closureq}
\end{equation}%
with the 5d matter contributions 
\begin{eqnarray}
\mathcal{M}_{kj} &=&\frac{\widetilde{\kappa }^{2}}{3}\dot{\widetilde{\pi }}%
_{\langle kj\rangle }+\frac{\widetilde{\kappa }^{2}}{3}D_{\langle k}%
\widetilde{q}_{j\rangle }+\frac{\widetilde{\kappa }^{2}}{3}\left( \widetilde{%
\rho }+\widetilde{p}\right) \sigma _{kj}+\frac{\widetilde{\kappa }^{2}%
\widetilde{q}}{3}\widehat{\sigma }_{kj}+\frac{\widetilde{\kappa }^{2}\Theta 
}{9}\widetilde{\pi }_{kj}  \notag \\
&&+\frac{2\widetilde{\kappa }^{2}}{3}\widetilde{q}_{\langle j}A_{k\rangle }+%
\frac{\widetilde{\kappa }^{2}}{3}\widetilde{\pi }_{\langle j}\left( 2%
\widehat{K}_{k\rangle }+L_{k\rangle }\right) +\frac{\widetilde{\kappa }^{2}}{%
3}\widetilde{\pi }_{\langle j}^{\,\,\,\,\,a}\left( \omega _{k\rangle
a}+\sigma _{k\rangle a}\right) \ .
\end{eqnarray}%
A particular solution of Eq. (\ref{closurecondgen}) would be%
\begin{equation}
\mathcal{F}_{kj}=\widehat{\mathcal{H}}_{kj}=\mathcal{E}_{j}=\mathcal{H}%
_{j}=K=\widehat{A}_{j}=0~.  \label{closurecond}
\end{equation}%
None of the quantities above appear in the brane equations presented in
subsection \ref{sym_sec}, thus those equations are not altered by the choice
(\ref{closurecond}), and the system becomes closed by Eq. (\ref{closureq}).%
\footnote{%
The no-go theorem for closure given in Ref. \cite{MukohyamaIntDiff} does not
apply here, as it was derived for perturbations of 5d Anti de Sitter
spacetimes.}

The 5d Schwarzschild-Anti de Sitter space-time containing a Friedmann brane
emerges as special case of the space-times obeying Eqs. (\ref{closureq}) and
(\ref{closurecond}), as they have $\mathcal{M}_{ab}=0=K$ and 
\begin{eqnarray}
0 &=&\mathcal{E}_{a}=\widehat{\mathcal{E}}_{a}=\mathcal{H}_{a}=L_{a}=K_{a}=%
\widehat{K}_{a}=\omega _{a}=\widehat{\omega }_{a}=A_{a}=\widehat{A}_{a}~, 
\notag \\
0 &=&\mathcal{E}_{ab}=\mathcal{F}_{ab}=\mathcal{H}_{ab}=\widehat{\mathcal{E}}%
_{ab}=\widehat{\mathcal{H}}_{ab}=\sigma _{ab}=\widehat{\sigma }_{ab}~.
\end{eqnarray}

At the end of this section we emphasize, that Eq. (\ref{closureq}) closing
the system of brane equations should allow for far more solutions than the
trivial one for $\widehat{\mathcal{E}}_{ab}$. We will construct a specific
example in Section \ref{astrophysical}.

\section{Cosmology\label{cosmological}}

In this section we consider generic embeddings. By employing the definitions
(\ref{kappa4d}) and (\ref{Lambdabrane}), also the conditions (\ref{brane}),
arising from the existence of the brane, we can derive average and
difference equations on the brane. We give here but the most relevant such
dynamical equations. The rest of the equations is straightforward to derive,
although it may be lengthy.

The average taken on the two sides of the brane of Eq. (\ref%
{localRicciscalar}) reduces to the \textit{generalized brane Friedmann
equation}. We also rewrite $\left\langle \widehat{\Theta }^{2}\right\rangle
=\left\langle \widehat{\Theta }\right\rangle ^{2}+\left( \Delta \widehat{%
\Theta }\right) ^{2}/4$ and use a similar relation for $\widehat{\sigma }%
_{ab}$. We take $\Delta \widehat{\Theta }$ from the Lanczos equations (\ref%
{Bound1})-(\ref{Bound4}). Eq. (\ref{localRicciscalar}) then becomes%
\begin{equation}
\frac{\mathcal{R}}{2}+\frac{\Theta ^{2}}{3}-\Lambda -\kappa ^{2}\rho \left(
1+\frac{\rho }{2\lambda }\right) -\frac{1}{2}\sigma _{ab}\sigma ^{ab}+\omega
_{a}\omega ^{a}=-\left\langle \mathcal{E}\right\rangle -\frac{\widetilde{%
\kappa }^{4}}{8}\pi _{ab}\pi ^{ab}+\frac{\left\langle \widehat{\Theta }%
\right\rangle ^{2}}{3}-\frac{1}{2}\left\langle \widehat{\sigma }%
_{ab}\right\rangle \left\langle \widehat{\sigma }^{ab}\right\rangle +\frac{%
\widetilde{\kappa }^{2}}{2}\left\langle \widetilde{\rho }+\widetilde{p}-%
\widetilde{\pi }\right\rangle \ .  \label{FriedmannGen}
\end{equation}

The \textit{generalized brane Raychaudhuri equation} is obtained from the
average of Eq. (\ref{Theta_dot}), by employing the same sequence of
simplifications, as for the Friedmann equation. We obtain: 
\begin{eqnarray}
&&\dot{\Theta}+\frac{\Theta ^{2}}{3}+\sigma _{ab}\sigma ^{ab}-2\omega
_{a}\omega ^{a}-D^{a}A_{a}-A^{a}A_{a}+\frac{\kappa ^{2}}{2}\left( \rho
+3p\right) -\Lambda  \notag \\
&&=\left\langle \mathcal{E}\right\rangle -\frac{\widetilde{\kappa }^{4}\rho 
}{12}\left( 2\rho +3p\right) +\frac{\widetilde{\kappa }^{4}}{4}%
q_{a}q^{a}-\left\langle \widehat{\Theta }\right\rangle \left\langle \widehat{%
K}\right\rangle +\left\langle \widehat{K}_{a}\right\rangle \left\langle 
\widehat{K}^{a}\right\rangle -\frac{\widetilde{\kappa }^{2}}{2}\left\langle 
\widetilde{\rho }+\widetilde{\pi }+\widetilde{p}\right\rangle \ .
\label{RaychaudhuriGen}
\end{eqnarray}

Finally we give the \textit{generalized brane} \textit{energy-balance
equation} from the jump across the brane of the Eq. (\ref{theta_tild_dot})
by the same procedure:%
\begin{equation}
\dot{\rho}+\Theta \left( \rho +p\right) +D_{a}q^{a}+2A_{a}q^{a}+\pi
_{ab}\sigma ^{ab}=-\Delta \widetilde{q}~.  \label{balanceGen}
\end{equation}%
These equations hold for arbitrary branes. Again, general relativistic
contributions are on the left hand sides, brane-world contributions on the
right hand sides.

\subsection{Friedmann brane with perfect fluid}

In this case the conditions $\omega _{a}=0=\sigma _{ab}=\Delta \widehat{%
\sigma }_{ab}$ hold, arising from the particular geometry and matter source
chosen, also $\mathcal{R}=6k/a^{2}$, where $a$ is the scale factor. Moreover

$\Theta /3=H\equiv \dot{a}/a$, where $H$ is the Hubble parameter. The
Friedmann, Raychaudhuri and energy-balance equations become:%
\begin{align}
3\left( H^{2}+\frac{k}{a^{2}}\right) & =\Lambda +\kappa ^{2}\rho \left( 1+%
\frac{\rho }{2\lambda }\right) -\left\langle \mathcal{E}\right\rangle  \notag
\\
& +\frac{\left\langle \widehat{\Theta }\right\rangle ^{2}}{3}-\frac{1}{2}%
\left\langle \widehat{\sigma }_{ab}\right\rangle \left\langle \widehat{%
\sigma }^{ab}\right\rangle +\frac{\widetilde{\kappa }^{2}}{2}\left\langle 
\widetilde{\rho }+\widetilde{p}-\widetilde{\pi }\right\rangle \ .
\label{Friedmann} \\
3\left( \dot{H}+H^{2}\right) +\frac{\kappa ^{2}}{2}\left( \rho +3p\right)
-\Lambda & =\left\langle \mathcal{E}\right\rangle -\frac{\widetilde{\kappa }%
^{4}\rho }{12}\left( 2\rho +3p\right) +\frac{\widetilde{\kappa }^{4}}{4}%
q_{a}q^{a}  \notag \\
& -\left\langle \widehat{\Theta }\right\rangle \left\langle \widehat{K}%
\right\rangle +\left\langle \widehat{K}_{a}\right\rangle \left\langle 
\widehat{K}^{a}\right\rangle -\frac{\widetilde{\kappa }^{2}}{2}\left\langle 
\widetilde{\rho }+\widetilde{\pi }+\widetilde{p}\right\rangle ~,
\label{Raychaudhuri} \\
\dot{\rho}+3H\left( \rho +p\right) & =-\Delta \widetilde{q}~.
\label{balance}
\end{align}%
For symmetric embedding the equations simplify by $\left\langle \widehat{%
\Theta }\right\rangle =0=\left\langle \widehat{\sigma }_{ab}\right\rangle $.
For generic asymmetric embedding\ Eq. (\ref{Friedmann}) can be rewritten as

\begin{equation}
3\left( H^{2}+\frac{k}{a^{2}}\right) =\Lambda +\kappa ^{2}\rho \left( 1+%
\frac{\rho }{2\lambda }\right) +\kappa ^{2}U-\frac{\left\langle
L\right\rangle }{4}-\frac{\widetilde{\kappa }^{2}}{2}\left\langle \widetilde{%
\pi }\right\rangle \ ,  \label{FriedU}
\end{equation}%
where $\left\langle L\right\rangle $ is given by Eq. (\ref{Lave}) and $U$ is
defined by%
\begin{equation}
\kappa ^{2}U=\frac{\left\langle \widehat{\Theta }\right\rangle ^{2}}{6}+%
\frac{1}{2}\left\langle \widehat{\Theta }\right\rangle \left\langle \widehat{%
K}\right\rangle -\frac{1}{2}\left\langle \widehat{K}_{b}\right\rangle
\left\langle \widehat{K}^{b}\right\rangle -\frac{1}{4}\left\langle \widehat{%
\sigma }_{ab}\right\rangle \left\langle \widehat{\sigma }^{ab}\right\rangle
-\left\langle \mathcal{E}\right\rangle +\frac{\widetilde{\kappa }^{2}}{2}%
\left\langle \widetilde{\rho }+\widetilde{p}\right\rangle ~.  \label{U}
\end{equation}%
The quantity $U$ is nothing but the \textit{effective energy density}
introduced in Ref. \cite{VarBraneTension}, encompassing the trace-free parts
of the Weyl, embedding and 5d matter sources in the effective Einstein
equation, given here in terms of the 3+1+1 kinematic,
gravito-electro-magnetic and 5d matter variables.

\subsection{Anisotropic brane-worlds\label{ani}}

Full brane-world solutions with homogeneous, but anisotropic 5d space-times
are also known.

In Ref. \cite{sopuerta} a vacuum 5d static and anisotropic space-time with
cosmological constant admitting a moving Bianchi I brane was analyzed. The $%
Z_{2}$-symmetric junction conditions could be obeyed only by anisotropic
stresses on the brane, hence the brane cannot support a perfect fluid.
Isotropy of the brane fluid could be achieved only for isotropic 5d
space-time and brane.

This setup was generalized in Ref. \cite{langlois} by allowing for a
non-static 5d space-time. Assuming separability of the metric components,
new 5d solutions combining the 4d Kasner solution with the static 5d
solutions of Ref. \cite{sopuerta} were obtained.

For the reader's convenience we give in Appendix \ref{BianchiBraneWorld} the
list of kinematical, gravito-electro-magnetic and matter variables for the
5-dimensional models presented in Ref. \cite{sopuerta}. Working out the
respective quantities for other metrics would be a similar straightforward
exercise.

\section{Stationary vacuum space-times with local rotational symmetry\label%
{astrophysical}}

In this section we discuss an application of our formalism, by assuming
vacuum in both 4d and 5d, but keeping the respective cosmological constants.
The embedding of the brane is symmetric. Then the effective Einstein
equation reduces to%
\begin{equation}
G_{ab}=\Lambda g_{ab}-\mathcal{E}\left( u_{a}u_{b}+\frac{1}{3}h_{ab}\right)
+2\widehat{\mathcal{E}}_{(a}u_{b)}-\widehat{\mathcal{E}}_{ab}\ .
\label{effective_LRS}
\end{equation}

We are interested in stationary space-times, therefore $\dot{f}=0$ for any
scalar field $f$. The stationarity implies a singled-out temporal Killing
vector, therefore the 3+1+1 decomposition of the gravitational dynamics
developed in this paper turns particularly useful for the study of
gravitational dynamics on the brane (which defines the other singled-out
direction).

We also specialize our treatment to space-times with local rotational
symmetry (LRS) on the brane. Such a symmetry singles out an additional
space-like unit vector field $e^{a}$, in the sense that there is a unique
preferred spatial direction at each point that assigns the local axis of
symmetry. Once such a special vector field is chosen, a further
decomposition of the spatial quantities would lead to a generic 2+1+1+1
formalism. For this, the metric $h_{ab}$ should be further decomposed as%
\begin{equation}
h_{ab}=e_{a}e_{b}+q_{ab}\ ,
\end{equation}%
where $q_{ab}$ is the induced 2-metric on the surface perpendicular to both $%
e^{a}$ and $u^{a}$, and lying in the brane.

In what follows we will see that these symmetries assure that the structure
of the space-time can be described solely in terms of scalars, thus no
vectors and tensorial quantities are needed. This is a powerful feature of
the formalism. Furthermore, the symmetries assure, that all scalars $f$
depend only on the coordinate parametrizing the integral curves of the
rotation axis field $e^{a}$. We denote this coordinate derivative as $%
f^{\star }\equiv e^{a}D_{a}f$ (a spatial covariant derivative along these
integral curves).

\subsection{Independent kinematic quantities related to the vector field $%
e^{a}$}

\subsubsection{Decomposition}

For the purposes of the present application it is enough to give here the
decomposition of the covariant brane derivative of the vector field $e^{a}$
in terms of kinematical quantities and extrinsic curvature type quantities
analogous to the ones appearing in the decomposition of the vector fields $%
u^{a}$ and $n^{a}$. We keep the familiar notations, with the remark that the
quantities belonging to the decomposition of $\nabla _{a}e_{b}$ will carry a
distinguishing $\widetilde{}$ mark. We thus have%
\begin{equation}
\nabla _{a}e_{b}=\widetilde{L}u_{a}u_{b}-\widetilde{L}_{a}u_{b}-u_{a}%
\widetilde{K}_{b}+e_{a}\widetilde{A}_{b}+\frac{\widetilde{\Theta }}{2}q_{ab}+%
\widetilde{\sigma }_{ab}+\widetilde{\omega }_{ab}\ ,
\end{equation}%
with%
\begin{eqnarray}
\widetilde{L} &=&u^{c}u^{d}\widetilde{\nabla }_{c}e_{d}\ \ ,\ \ \widetilde{%
\Theta }=q^{ab}D_{a}e_{b}\ \ ,  \notag \\
\widetilde{L}_{a} &=&u^{d}h_{a}^{\,\,\,\,c}\widetilde{\nabla }_{c}e_{d}\ \
,\ \ \widetilde{K}_{b}=u^{c}h_{b}^{\,\,\,\,d}\widetilde{\nabla }_{c}e_{d}\ \
,\ \ \ \widetilde{A}_{b}=e^{c}q_{b}^{\,\,\,\,d}\widetilde{\nabla }_{c}e_{d}\
\ ,  \notag \\
\widetilde{\omega }_{ab} &=&q_{[a}^{\,\,\,\,\,\,c}q_{b]}^{\,\,\,\,\,d}%
\widetilde{\nabla }_{c}e_{d}\ \ ,\ \ \widetilde{\sigma }_{ab}=q_{a}^{\,\,\,%
\,c}q_{b}^{\,\,\,d}\widetilde{\nabla }_{c}e_{d}-\frac{\widetilde{\Theta }}{2}%
q_{ab}\ \ .
\end{eqnarray}%
We remark that $\widetilde{L}$ and $\widetilde{L}_{a}$ are not independent
from the previously introduced sets of variables, they can be expressed in
term of projections of the kinematic quantities related to $u^{a}$:%
\begin{equation}
\widetilde{L}=-e^{a}A_{a}\ \ ,\ \ \widetilde{L}_{a}=-e^{d}\left( \frac{%
\Theta }{3}h_{ad}+\sigma _{ad}+\omega _{ad}\right) \ \ .
\end{equation}%
By contrast, the quantities $\widetilde{\Theta }$, $\widetilde{K}_{b}$, $%
\widetilde{A}_{b}$, $\widetilde{\sigma }_{ab}$ and $\widetilde{\omega }_{ab}$
are independent of the rest of the kinematical and extrinsic curvature-type
variables. Similarly to $\omega _{a}$ and $\widehat{\omega }_{a}$, we can
also define a rotation vector 
\begin{equation}
\widetilde{\omega }_{a}=\frac{1}{2}\varepsilon _{abc}\widetilde{\omega }%
^{bc}~.  \label{omegatilde}
\end{equation}

\subsubsection{LRS symmetry}

The preferred spacelike unit vector field $e_{a}$ satisfies $u^{a}e_{a}=0$,$%
\ \ e^{a}e_{a}=1$. We employ here various results following from the LRS
symmetry, following Ref. \cite{LRS}. The symmetry and normalization implies:%
\begin{equation}
e^{a}D_{a}e_{b}=0\ \ ,\ \ \ \dot{e}_{\langle b\rangle }=0\ \ ,  \label{rot1}
\end{equation}%
i.e. $e_{a}$ is geodesic with respect to the connection compatible with $%
h_{ab}$ and is Fermi propagated along the world line of a brane observer.
The above equations and normalization further imply that Eq. (\ref%
{omegatilde}) can be rewritten into the standard form%
\begin{equation}
\widetilde{\omega }_{a}=\varepsilon _{a}^{~\
bc}q_{[b}^{\,\,\,\,\,\,e}q_{c]}^{\,\,\,\,\,d}\widetilde{\nabla }%
_{e}e_{d}=\varepsilon _{abc}D^{b}e^{c}\ .  \label{omtildea}
\end{equation}%
Due to LRS, all spacelike vector fields must be proportional to $e^{a}$,
thus 
\begin{eqnarray}
A^{a} &=&Ae^{a}\ ,\quad \omega _{a}=\omega e_{a}\ ,\quad \widetilde{\omega }%
_{a}=\widetilde{\omega }e_{a}\ .  \notag \\
\widehat{\mathcal{E}}^{a} &=&\widehat{\mathcal{E}}^{V}e^{a}\ ,\quad
D^{a}\Theta =\Theta ^{\star }e_{a}\ ,\quad D_{a}\mathcal{E}=\mathcal{E}%
^{\star }e_{a}\ \ ,  \label{def1}
\end{eqnarray}

The vector field $e^{a}$ and the induced metric $q_{ab}$ define a unique
spacelike tracefree symmetric tensor field $e_{ab}$ as%
\begin{equation}
e_{ab}=e_{a}e_{b}-\frac{q_{ab}}{2}\ ,
\end{equation}%
satisfying:%
\begin{eqnarray}
u^{a}e_{ab} &=&0\ \ \ ,\ \ e^{a}e_{ab}=e_{b}\ \ ,\ \ e_{\,\,\,\,a}^{a}=0\ \
,\ \ 2e_{ac}e_{\,\,\,b}^{c}=e_{ab}+h_{ab}\ \ ,  \notag \\
e_{ab}e^{ab} &=&\frac{3}{2}\ ,\ \ D^{b}e_{ab}=\frac{3\widetilde{\Theta }}{2}%
e_{a}\ \ ,\ \ \dot{e}_{\langle ab\rangle }=0\ \ .  \label{Id1}
\end{eqnarray}%
Again, due to LRS all 3d tracefree symmetric tensor fields are proportional
to $e_{ab}$:%
\begin{equation}
\sigma _{ab}=\frac{2\sigma }{\sqrt{3}}e_{ab}\ \ ,\ \ E_{ab}=\frac{2E}{\sqrt{3%
}}e_{ab}\ \ ,\ \ H_{ab}=\frac{2H}{\sqrt{3}}e_{ab}\ \ ,\ \ \widehat{\mathcal{E%
}}_{ab}=\frac{2\widehat{\mathcal{E}}}{\sqrt{3}}e_{ab}\ \ .\   \label{def2}
\end{equation}%
In Eqs. (\ref{def1}) and (\ref{def2}) we have introduced suitably normalized
scalars $A$, $\omega $, $\widetilde{\omega }$, $\sigma $, $\widehat{\mathcal{%
E}}^{V}$, $\widehat{\mathcal{E}}$, $E$ and $H$, replacing the vectorial and
tensorial variables.

From LRS, by use of Eqs. (\ref{rot1}), (\ref{def1}), (\ref{def2}) and (\ref%
{Id1}) also follows that 
\begin{eqnarray}
\widetilde{L} &=&-A\ \ ,\ \ \widetilde{L}_{a}=-\left( \frac{\Theta }{3}+%
\frac{2\sigma }{\sqrt{3}}\right) e_{a}\ \ , \\
\widetilde{K}_{a} &=&\widetilde{A}_{a}=0\ \ ,\ \ \widetilde{\sigma }_{ab}=0\
\ .
\end{eqnarray}%
Thus there are only two kinematic quantities related to $e^{a}$ left, which
are both non-trivial and independent from those introduced the Sec. \ref%
{decomp_sec}. These are $\widetilde{\Theta }$ and $\widetilde{\omega }$.

\subsection{LRS class I type conditions}

The general relativistic classification of the LRS models presented in \cite%
{LRS} is recovered for $\widehat{\mathcal{E}}_{a}=0=\widehat{\mathcal{E}}%
_{ab}$ and $\mathcal{E}<0$. For brane-worlds, when the above conditions do
not hold even under the simplifying assumptions of this chapter, this
classification should be refined, however for the application we are
interested in, we shall still assume the conditions $\widetilde{\omega }%
=\Theta =\sigma =0$ characterizing the LRS class I of general relativity.
From these conditions, Eqs. (\ref{Dsigma_brane}), (\ref{omtildea}), (\ref%
{def1}) and $\varepsilon _{aij}e^{i}e^{j}=0$ we also find $\widehat{\mathcal{%
E}}^{V}=0$. Therefore we have verified, that Eq. (\ref{closureq}), which
closes the system of brane equations for the considered stationary vacuum
space-times containing LRS class I type branes, trivially holds.

\subsubsection{Dynamics}

The evolution of the single kinematic quantity $\widetilde{\Theta }$
characterizing $e^{a}$ can be inferred from the Ricci identity for $e^{a}$:%
\begin{equation}
\widetilde{\Theta }^{\star }+\frac{\widetilde{\Theta }^{2}}{2}=-\frac{2E}{%
\sqrt{3}}+\frac{2\mathcal{E}}{3}\ +\frac{\widehat{\mathcal{E}}}{\sqrt{3}}-%
\frac{2\Lambda }{3}\ .  \label{Eq1}
\end{equation}%
Other nontrivial brane Eqs. are (\ref{CallEa_brane_dot}), (\ref%
{thetadot_brane}), (\ref{sigmadot_brane}), (\ref{Hab_brane}), (\ref%
{Domega_brane}), (\ref{DElectric_brane}), and (\ref{DMagnetic_brane}). Under
the assumptions of this section they simplify to:%
\begin{eqnarray}
\mathcal{E}^{\star }+4A\mathcal{E}+2\sqrt{3}\left[ \widehat{\mathcal{E}}%
^{\star }+\left( \frac{3\widetilde{\Theta }}{2}+A\right) \widehat{\mathcal{E}%
}\right] &=&0\ ,  \label{Eq2} \\
A^{\star }+\left( \widetilde{\Theta }+A\right) A+2\omega ^{2}+\Lambda &=&-%
\mathcal{E}\ ,  \label{Eq3} \\
A^{\star }+\left( A-\frac{\widetilde{\Theta }}{2}\right) A-\omega ^{2}-\sqrt{%
3}E &=&\frac{\sqrt{3}\widehat{\mathcal{E}}}{2}\ ,  \label{Eq4} \\
H+\left( 2A-\widetilde{\Theta }\right) \frac{\sqrt{3}\omega }{2} &=&0\ ,
\label{Eq5} \\
\omega ^{\star }+\left( \widetilde{\Theta }-A\right) \omega &=&0\ ,
\label{Eq6} \\
E^{\star }+\frac{3\widetilde{\Theta }}{2}E-3\omega H &=&\frac{\widehat{%
\mathcal{E}}^{\star }}{2}+\frac{3\widetilde{\Theta }\widehat{\mathcal{E}}}{4}%
-\frac{\mathcal{E}^{\star }}{2\sqrt{3}}\ ,  \label{Eq7} \\
2H^{\star }+3\widetilde{\Theta }H+6E\omega &=&\frac{4\omega \mathcal{E}}{%
\sqrt{3}}-\omega \widehat{\mathcal{E}}\ .  \label{Eq8}
\end{eqnarray}%
Eliminating $A^{\star }$ from Eqs. (\ref{Eq3}) and (\ref{Eq4}) we obtain the
algebraic equation%
\begin{equation}
0=\frac{3\widetilde{\Theta }A}{2}+3\omega ^{2}+\Lambda +\sqrt{3}E+\mathcal{E}%
+\frac{\sqrt{3}\widehat{\mathcal{E}}}{2}\ .  \label{Eq9}
\end{equation}%
Also for any solution of the system (\ref{Eq1})-(\ref{Eq6}), Eqs. (\ref{Eq7}%
) and (\ref{Eq8}) are identically obeyed. This can be seen as follows. By
taking the $\star $-derivative of Eq. (\ref{Eq9}) and employing Eqs. (\ref%
{Eq1})-(\ref{Eq6}) we obtain Eq. (\ref{Eq7}). Similarly, the $\star $%
-derivative of Eq. (\ref{Eq5}), combined with Eqs. (\ref{Eq1})-(\ref{Eq6})
gives (\ref{Eq8}).

As we have 6 independent (2 algebraic and 4 first order differential)
equations left for the 7 variables ($\mathcal{E}$, $\widehat{\mathcal{E}}$, $%
\widetilde{\Theta }$, $A$, $\omega $, $E$, $H$), we need to impose an
additional ansatz, chosen as 
\begin{equation}
\widehat{\mathcal{E}}=-\frac{2\mathcal{E}}{\sqrt{3}}\ .  \label{Eq10}
\end{equation}%
This condition will considerably simplify the algebraic equation Eq. (\ref%
{Eq9}).

\subsubsection{Discussion}

The algebraic Eqs. (\ref{Eq5}) and (\ref{Eq9}) give $H$ and $E$ in terms of
the rest of variables. By Eq. (\ref{Eq10}) the system (\ref{Eq1})-(\ref{Eq3}%
) and (\ref{Eq6}) reduces to 
\begin{eqnarray}
\widetilde{\Theta }^{\star }+\frac{\widetilde{\Theta }^{2}}{2}-\widetilde{%
\Theta }A-2\omega ^{2} &=&0\ ,  \label{Eq11} \\
\mathcal{E}^{\star }+2\mathcal{E}\widetilde{\Theta } &=&0\ ,  \label{Eq22} \\
A^{\star }+\left( \widetilde{\Theta }+A\right) A+2\omega ^{2}+\Lambda &=&-%
\mathcal{E}\ ,  \label{Eq33} \\
\omega ^{\star }+\left( \widetilde{\Theta }-A\right) \omega &=&0\ .
\label{Eq44}
\end{eqnarray}%
From the newly arised two algebraic Eqs. (\ref{Eq22}) and (\ref{Eq44}) we
express 
\begin{eqnarray}
\widetilde{\Theta } &=&\left( \ln \mathcal{E}^{-1/2}\right) ^{\star }\ ,
\label{thetatildescalar} \\
A &=&\left( \ln \omega \mathcal{E}^{-1/2}\right) ^{\star }\ .
\label{Ascalar}
\end{eqnarray}%
In terms of the auxiliary variables%
\begin{eqnarray}
x &=&\ln \mathcal{E}^{-1/2}\ , \\
y &=&\ln \omega \mathcal{E}^{-1/2}\ ,
\end{eqnarray}%
the remaining Eqs. (\ref{Eq11}) and (\ref{Eq33}) become%
\begin{eqnarray}
x^{\star \star }+\frac{\left( x^{\star }\right) ^{2}}{2}-x^{\star }y^{\star
} &=&2e^{y-x}\ ,  \label{syst1} \\
y^{\star \star }+\left( y^{\star }\right) ^{2}+x^{\star }y^{\star }
&=&-\left( 2e^{y-x}+e^{-2x}+\Lambda \right) \ .  \label{syst2}
\end{eqnarray}%
They form a coupled second order system, which would eventually give $\omega 
$ and $\mathcal{E}$ in full generality. The solution of this system is
however not immediate, therefore in what follows we will employ a metric
ansatz in order to find a particular solution.

\subsection{Taub-NUT-(A)dS solution with tidal charge}

We take the following metric ansatz, compatible with the chosen symmetries:%
\begin{equation}
ds^{2}=-\frac{f\left( r\right) }{g\left( r\right) }\left( dt+\omega _{%
\mathrm{k}}d\varphi \right) ^{2}+\frac{g\left( r\right) }{f\left( r\right) }%
dr^{2}+g\left( r\right) \left( d\theta ^{2}+\Omega _{\mathrm{k}}^{2}d\varphi
^{2}\right) \ ,  \label{ansatz}
\end{equation}%
where 
\begin{equation}
\Omega _{\mathrm{k}}=\left\{ 
\begin{array}{c}
\sin \theta ~,\qquad k=1 \\ 
\quad 1~~~,\qquad \ k=0 \\ 
~\sinh \theta ~,\qquad k=-1%
\end{array}%
\right. ~,  \label{ansatz1}
\end{equation}%
and $\omega _{\mathrm{k}}$\ is another function of $\theta $. The axis of
LRS is given by%
\begin{equation}
e^{a}=\frac{f^{1/2}}{g^{1/2}}\left( \frac{\partial }{\partial r}\right)
^{a}\ \ .  \label{LRSaxis}
\end{equation}%
Choosing the\ 1-form $u_{a}$ as%
\begin{equation}
u=-\frac{f^{1/2}}{g^{1/2}}\left( dt+2\omega _{\mathrm{k}}d\varphi \right) \ ,
\end{equation}%
and employing $\widehat{\mathcal{E}}^{V}=0$, Eqs. (\ref{def2}), (\ref{Eq10}%
), (\ref{LRSaxis}), the electric part of the 5d Weyl tensor is found as%
\begin{equation}
^{\left( 4\right) }\!\mathcal{E}_{\,\,\,\,t}^{t}=\,^{\left( 4\right) }\!%
\mathcal{E}_{\,\,\,\,r}^{r}=-\,^{\left( 4\right) }\!\mathcal{E}%
_{\,\,\,\,\theta }^{\theta }=-\,^{\left( 4\right) }\!\mathcal{E}%
_{\,\,\,\,\varphi }^{\varphi }=-\mathcal{E}\ \ ,\ \,^{\left( 4\right) }\!%
\mathcal{E}_{\,\,\,\,\varphi }^{t}=-2\mathcal{E}\omega _{\mathrm{k}}\ \ .
\label{epsilons}
\end{equation}

Both from $\widehat{\mathcal{E}}_{a}\propto \widehat{\mathcal{E}}^{V}=0$\
and from $H_{ab}$ being proportional to\ $e_{ab}$ we find%
\begin{equation}
\Omega _{\mathrm{k}}\frac{d^{2}\omega _{\mathrm{k}}}{d\theta ^{2}}-\frac{%
d\Omega _{\mathrm{k}}}{d\theta }\frac{d\omega _{\mathrm{k}}}{d\theta }=0\ .
\label{cond1}
\end{equation}%
Equivalently, by an integration:\textbf{\ }%
\begin{equation}
\Omega _{\mathrm{k}}^{-1}\frac{d\omega _{\mathrm{k}}}{d\theta }=-2l
\label{diffomegak}
\end{equation}%
where $l$\ is constant\textbf{.\ }A second integration gives%
\begin{equation}
\omega _{\mathrm{k}}\left( \theta \right) =\left\{ 
\begin{array}{c}
2l\cos \theta +L~,\qquad k=1 \\ 
\quad -2l\theta +L~~~,\qquad k=0 \\ 
~-2l\cosh \theta +L~,\qquad k=-1%
\end{array}%
\right. ~,  \label{kisomega}
\end{equation}%
where $L$\ is another integration constant\textbf{. }Locally, this constant
can be absorbed in a new time variable $t+L\varphi \rightarrow t$, which
translates to the choice $L=0$.

Direct computation, employing Eq. (\ref{kisomega}) shows that the metric
ansatz is compatible with the chosen LRS class I conditions%
\begin{equation}
\Theta =\sigma =a=\widehat{\mathcal{E}}^{V}=0\ \ .
\end{equation}

The nontrivial kinematic and Weyl quantities, by employing Eqs. (\ref{ansatz}%
) and (\ref{kisomega}) are 
\begin{eqnarray}
\ \ \widetilde{\Theta } &=&\frac{f^{1/2}}{g^{3/2}}\frac{dg}{dr}\ ,\quad A=%
\frac{d}{dr}\left( \frac{f}{g}\right) ^{1/2}\ ,\quad \omega =\frac{lf^{1/2}}{%
g^{3/2}}\ ,  \notag \\
\mathcal{E} &=&\frac{f^{1/2}}{g^{5/4}}\frac{d}{dr}\left( \frac{f^{1/2}}{%
g^{3/4}}\frac{dg}{dr}\right) -\frac{3l^{2}f}{g^{3}}-\frac{k}{g}+\Lambda \ , 
\notag \\
\frac{2\widehat{\mathcal{E}}}{\sqrt{3}} &=&\frac{1}{3g}\frac{d^{2}f}{dr^{2}}-%
\frac{2}{3g^{2}}\frac{df}{dr}\frac{dg}{dr}+\frac{f}{3g^{3}}\left( \frac{dg}{%
dr}\right) ^{2}+\frac{4l^{2}f}{3g^{3}}+\frac{2k}{3g}\ ,  \notag \\
\frac{2E}{\sqrt{3}} &=&\frac{1}{6g}\frac{d^{2}f}{dr^{2}}-\frac{1}{3f^{1/2}}%
\frac{d}{dr}\left( \frac{f^{3/2}}{g^{2}}\frac{dg}{dr}\right) -\frac{4l^{2}f}{%
3g^{3}}-\frac{k}{3g}\ ,  \notag \\
\frac{2H}{\sqrt{3}} &=&-l\frac{d}{dr}\left( fg^{-2}\right) \ .
\label{scalarsmetricfunc}
\end{eqnarray}%
These quantities are constrained by Eqs. (\ref{Eq5}), (\ref{Eq9}), (\ref%
{Eq10}), and (\ref{Eq11})-(\ref{Eq44}). From here by straightforward algebra
we find two independent equations for the metric functions $f\left( r\right) 
$\ and $g\left( r\right) $:%
\begin{eqnarray}
g^{3/2}\frac{d^{2}g^{1/2}}{dr^{2}} &=&l^{2}\ ,  \label{diffeq1} \\
\frac{d^{2}f}{dr^{2}} &=&2k-4\Lambda g\ .  \label{diffeq2}
\end{eqnarray}%
By multiplying the first equation with $dg/dr$ and integrating, we get%
\begin{equation}
\left( \frac{dg}{dr}\right) ^{2}+4l^{2}=C_{1}g\ ,
\end{equation}%
with $C_{1}>0$\textbf{\ }an integration constant. A further integration then
gives%
\begin{equation}
g\left( r\right) =\frac{C_{1}}{4}\left( r+C_{2}\right) ^{2}+\frac{4l^{2}}{%
C_{1}}\ ,
\end{equation}%
with $C_{2}$\textbf{\ }another integration constant. The constant $C_{2}$
can be absorbed into $r$ by a redefinition of its origin, therefore, without
restricting generality we choose $C_{2}=0$. Also the constant $C_{1}$
disappears in the following rescaling of the coordinates and parameters: $%
\left( 2t/C_{1}^{1/2},~C_{1}^{1/2}r/2,~4l^{2}/C_{1}\right) \rightarrow
\left( t,~r,~l^{2}\right) $. Formally this is the same as choosing $C_{1}=4$.

Eq. (\ref{diffeq2}) then gives the other metric function as%
\begin{equation}
f\left( r\right) =C_{4}+C_{3}r+\left( k-2l^{2}\Lambda \right) r^{2}-\frac{%
\Lambda }{3}r^{4}\ ,
\end{equation}%
with $C_{3}$ and $C_{4}$ emerging as integration constants. With the
reparametrizations $C_{3}=-2m$ and $C_{4}=q-kl^{2}+\Lambda l^{4}$ we find
the brane solution given by Eqs. (\ref{ansatz}), (\ref{ansatz1}) and 
\begin{eqnarray}
f\left( r\right) &=&k\left( r^{2}-l^{2}\right) -2mr+q-\Lambda \left( \frac{%
r^{4}}{3}+2l^{2}r^{2}-l^{4}\right) \ ,  \notag \\
g\left( r\right) &=&r^{2}+l^{2}\ ,  \notag \\
\omega _{k}\left( \theta \right) &=&\left\{ 
\begin{array}{c}
2l\cos \theta ~,\qquad k=1 \\ 
\quad -2l\theta ~~~,\qquad k=0 \\ 
~-2l\cosh \theta ~,\qquad k=-1%
\end{array}%
\right. ~.  \label{ansatz2}
\end{eqnarray}%
This is quite similar to the charged-Taub-NUT-(A)dS solution of a general
relativistic Einstein-Maxwell system with mass $m$, electric charge $Q$, NUT
charge $l$, and cosmological constant $\Lambda $, however the constant $q$
replacing $Q^{2}$ is not restricted to positive values. A glance at the
effective Einstein equation (\ref{effective_LRS}) shows that this constant
could possibly arise only from the electric part of the 5d Weyl tensor, Eq. (%
\ref{epsilons}). Indeed from the fourth Eq. (\ref{scalarsmetricfunc}) we get

\begin{equation}
\mathcal{E}=-\frac{q}{\left( l^{2}+r^{2}\right) ^{2}}\ \ .
\end{equation}%
As $q$ originates in the Weyl sector of the higher dimensional space-time,
the derived solution has the interpretation of a \textit{Taub-NUT-(A)dS
brane with tidal charge.}

\section{Concluding remarks\label{concl_sec}}

In this paper we have developed a generic 3+1+1 covariant formalism for
characterizing 5d gravitational dynamics on and outside a brane. The
singled-out directions are the off-brane and temporal directions, thus the
3-spaces are constant time sections of the brane. Generalizing previous
approaches, like 3+1+1 with the extra requirement of double foliability \cite%
{KOZO1}, \cite{KOZO2}, 3+1 in general relativity \cite{Ehlers}-\cite%
{Bonometto} and in brane-worlds \cite{Maartens eqs}, finally 2+1+1 in
general relativity \cite{Clarkson Barrett 211}, \cite{Clarkson 211}, we
presented gravitational evolution and constraint equations in terms of
kinematic, gravito-electro-magnetic and matter variables, defined as
scalars, 3-vectors or 3-tensors. The number of variables being higher than
for other lower dimensional formalisms, the 5d matter and especially the 5d
Weyl tensor leads to a multitude of projections without counterpart in the
mentioned approaches. Only the kinematical set of variables is similar. We
have compared and checked our results with those presented in a recent work
by Nzioki, Carloni, Goswami and Dunsby \cite{NCGD 211} on the 2+1+1
decomposition of $f(R)$ gravity and we give the correspondence between our
and the notations of Ref. \cite{NCGD 211} in Table 1. 
\begin{table}[t]
\caption{Comparison of the notations for the kinematic quantities with Ref. 
\protect\cite{NCGD 211}.}
\label{Table1}
\begin{center}
\begin{tabular}{lll}
$n_{a}\rightarrow e_{a}$ & $\quad K\rightarrow \frac{1}{3}\Theta +\Sigma $ & 
$\,\quad \widehat{K}\rightarrow -A$ \\ 
$A_{a}\rightarrow A_{a}$ & $\,\quad K_{a}\rightarrow \Sigma _{a}+\varepsilon
_{ab}\Omega ^{b}$ & $\,\quad \widehat{K}_{a}\rightarrow \alpha _{a}$ \\ 
$\widehat{A}_{a}\rightarrow a_{a}$ & $\,\quad \Theta \rightarrow \Theta -%
\frac{3}{2}\Sigma $ & $\,\quad \widehat{\Theta }\rightarrow \phi $ \\ 
$L_{a}\rightarrow \Sigma _{a}-\varepsilon _{ab}\Omega ^{b}$ & $\,\quad
\omega _{ab}\rightarrow \Omega \varepsilon _{ab}$ & $\,\quad \widehat{\omega 
}_{ab}\rightarrow \xi \varepsilon _{ab}$ \\ 
$h_{ab}\rightarrow N_{ab}$ & $\,\quad \sigma _{ab}\rightarrow \Sigma _{ab}$
& $\,\quad \widehat{\sigma }_{ab}\rightarrow \zeta _{ab}$%
\end{tabular}%
\end{center}
\end{table}

Our generic formalism contains the complete set of dynamical equations in
the 5d space-time. All equations, with the exception of those in IV to VI
are independent of the particular form of the dynamics on the brane. As
such, they can be employed both to discuss DGP / induced gravity type \cite%
{Induced} branes (a project deferred for future work) and one-brane
Randall-Sundrum type branes. For the latter, we have given both the full set
of equations on the brane, and those containing off-brane evolutions.

The brane equations are more general than previously published results by
the inclusion of arbitrary 5d sources. Although we have recovered the known
fact that in the generic case this system does not close on the brane
(excepting the trivial case $\widehat{\mathcal{E}}_{ab}=0$), by deriving the
complementary system of equations of the 5d dynamics we could establish a
more generic \textit{condition for closure}. This is given by either of Eqs.
(\ref{closurecondgen}), (\ref{closurecond}) which carry much richer
possibilities than the previously known $\widehat{\mathcal{E}}_{ab}=0$ case.

The initial value problem in general relativity is usually discussed in
ADM-like variables (including modifications \ enhancing stability, like the
use of variables with factored-out conformal factors \cite{modADM}),
therefore a similar treatment would be possible to develop in the framework
of the Hamiltonian approach presented in \cite{KOZO1,KOZO2}. A 3+1 covariant
approach for discussing the evolution of cosmic microwave background
anisotropies in the cold dark matter model was employed in Ref. \cite%
{Challinor Lasenby}. The complete set of infinitesimal frame transformations
given in the present paper may turn useful in the study of perturbations in
a 3+1+1 setup.

We have decomposed the Lanczos equations and all source terms of the
effective Einstein equation in terms of the 3+1+1 covariant variables. The
ensemble of these results opens the possibility for applications, with both
cosmological and other symmetries.

We have given generalized Friedmann, Raychaudhuri and energy balance
equations for a generic brane, and by specifying for cosmological symmetries
we have established correspondence with previous related work \cite%
{VarBraneTension}. We have also established the correspondence with the
anisotropic brane-world presented in Ref. \cite{sopuerta}.

We have also employed the 3+1+1 covariant formalism to discuss stationary
space-times with local rotation symmetry of class I, imposed on the brane.
We have shown that such space-times obey the closure condition presented
here. The symmetries and metric ansatz (\ref{ansatz}) implemented in the
3+1+1 covariant formalism led to a simple decoupled system of second order
differential equations for the metric functions, the solution of which gave 
\textit{a new exact solution of the effective Einstein equation, the tidal
charged Taub-NUT-(A)dS brane}, given by Eqs. (\ref{ansatz1}) and (\ref%
{ansatz2}).

In the spherically symmetric and rotating cases tidal charged brane
solutions were already found \cite{tidal}-\cite{tidal Aliev}, which
correspond to electrically charged general relativistic Einstein-Maxwell
solutions, when a formal identification of the electric charge squared with
the tidal charge is carried on. Here we also found that replacing the
electric charge squared in the electrically charged Taub-NUT-(A)dS solution
of an Einstein-Maxwell system with a tidal charge originating in the Weyl
curvature of the 5d space-time leads to a brane solution with the same
symmetries. Unless the electric charge squared, the tidal charge however is
allowed to have both positive and negative values, thus allowing to either
weaken gravity, or contribute towards its confinement on the brane.

\section{Acknowledgements}

We thank Roy Maartens and Chris Clarkson for discussions on the paper and
for suggesting ways to simplify the notations. This work was supported by
the Hungarian Scientific Research Fund (OTKA) grant no. 69036. L\'{A}G was
further supported by the Pol\'{a}nyi and Sun Programs of the Hungarian
National Office for Research and Technology (NKTH) and Collegium Budapest.

\appendix

\section{Commutation relations\label{comm_app}}

In this appendix we enlists some useful differential identities, obtained by
computing the commutators among the derivatives $u^{a}\widetilde{\nabla }%
_{a}\equiv D/d\tau $ (dot), $n^{a}\widetilde{\nabla }_{a}\equiv D/dy$
(prime), and $D_{a}$ (induced metric compatible covariant derivative) on
scalars, brane-vectors and symmetric brane-tensors of second rank.

For a scalar field $\phi $ the following commutation relations hold:%
\begin{eqnarray}
n^{a}\widetilde{\nabla }_{a}(\dot{\phi})-u^{c}\widetilde{\nabla }_{c}(\phi
^{\prime }) &=&K\phi ^{\prime }+\widehat{K}\dot{\phi}+\left( K^{a}-\widehat{K%
}^{a}\right) D_{a}\phi \ , \\
D_{a}\phi ^{\prime }-h_{a}^{\,\,\,\,i}n^{b}\widetilde{\nabla }_{b}\left(
D_{i}\phi \right) &=&-\left( K_{a}-L_{a}\right) \dot{\phi}+\widehat{A}%
_{a}\phi ^{\prime }+\frac{\widehat{\Theta }}{3}D_{a}\phi +\left( \widehat{%
\omega }_{ab}+\widehat{\sigma }_{ab}\right) D^{b}\phi \ ,  \label{Daphiprime}
\\
D_{a}\dot{\phi}-h_{a}^{\,\,\,\,i}u^{b}\widetilde{\nabla }_{b}\left(
D_{i}\phi \right) &=&-A_{a}\dot{\phi}+\left( \widehat{K}_{a}+L_{a}\right)
\phi ^{\prime }+\frac{\Theta }{3}D_{a}\phi +\left( \omega _{ab}+\sigma
_{ab}\right) D^{b}\phi \ ,  \label{Daphidot} \\
D_{[a}D_{b]}\phi &=&\omega _{ab}\dot{\phi}-\widehat{\omega }_{ab}\phi
^{\prime }\ .
\end{eqnarray}

For a 3-vector field $V^{a}$ the following commutation relations hold:%
\begin{eqnarray}
h_{b}^{\,\,\,\,\,j}n^{a}\widetilde{\nabla }_{a}(\dot{V}_{\langle j\rangle
})-h_{b}^{\,\,\,\,\,j}u^{c}\widetilde{\nabla }_{c}(V_{\langle j\rangle
}^{\prime }) &=&-\varepsilon _{kab}\mathcal{H}^{a}V^{b}+KV_{\langle b\rangle
}^{\prime }+\widehat{K}\dot{V}_{\langle b\rangle }+\left( K^{a}-\widehat{K}%
^{a}\right) D_{a}V_{b}  \notag \\
&&-A_{a}V^{a}K_{b}+\widehat{K}_{a}V^{a}\widehat{A}_{b}+K_{a}V^{a}A_{b}-%
\widehat{A}_{a}V^{a}\widehat{K}_{b}\ ,
\end{eqnarray}%
\begin{eqnarray}
h_{c}^{\,\,\,\,i}h_{d}^{\,\,\,\,\,\,j}u^{a}\widetilde{\nabla }_{a}\left(
D_{i}V_{j}\right) -D_{c}\left( \dot{V}_{\langle d\rangle }\right) &=&-\left( 
\widehat{K}_{c}+L_{c}\right) V_{\langle d\rangle }^{\prime }+A_{c}\dot{V}%
_{\langle d\rangle }-\frac{\Theta }{3}D_{c}V_{d}+\frac{\widehat{\Theta }}{3}%
\widehat{K}_{d}V_{c}-\frac{\Theta }{3}A_{d}V_{c}  \notag \\
&&-\left( \sigma _{ac}-\omega _{ac}\right) D^{a}V_{d}+\frac{1}{2}\widehat{%
\mathcal{E}}_{a}V^{a}h_{cd}-\frac{1}{2}\widehat{\mathcal{E}}%
_{d}V_{c}-\varepsilon _{dab}\mathcal{H}_{c}^{\,\,\,b}V^{a}+\frac{\widetilde{%
\kappa }^{2}}{3}\widetilde{q}_{a}V^{a}h_{cd}  \notag \\
&&-\frac{\widetilde{\kappa }^{2}}{3}V_{c}\widetilde{q}_{d}-\left( \frac{%
\widehat{\Theta }}{3}h_{cd}+\widehat{\omega }_{cd}+\widehat{\sigma }%
_{cd}\right) \widehat{K}_{a}V^{a}+\widehat{K}_{d}\left( \widehat{\omega }%
_{ca}+\widehat{\sigma }_{ca}\right) V^{a}  \notag \\
&&+\left( \frac{\Theta }{3}h_{cd}+\omega _{cd}+\sigma _{cd}\right)
A_{a}V^{a}-A_{d}\left( \omega _{ca}+\sigma _{ca}\right) V^{a}\ ,
\label{DaVbdot} \\
h_{c}^{\,\,\,\,i}h_{d}^{\,\,\,\,\,\,j}n^{a}\widetilde{\nabla }_{a}\left(
D_{i}V_{j}\right) -D_{c}\left( V_{\langle d\rangle }^{\prime }\right)
&=&-\left( L_{c}-K_{c}\right) \dot{V}_{\langle d\rangle }-\widehat{A}%
_{c}V_{\langle d\rangle }^{\prime }-\frac{\widehat{\Theta }}{3}D_{c}V_{d}-%
\frac{\Theta }{3}K_{d}V_{c}+\frac{\widehat{\Theta }}{3}\widehat{A}_{d}V_{c} 
\notag \\
&&-\left( \widehat{\sigma }_{ac}-\widehat{\omega }_{ac}\right) D^{a}V_{d}-%
\frac{1}{2}\mathcal{E}_{a}V^{a}h_{cd}+\frac{1}{2}\mathcal{E}%
_{d}V_{c}-\varepsilon _{dab}\widehat{\mathcal{H}}_{c}^{\,\,\,b}V^{a}-\frac{%
\widetilde{\kappa }^{2}}{3}\widetilde{\pi }_{a}V^{a}h_{cd}  \notag \\
&&+\frac{\widetilde{\kappa }^{2}}{3}V_{c}\widetilde{\pi }_{d}+\left( \frac{%
\Theta }{3}h_{cd}+\omega _{cd}+\sigma _{cd}\right) K_{a}V^{a}-K_{d}\left(
\omega _{ca}+\sigma _{ca}\right) V^{a}  \notag \\
&&-\left( \frac{\widehat{\Theta }}{3}h_{cd}+\widehat{\omega }_{cd}+\widehat{%
\sigma }_{cd}\right) \widehat{A}_{a}V^{a}+\widehat{A}_{d}\left( \widehat{%
\omega }_{ca}+\widehat{\sigma }_{ca}\right) V^{a}\ ,  \label{DaVbprime}
\end{eqnarray}%
\begin{eqnarray}
D_{[a}D_{b]}V_{c} &=&\omega _{ab}\dot{V}_{\langle c\rangle }-\widehat{\omega 
}_{ab}V_{\langle c\rangle }^{\prime }+h_{c[a}\mathcal{E}_{b]d}V^{d}+\mathcal{%
E}_{c[a}V_{b]}-h_{c[a}\widehat{\mathcal{E}}_{b]d}V^{d}-\widehat{\mathcal{E}}%
_{c[a}V_{b]}  \notag \\
&&-\frac{1}{9}\left( \Theta ^{2}-\widehat{\Theta }^{2}\right) h_{c[a}V_{b]}-%
\frac{\Theta }{3}\left( \sigma _{c[a}-\omega _{c[a}\right) V_{b]}+\frac{%
\widehat{\Theta }}{3}\left( \widehat{\sigma }_{c[a}-\widehat{\omega }%
_{c[a}\right) V_{b]}  \notag \\
&&-\frac{\Theta }{3}h_{c[a}\left( \sigma _{b]d}+\omega _{b]d}\right) V^{d}+%
\frac{\widehat{\Theta }}{3}h_{c[a}\left( \widehat{\sigma }_{b]d}+\widehat{%
\omega }_{b]d}\right) V^{d}-\frac{1}{3}\mathcal{E}h_{c[a}V_{b]}  \notag \\
&&-\left( \sigma _{c[a}-\omega _{c[a}\right) \left( \omega _{b]d}+\sigma
_{b]d}\right) V^{d}+\left( \widehat{\sigma }_{c[a}-\widehat{\omega }%
_{c[a}\right) \left( \widehat{\omega }_{b]d}+\widehat{\sigma }_{b]d}\right)
V^{d}  \notag \\
&&+\frac{\widetilde{\Lambda }}{6}h_{c[a}V_{b]}+\frac{\widetilde{\kappa }^{2}%
}{6}\left( \widetilde{\rho }-\widetilde{\pi }+\widetilde{p}\right)
h_{c[a}V_{b]}+\frac{\widetilde{\kappa }^{2}}{3}h_{c[a}\widetilde{\pi }%
_{b]d}V^{d}+\frac{\widetilde{\kappa }^{2}}{3}\widetilde{\pi }_{c[a}V_{b]}\ .
\end{eqnarray}%
For a symmetric trace-free 3-tensor field $T_{ab}$ the following commutation
relations hold:%
\begin{eqnarray}
h_{\langle c}^{\,\,\,\,\,\,i}h_{d\rangle }^{\,\,\,\,\,\,j}n^{a}\widetilde{%
\nabla }_{a}(\dot{T}_{\langle ij\rangle })-h_{\langle
c}^{\,\,\,\,\,\,i}h_{d\rangle }^{\,\,\,\,\,\,j}u^{b}\widetilde{\nabla }%
_{b}(T_{\langle ij\rangle }^{\prime }) &=&KT_{\langle cd\rangle }^{\prime }+%
\widehat{K}\dot{T}_{\langle cd\rangle }+K^{a}D_{a}T_{cd}-\widehat{K}%
^{a}D_{a}T_{cd}  \notag \\
&&-2K_{\langle c}T_{d\rangle a}A^{a}+2\widehat{A}_{\langle c}T_{d\rangle a}%
\widehat{K}^{a}+2A_{\langle c}T_{d\rangle a}K^{a}  \notag \\
&&-2\widehat{K}_{\langle c}T_{d\rangle a}\widehat{A}^{a}-2\varepsilon
_{ab\langle d}T_{c\rangle }^{\,\,\,\,\,a}\mathcal{H}^{b}\ ,
\end{eqnarray}%
\begin{eqnarray}
h_{a}^{\,\,\,\,\,k}h_{\langle b}^{\,\,\,\,\,\,i}h_{c\rangle
}^{\,\,\,\,\,\,j}n^{d}\widetilde{\nabla }_{d}\left( D_{k}T_{ij}\right)
-D_{a}\left( T_{\langle bc\rangle }^{\prime }\right) &=&-\left(
L_{a}-K_{a}\right) \dot{T}_{\langle bc\rangle }-\widehat{A}_{a}T_{\langle
bc\rangle }^{\prime }-\frac{\widehat{\Theta }}{3}D_{d}T_{bc}+\mathcal{E}%
_{\langle b}T_{c\rangle a}  \notag \\
&&-\left( \widehat{\omega }_{ad}+\widehat{\sigma }_{ad}\right)
D^{d}T_{bc}-2T_{\langle b}^{\,\,\,\,\,d}\varepsilon _{c\rangle di}\widehat{%
\mathcal{H}}_{a}^{\,\,\,\,i}+\frac{2\Theta }{3}h_{a\langle b}T_{c\rangle
d}K^{d}  \notag \\
&&-h_{a\langle b}T_{c\rangle }^{\,\,\,\,\,d}\mathcal{E}_{d}+\frac{2\widehat{%
\Theta }}{3}\widehat{A}_{\langle b}T_{c\rangle a}-\frac{2\widehat{\Theta }}{3%
}h_{a\langle b}T_{c\rangle d}\widehat{A}^{d}  \notag \\
&&-2\left( \omega _{ad}+\sigma _{ad}\right) K_{\langle b}T_{c\rangle
}^{\,\,\,\,\,d}+2\left( \widehat{\omega }_{ad}+\widehat{\sigma }_{ad}\right) 
\widehat{A}_{\langle b}T_{c\rangle }^{\,\,\,\,\,d}  \notag \\
&&+2\left( \omega _{a\langle b}+\sigma _{a\langle b}\right) T_{c\rangle
d}K^{d}-2\left( \widehat{\omega }_{a\langle b}+\widehat{\sigma }_{a\langle
b}\right) T_{c\rangle d}\widehat{A}^{d}  \notag \\
&&-\frac{2\Theta }{3}K_{\langle b}T_{c\rangle a}-\frac{2\widetilde{\kappa }%
^{2}}{3}h_{a\langle b}T_{c\rangle }^{\,\,\,\,\,d}\widetilde{\pi }_{d}+\frac{2%
\widetilde{\kappa }^{2}}{3}\widetilde{\pi }_{\langle b}T_{c\rangle a}\ ,
\end{eqnarray}%
\begin{eqnarray}
h_{a}^{\,\,\,\,\,k}h_{\langle b}^{\,\,\,\,\,\,i}h_{c\rangle
}^{\,\,\,\,\,\,j}u^{d}\widetilde{\nabla }_{d}\left( D_{k}T_{ij}\right)
-D_{a}\left( \dot{T}_{\langle bc\rangle }\right) &=&-\left( \widehat{K}%
_{a}+L_{a}\right) T_{\langle bc\rangle }^{\prime }+A_{a}\dot{T}_{\langle
bc\rangle }-\frac{\Theta }{3}D_{d}T_{bc}-\widehat{\mathcal{E}}_{\langle
b}T_{c\rangle a}  \notag \\
&&-\left( \omega _{ad}+\sigma _{ad}\right) D^{d}T_{bc}-2T_{\langle
b}^{\,\,\,\,\,\,d}\varepsilon _{c\rangle di}\mathcal{H}_{a}^{\,\,\,\,i}+h_{a%
\langle b}T_{c\rangle }^{\,\,\,\,\,d}\widehat{\mathcal{E}}_{d}  \notag \\
&&-\frac{2\widehat{\Theta }}{3}h_{a\langle b}T_{c\rangle d}\widehat{K}^{d}+%
\frac{2\widehat{\Theta }}{3}\widehat{K}_{\langle b}T_{c\rangle a}-\frac{%
2\Theta }{3}A_{\langle b}T_{c\rangle a}  \notag \\
&&+2\left( \widehat{\omega }_{ad}+\widehat{\sigma }_{ad}\right) \widehat{K}%
_{\langle b}T_{c\rangle }^{\,\,\,\,\,d}-2\left( \omega _{ad}+\sigma
_{ad}\right) A_{\langle b}T_{c\rangle }^{\,\,\,\,\,\,d}  \notag \\
&&-2\left( \widehat{\omega }_{a\langle b}+\widehat{\sigma }_{a\langle
b}\right) T_{c\rangle d}\widehat{K}^{d}+2\left( \omega _{a\langle b}+\sigma
_{a\langle b}\right) T_{c\rangle d}A^{d}  \notag \\
&&+\frac{2\Theta }{3}h_{a\langle b}T_{c\rangle d}A^{d}+\frac{2\widetilde{%
\kappa }^{2}}{3}h_{a\langle b}T_{c\rangle }^{\,\,\,\,\,d}\widetilde{q}_{d}-%
\frac{2\widetilde{\kappa }^{2}}{3}\widetilde{q}_{\langle b}T_{c\rangle a}\ .
\end{eqnarray}%
These results apply for arbitrary (not necessarily small) scalar, vector and
tensor fields.

\section{Infinitesimal frame transformations\label{frame_app}}

An infinitesimal frame transformation from the diad ($u^{a}$, $n^{a}$) to
the diad ($\overline{u}^{a}$, $\overline{n}^{a}$) can be defined as a
generalization of the corresponding general relativistic procedure \cite%
{Maartens nonlin} as:%
\begin{eqnarray}
\overline{u}_{a} &=&u_{a}+\upsilon _{a}+\nu n_{a}\ ,\ \text{with\ \ }%
u^{a}\upsilon _{a}=n^{a}\upsilon _{a}=0\ ,  \label{ua_frame_change} \\
\overline{n}_{a} &=&n_{a}+l_{a}+mu_{a}\ ,\ \text{with\ \ }%
u^{a}l_{a}=n^{a}l_{a}=0\ ,  \label{na_frame_change}
\end{eqnarray}%
where $\upsilon _{a},~l_{a},~\nu ,~m$ are all $\mathcal{O}\left( 1\right) $.%
\footnote{%
The quantities denoted $\mathcal{O}\left( 1\right) $ all vanish for
identical transformation. We also assume a first order accuracy, thus all
quantities $\mathcal{O}\left( 2\right) =\mathcal{O}\left( 1\right) ^{2}$
will be dropped.} The new diad also obeys 
\begin{equation}
\overline{u}^{a}\overline{u}_{a}=-1\ ,\ \overline{n}^{a}\overline{n}_{a}=1\
,\ \overline{u}^{a}\overline{n}_{a}=0\ ,
\end{equation}%
which implies%
\begin{equation}
\nu =m\ .
\end{equation}%
For $\upsilon _{a}=l_{a}=0$ the above parameters define an infinitesimal
orthogonal transformation (this is a 2-dimensional infinitesimal Lorentz
boost). The parameters $\upsilon _{a}$ and $l_{a}$ represent infinitesimal
translations. These transformations represent gauge degrees of freedom,
worth to explore in order to achieve particular tasks, for example to
fulfill physical conditions, to conveniently close subsets of differential
equations etc.

The fundamental algebraic tensors $h_{ab}$ and $\varepsilon _{abc}$ change
accordingly:%
\begin{eqnarray}
\overline{h}_{ab} &=&h_{ab}+2u_{(a}\upsilon _{b)}-2n_{(a}l_{b)}\ ,
\label{hbar} \\
\overline{\varepsilon }_{abc} &=&\varepsilon _{abc}-\left( n_{c}\varepsilon
_{abe}+2n_{[a}\varepsilon _{b]ce}\right) l^{e}+\left( u_{c}\varepsilon
_{abd}+2u_{[a}\varepsilon _{b]cd}\right) \upsilon ^{d}\ .
\end{eqnarray}%
The new 3-metric obeys $\overline{h}_{ab}\overline{n}^{a}=\overline{h}_{ab}%
\overline{u}^{a}=0$.

The covariant derivatives of the new diad vectors can be invariantly
decomposed as%
\begin{eqnarray}
\widetilde{\nabla }_{a}\overline{u}_{b} &=&-\overline{u}_{a}\overline{A}_{b}+%
\overline{\widehat{K}}\overline{u}_{a}\overline{n}_{b}+\overline{K}\overline{%
n}_{a}\overline{n}_{b}+\overline{n}_{a}\overline{K}_{b}+\overline{L}_{a}%
\overline{n}_{b}+\frac{\overline{\Theta }}{3}\overline{h}_{ab}+\overline{%
\omega }_{ab}+\overline{\sigma }_{ab}\ , \\
\widetilde{\nabla }_{a}\overline{n}_{b} &=&\overline{n}_{a}\overline{%
\widehat{A}}_{b}+\overline{K}\overline{n}_{a}\overline{u}_{b}+\overline{%
\widehat{K}}\overline{u}_{a}\overline{u}_{b}-\overline{u}_{a}\overline{%
\widehat{K}}_{b}+\overline{L}_{a}\overline{u}_{b}+\frac{\overline{\widehat{%
\Theta }}}{3}\overline{h}_{ab}+\overline{\widehat{\omega }}_{ab}+\overline{%
\widehat{\sigma }}_{ab}\ ,
\end{eqnarray}%
which imply the following transformations for the kinematic quantities:

\begin{eqnarray}
\overline{K} &=&K-\widehat{A}_{a}\upsilon ^{a}+\left( K_{a}+L_{a}\right)
l^{a}-m\widehat{K}+m^{\prime }\ , \\
\overline{\widehat{K}} &=&\widehat{K}-A_{a}l^{a}+\left( \widehat{K}%
_{a}-L_{a}\right) \upsilon ^{a}-mK-\dot{m}\ , \\
\overline{\widehat{\Theta }} &=&\widehat{\Theta }+m\Theta +D^{a}l_{a}-l^{a}%
\widehat{A}_{a}-\upsilon ^{a}\left( L_{a}-\widehat{K}_{a}\right) \ , \\
\overline{\Theta } &=&\Theta +m\widehat{\Theta }+D^{a}\upsilon _{a}+\upsilon
^{a}A_{a}-l^{a}\left( K_{a}+L_{a}\right) \ , \\
\overline{A}_{c} &=&A_{c}+\upsilon ^{b}A_{b}u_{c}+\widehat{K}%
l_{c}-l^{b}A_{b}n_{c}+\frac{\Theta }{3}\upsilon _{c}+\left( \omega
_{ac}+\sigma _{ac}\right) \upsilon ^{a}+m\left( K_{c}+\widehat{K}_{c}\right)
+\dot{\upsilon}_{\langle c\rangle }\ , \\
\overline{\widehat{A}}_{c} &=&\widehat{A}_{c}-l^{b}\widehat{A}%
_{b}n_{c}-K\upsilon _{c}+\upsilon ^{b}\widehat{A}_{b}u_{c}+\frac{\widehat{%
\Theta }}{3}l_{c}+\left( \widehat{\omega }_{ac}+\widehat{\sigma }%
_{ac}\right) l^{a}+m\left( K_{c}+\widehat{K}_{c}\right) +l_{\langle c\rangle
}^{\prime }\ , \\
\overline{K}_{c} &=&K_{c}-Kl_{c}+u_{c}\upsilon ^{b}K_{b}-n_{c}l^{b}K_{b}+%
\frac{\Theta }{3}l_{c}+\left( \sigma _{ac}+\omega _{ac}\right) l^{a}+m\left(
A_{c}+\widehat{A}_{c}\right) +\upsilon _{\langle c\rangle }^{\prime }\ ,
\label{Framechange_Ka} \\
\overline{\widehat{K}}_{c} &=&\widehat{K}_{c}+\widehat{K}\upsilon
_{c}-n_{c}l^{b}\widehat{K}_{b}+u_{c}\upsilon ^{b}\widehat{K}_{b}+\frac{%
\widehat{\Theta }}{3}\upsilon _{c}+\left( \widehat{\sigma }_{ac}+\widehat{%
\omega }_{ac}\right) \upsilon ^{a}+m\left( A_{c}+\widehat{A}_{c}\right) +%
\dot{l}_{\langle c\rangle }\ , \\
\overline{L}_{c} &=&L_{c}-\widehat{K}\upsilon _{c}-Kl_{c}+u_{c}\upsilon
^{a}L_{a}-n_{c}l^{a}L_{a}-\frac{\widehat{\Theta }}{3}\upsilon _{c}  \notag \\
&&+\frac{\Theta }{3}l_{c}+D_{c}m-\left( \widehat{\sigma }_{cb}+\widehat{%
\omega }_{cb}\right) \upsilon ^{b}+\left( \sigma _{ca}+\omega _{ca}\right)
l^{a}\ , \\
\overline{\widehat{\sigma }}_{cd} &=&\widehat{\sigma }_{cd}+m\sigma
_{cd}+D_{\langle c}l_{d\rangle }+\upsilon _{\langle c}\left( \widehat{K}%
_{d\rangle }-L_{d\rangle }\right) -l_{\langle c}\widehat{A}_{d\rangle
}+2\upsilon ^{a}\widehat{\sigma }_{a(d}u_{c)}-2l^{a}\widehat{\sigma }%
_{a(d}n_{c)}\ ,  \label{Framechange_sigmahatab} \\
\overline{\sigma }_{cd} &=&\sigma _{cd}+m\widehat{\sigma }_{cd}+D_{\langle
c}\upsilon _{d\rangle }-l_{\langle c}\left( K_{d\rangle }+L_{d\rangle
}\right) +\upsilon _{\langle c}A_{d\rangle }-2l^{a}\sigma
_{a(d}n_{c)}+2\upsilon ^{a}\sigma _{a(d}u_{c)}\ , \\
\overline{\widehat{\omega }}_{cd} &=&\widehat{\omega }_{cd}+m\omega
_{cd}+D_{[c}l_{d]}+\upsilon _{\lbrack c}\left( \widehat{K}%
_{d]}-L_{d]}\right) -l_{[c}\widehat{A}_{d]}+2\upsilon ^{a}\widehat{\omega }%
_{a[d}u_{c]}-2l^{a}\widehat{\omega }_{a[d}n_{c]}\ , \\
\overline{\omega }_{cd} &=&\omega _{cd}+m\widehat{\omega }%
_{cd}+D_{[c}\upsilon _{d]}-l_{[c}\left( K_{d]}+L_{d]}\right) +\upsilon
_{\lbrack c}A_{d]}-2l^{a}\omega _{a[d}n_{c]}+2\upsilon ^{a}\omega
_{a[d}u_{c]}\ .
\end{eqnarray}%
Similarly, the transformation laws for the gravito-electro-magnetic
quantities are%
\begin{eqnarray}
\overline{\mathcal{E}} &=&\mathcal{E}+2\mathcal{E}_{a}l^{a}+2\widehat{%
\mathcal{E}}_{a}l^{a}\ , \\
\overline{\mathcal{E}}_{k} &=&\mathcal{E}_{k}-m\widehat{\mathcal{E}}_{k}-%
\frac{4}{3}\mathcal{E}l_{k}+\mathcal{E}_{a}\upsilon ^{a}u_{k}-\mathcal{E}%
_{a}l^{a}n_{k}+\varepsilon _{kab}\mathcal{H}^{a}\upsilon ^{b}+\mathcal{E}%
_{ka}l^{a}-\mathcal{F}_{ka}\upsilon ^{a}\ , \\
\overline{\widehat{\mathcal{E}}}_{k} &=&\widehat{\mathcal{E}}_{k}-m\mathcal{E%
}_{k}+\frac{4}{3}\mathcal{E}\upsilon _{k}-\widehat{\mathcal{E}}%
_{a}l^{a}n_{k}+\widehat{\mathcal{E}}_{a}\upsilon ^{a}u_{k}-\varepsilon _{kab}%
\mathcal{H}^{a}l^{b}+\widehat{\mathcal{E}}_{ka}\upsilon ^{a}-\mathcal{F}%
_{ka}l^{a}\ , \\
\overline{\mathcal{H}}_{k} &=&\mathcal{H}_{k}-\frac{1}{2}\varepsilon _{k}^{\
\ ab}\widehat{\mathcal{E}}_{a}l_{b}-\frac{1}{2}\varepsilon _{k}^{\ \ ab}%
\mathcal{E}_{a}\upsilon _{b}-n_{k}\mathcal{H}_{a}l^{a}+u_{k}\mathcal{H}%
_{a}\upsilon ^{a}-\varepsilon _{k}^{\ \ ab}\mathcal{E}_{a}\upsilon _{b}+%
\widehat{\mathcal{H}}_{ka}\upsilon ^{a}-\mathcal{H}_{ka}l^{a}\ , \\
\overline{\mathcal{F}}_{kl} &=&\mathcal{F}_{kl}+2u_{(k}\mathcal{F}%
_{l)a}\upsilon ^{a}-2n_{(k}\mathcal{F}_{l)a}l^{a}-\frac{3}{2}\mathcal{E}%
_{\langle k}\upsilon _{l\rangle }  \notag \\
&&+\frac{3}{2}\widehat{\mathcal{E}}_{\langle k}l_{l\rangle }+m\widehat{%
\mathcal{E}}_{kl}+m\mathcal{E}_{kl}-\varepsilon _{ab(k}\widehat{\mathcal{H}}%
_{l)}^{\,\,\,a}\upsilon ^{b}-\varepsilon _{ab(k}\mathcal{H}%
_{l)}^{\,\,\,a}l^{b}\ , \\
\overline{\widehat{\mathcal{E}}}_{kl} &=&\widehat{\mathcal{E}}_{kl}+2m%
\mathcal{F}_{kl}+2\upsilon _{\langle k}\widehat{\mathcal{E}}_{l\rangle }-%
\mathcal{E}_{\langle k}l_{l\rangle }+2u_{(k}\widehat{\mathcal{E}}%
_{l)a}\upsilon ^{a}-2n_{(k}\widehat{\mathcal{E}}_{l)a}l^{a}-2\varepsilon
_{ab\langle k}\widehat{\mathcal{H}}_{l\rangle }^{\,\,\,\,\,a}l^{b}\ ,
\label{eptranf} \\
\overline{\mathcal{E}}_{kl} &=&\mathcal{E}_{kl}+2m\mathcal{F}%
_{kl}-2l_{\langle k}\mathcal{E}_{l\rangle }+\widehat{\mathcal{E}}_{\langle
k}\upsilon _{l\rangle }+2u_{\langle k}\mathcal{E}_{l\rangle a}\upsilon
^{a}-2n_{\langle k}\mathcal{E}_{l\rangle a}l^{a}-2\varepsilon _{ab\langle k}%
\mathcal{H}_{l\rangle }^{\,\,\,\,a}\upsilon ^{b}\ , \\
\overline{\mathcal{H}}_{kl} &=&\mathcal{H}_{kl}+m\widehat{\mathcal{H}}_{kl}+%
\frac{3}{2}\mathcal{H}_{\langle k}l_{l\rangle }-\varepsilon _{(k}^{\ \ \ \
ab}\mathcal{F}_{l)a}l_{b}+2\varepsilon _{(k}^{\ \ \ \ ab}\mathcal{E}%
_{l)a}\upsilon _{b}  \notag \\
&&-\varepsilon _{(k}^{\ \ \ \ ab}\widehat{\mathcal{E}}_{l)a}\upsilon
_{b}-2n_{(k}\mathcal{H}_{l)a}l^{a}+2u_{(k}\mathcal{H}_{l)a}\upsilon ^{a}\ ,
\\
\overline{\widehat{\mathcal{H}}}_{kl} &=&\widehat{\mathcal{H}}_{kl}+m%
\mathcal{H}_{kl}+\frac{3}{2}\mathcal{H}_{\langle k}\upsilon _{l\rangle
}+\varepsilon _{(k}^{\ \ \ \ ab}\mathcal{F}_{l)a}\upsilon _{b}-2\varepsilon
_{(k}^{\ \ \ \ ab}\widehat{\mathcal{E}}_{l)a}l_{b}  \notag \\
&&+\varepsilon _{(k}^{\ \ \ \ ab}\mathcal{E}_{l)a}l_{b}+2u_{(k}\widehat{%
\mathcal{H}}_{l)a}\upsilon ^{a}-2n_{(k}\widehat{\mathcal{H}}_{l)a}l^{a}\ .
\end{eqnarray}%
The matter variables transform as%
\begin{eqnarray}
\overline{\widetilde{\rho }} &=&\widetilde{\rho }-2\upsilon ^{a}\widetilde{q}%
_{a}-2m\widetilde{q}\ , \\
\overline{\widetilde{\pi }} &=&\widetilde{\pi }+2l^{a}\widetilde{\pi }_{a}-2m%
\widetilde{q}\ , \\
\overline{\widetilde{p}} &=&\widetilde{p}-\frac{2}{3}\upsilon ^{a}\widetilde{%
q}_{a}-\frac{2}{3}l^{a}\widetilde{\pi }_{a}\ , \\
\overline{\widetilde{q}} &=&\widetilde{q}+l^{a}\widetilde{q}_{a}-\upsilon
^{a}\widetilde{\pi }_{a}-m\left( \widetilde{\rho }+\widetilde{\pi }\right) \
, \\
\overline{\widetilde{q}}_{a} &=&\widetilde{q}_{a}-\widetilde{\rho }\upsilon
_{a}-\widetilde{q}l_{a}-2\widetilde{p}\upsilon _{a}-2\widetilde{\pi }%
_{ab}\upsilon ^{b}+u_{a}\upsilon ^{c}\widetilde{q}_{c}-n_{a}l^{c}\widetilde{q%
}_{c}-m\widetilde{\pi }_{a}\ , \\
\overline{\widetilde{\pi }}_{a} &=&\widetilde{\pi }_{a}-\widetilde{q}%
\upsilon _{a}+2\widetilde{p}l_{a}-\widetilde{\pi }l_{a}+2\widetilde{\pi }%
_{ab}l^{b}-n_{a}l^{c}\widetilde{\pi }_{c}+u_{a}\upsilon ^{c}\widetilde{\pi }%
_{c}-m\widetilde{q}_{a}\ , \\
\widetilde{\pi }_{ab} &=&\pi _{ab}+2\widetilde{p}\upsilon _{\langle
a}u_{b\rangle }-2\widetilde{p}l_{\langle a}n_{b\rangle }-2\widetilde{q}%
_{\langle a}\upsilon _{b\rangle }+2\pi _{\langle
a}^{\,\,\,\,\,\,d}u_{b\rangle }\upsilon _{d}-2\pi _{\langle
a}^{\,\,\,\,\,\,c}n_{b\rangle }l_{c}-2\widetilde{\pi }_{\langle
a}l_{b\rangle }\ .
\end{eqnarray}

We have checked that in the particular case~$l^{a}=0=m$, by applying the
Lanczos equations (\ref{Bound1})-(\ref{Bound4}) for eliminating the
quantities $\widehat{K}$, $\widehat{\Theta }$, $L_{a}=-\widehat{K}_{a}$ and $%
\widehat{\sigma }_{ab}$, by suppressing the quantities related to the brane
normal, in particular imposing $\widehat{\omega }_{ab}=0$, we recover the
linearized form of the transformation laws for the kinematical and dynamical
quantities $\left( \Theta ,~\sigma _{ab},~\omega _{ab},~A_{a}\right) $ and $%
\left( \rho ,~p,~q_{a},~\pi _{ab}\right) $ of Ref. \cite{Maartens nonlin}.
Similarly, by employing Eqs. (\ref{Eab}) and (\ref{Hab}), we obtain the
required transformations for the Weyl projections $\left(
E_{ab},~H_{ab}\right) $.

If we would like to apply the generic transformations derived in this
Appendix in a brane-world scenario, we have to impose $l^{a}=0=m$ on the
brane (in order to preserve the vector field $n^{a}$ at $y=0$, which defines
the brane), however the derivatives along the off-brane direction (the
derivatives denoted by prime) of these quantities can be different from zero
even at $y=0$.

\section{Gravitational evolution and constraint equations on an
asymmetrically embedded brane\label{asym_app}}

The brane equations describing the gravitational dynamics are

\begin{eqnarray}
0 &=&\dot{\widehat{\Theta }}-D^{a}\widehat{K}_{a}+\left( \widehat{K}+\frac{%
\widehat{\Theta }}{3}\right) \Theta -2\widehat{K}^{a}A_{a}+\widehat{\sigma }%
_{ab}\sigma ^{ab}-\widetilde{\kappa }^{2}\widetilde{q}\ ,
\label{Thetahat_brane_dot} \\
0 &=&\dot{\widehat{K}}_{\langle a\rangle }-D_{a}\left( \widehat{K}-\frac{2}{3%
}\widehat{\Theta }\right) -D^{b}\widehat{\sigma }_{ab}+\frac{4\Theta }{3}%
\widehat{K}_{a}-\left( \widehat{K}+\frac{\widehat{\Theta }}{3}\right) A_{a}-%
\widehat{\sigma }_{ab}A^{b}-\omega _{ab}\widehat{K}^{b}+\sigma _{ab}\widehat{%
K}^{b}+\widetilde{\kappa }^{2}\widetilde{\pi }_{a}\ , \\
0 &=&\dot{\mathcal{E}}-D^{a}\widehat{\mathcal{E}}_{a}+\frac{4}{3}\Theta 
\mathcal{E}+\widehat{\mathcal{E}}_{ab}\sigma ^{ab}-2\widehat{\mathcal{E}}%
_{a}A^{a}+\widehat{\sigma }^{ab}\dot{\widehat{\sigma }}_{\langle ab\rangle }-%
\widehat{\sigma }^{ab}D_{a}\widehat{K}_{b}-\widehat{K}^{a}D^{b}\widehat{%
\sigma }_{ab}+\frac{2}{3}\widehat{K}^{a}D_{a}\widehat{\Theta }  \notag \\
&&-2A_{b}\widehat{K}_{a}\widehat{\sigma }^{ab}+\frac{\Theta }{3}\widehat{%
\sigma }_{ab}\widehat{\sigma }^{ab}+\left( \widehat{K}+\frac{\widehat{\Theta 
}}{3}\right) \sigma _{ab}\widehat{\sigma }^{ab}+\sigma _{ca}\widehat{\sigma }%
_{b}^{\,\,\,\,c}\widehat{\sigma }^{ab}+\frac{2\Theta }{3}\widehat{K}_{a}%
\widehat{K}^{a}-\sigma _{ab}\widehat{K}^{a}\widehat{K}^{b}  \notag \\
&&-\frac{\widetilde{\kappa }^{2}}{2}\left( \widetilde{\rho }-\widetilde{\pi }%
+\widetilde{p}\right) ^{\cdot }-\frac{2\widetilde{\kappa }^{2}}{3}D^{a}%
\widetilde{q}_{a}-\frac{2\widetilde{\kappa }^{2}}{3}\Theta \left( \widetilde{%
\rho }+\widetilde{p}\right) -\frac{2\widetilde{\kappa }^{2}}{3}\widehat{%
\Theta }\widetilde{q}-\frac{4\widetilde{\kappa }^{2}}{3}\widetilde{q}%
_{a}A^{a}-\frac{2\widetilde{\kappa }^{2}}{3}\widetilde{\pi }_{ab}\sigma
^{ab}\ ,
\end{eqnarray}%
\begin{eqnarray}
0 &=&\dot{\widehat{\mathcal{E}}}_{\langle k\rangle }+\frac{4}{3}\Theta 
\widehat{\mathcal{E}}_{k}-\frac{1}{3}D_{k}\mathcal{E}-\frac{4\mathcal{E}}{3}%
A_{k}-D^{a}\widehat{\mathcal{E}}_{ka}-\widehat{\mathcal{E}}_{ka}A^{a}-\left(
\omega _{ka}-\sigma _{ka}\right) \widehat{\mathcal{E}}^{a}+\dot{\widehat{%
\sigma }}_{\langle ka\rangle }\widehat{K}^{a}  \notag \\
&&-\left( \widehat{K}+\frac{\widehat{\Theta }}{3}\right) D^{b}\widehat{%
\sigma }_{kb}+\widehat{K}^{a}\sigma _{ck}\widehat{\sigma }_{a}^{\,\,\,\,c}+%
\frac{2}{3}\left( \widehat{K}+\frac{\widehat{\Theta }}{3}\right) D_{k}%
\widehat{\Theta }+\frac{2\Theta }{3}\left( \widehat{K}+\frac{\widehat{\Theta 
}}{3}\right) \widehat{K}_{k}-2\widehat{K}^{a}D_{k}\widehat{K}_{a}+\widehat{K}%
^{a}D_{a}\widehat{K}_{k}  \notag \\
&&+\frac{1}{3}\widehat{K}_{k}D^{a}\widehat{K}_{a}-2\widehat{K}^{a}A_{\langle
k}\widehat{K}_{a\rangle }-\sigma _{ba}\widehat{\sigma }_{k}^{\,\,\,\,b}%
\widehat{K}^{a}+\frac{2}{3}\widehat{K}_{k}\sigma _{ab}\widehat{\sigma }^{ab}+%
\widehat{\sigma }^{ab}D_{k}\widehat{\sigma }_{ab}-\widehat{\sigma }%
_{b}^{\,\,\,\,a}D^{b}\widehat{\sigma }_{ka}-\frac{1}{3}\widehat{\sigma }%
_{k}^{\,\,\,\,a}D_{a}\widehat{\Theta }  \notag \\
&&+\varepsilon _{cab}\widehat{K}^{a}\omega ^{b}\widehat{\sigma }%
_{k}^{\,\,\,\,c}+\varepsilon _{k}^{\ \ ab}\widehat{K}_{c}\omega _{a}\widehat{%
\sigma }_{b}^{\,\,\,\,c}-\frac{2\widetilde{\kappa }^{2}}{3}\widetilde{\pi }%
_{\langle k\rangle }^{\prime }+\frac{\widetilde{\kappa }^{2}}{6}D_{k}\left( 
\widetilde{\rho }+3\widetilde{\pi }-3\widetilde{p}\right) +\frac{2\widetilde{%
\kappa }^{2}}{3}\left( \widehat{K}-\frac{\widehat{\Theta }}{3}\right) 
\widetilde{\pi }_{k}  \notag \\
&&-\frac{2\widetilde{\kappa }^{2}}{3}K\widetilde{q}_{k}-\frac{2\widetilde{%
\kappa }^{2}}{3}\widetilde{q}\left( 2\widehat{K}_{k}+K_{k}\right) +\frac{2%
\widetilde{\kappa }^{2}}{3}\widetilde{\pi }_{ka}\widehat{A}^{a}-\frac{2%
\widetilde{\kappa }^{2}}{3}\left( \widetilde{\pi }-\widetilde{p}\right) 
\widehat{A}_{k}-\frac{5\widetilde{\kappa }^{2}}{3}\widehat{\sigma }_{ka}%
\widetilde{\pi }^{a}\ , \\
0 &=&\dot{\Theta}-D^{a}A_{a}+\frac{\Theta ^{2}}{3}+\widehat{\Theta }\widehat{%
K}-A^{a}A_{a}-2\omega _{a}\omega ^{a}+\sigma _{ab}\sigma ^{ab}-\widehat{K}%
^{a}\widehat{K}_{a}-\mathcal{E}-\frac{1}{2}\widetilde{\Lambda }+\frac{%
\widetilde{\kappa }^{2}}{2}\left( \widetilde{\rho }+\widetilde{\pi }+%
\widetilde{p}\right) \ , \\
0 &=&\dot{\omega}_{\langle a\rangle }-\frac{1}{2}\varepsilon _{a}^{\ \
cd}D_{c}A_{d}+\frac{2\Theta }{3}\omega _{a}-\sigma _{ab}\omega ^{b}\ ,
\end{eqnarray}%
\begin{eqnarray}
0 &=&\dot{\sigma}_{\langle ab\rangle }-D_{\langle a}A_{b\rangle }+\frac{%
2\Theta }{3}\sigma _{ab}+\frac{1}{2}\left( \widehat{K}-\frac{\widehat{\Theta 
}}{3}\right) \widehat{\sigma }_{ab}-A_{\langle a}A_{b\rangle }-\frac{1}{2}%
\widehat{K}_{\langle a}\widehat{K}_{b\rangle }  \notag \\
&&+\omega _{\langle a}\omega _{b\rangle }+\sigma _{c\langle a}\sigma
_{b\rangle }^{\,\,\,\,\,\,c}+\frac{1}{2}\widehat{\sigma }_{c\langle a}%
\widehat{\sigma }_{b\rangle }^{\,\,\,\,\,\,c}+E_{ab}+\frac{1}{2}\widehat{%
\mathcal{E}}_{ab}-\frac{\widetilde{\kappa }^{2}}{3}\widetilde{\pi }_{ab}\ ,
\\
0 &=&D^{a}\omega _{a}-A_{a}\omega ^{a}\ , \\
0 &=&D_{\langle c}\omega _{k\rangle }+\varepsilon _{ab\langle k}D^{b}\sigma
_{c\rangle }^{\,\,\,\,\,a}+2A_{\langle c}\omega _{k\rangle }+H_{ab}\ , \\
0 &=&D^{b}\sigma _{ab}-\frac{2}{3}D_{a}\Theta +\varepsilon _{a}^{\ \
ck}D_{c}\omega _{k}-\frac{2\widehat{\Theta }}{3}\widehat{K}_{a}+2\varepsilon
_{a}^{\ \ ck}A_{c}\omega _{k}+\widehat{\sigma }_{ab}\widehat{K}^{b}+\widehat{%
\mathcal{E}}_{a}+\frac{2\widetilde{\kappa }^{2}}{3}\widetilde{q}_{a}\ ,
\end{eqnarray}%
\begin{eqnarray}
0 &=&\dot{E}_{\langle kj\rangle }-\frac{1}{2}\dot{\widehat{\mathcal{E}}}%
_{\langle kj\rangle }-\varepsilon _{ab\langle k}D^{a}H_{j\rangle
}^{\,\,\,\,b}+\frac{1}{2}D_{\langle k}\widehat{\mathcal{E}}_{j\rangle
}+\Theta E_{kj}-\frac{\Theta }{6}\widehat{\mathcal{E}}_{kj}+\widehat{%
\mathcal{E}}_{\langle k}A_{j\rangle }-\frac{2\mathcal{E}}{3}\sigma _{kj} 
\notag \\
&&-\frac{1}{2}\widehat{\mathcal{E}}_{\langle j}^{\,\,\,\,a}\left( \omega
_{k\rangle a}+\sigma _{k\rangle a}\right) +E_{\langle k}^{\,\,\,\,\,a}\left(
\omega _{j\rangle a}-3\sigma _{j\rangle a}\right) +2\varepsilon _{\langle
k}^{\ \ \ \ ab}H_{j\rangle a}A_{b}-\frac{1}{2}\left( \widehat{K}-\frac{%
\widehat{\Theta }}{3}\right) \dot{\widehat{\sigma }}_{\langle kj\rangle } 
\notag \\
&&-\widehat{\sigma }_{\langle j}^{\,\,\,\,\,a}\dot{\widehat{\sigma }}%
_{k\rangle a}-\frac{1}{2}\left( \widehat{K}-\frac{\widehat{\Theta }}{3}%
\right) ^{\cdot }\widehat{\sigma }_{kj}+\widehat{K}_{\langle k}D_{j\rangle
}\left( \widehat{K}-\widehat{\Theta }\right) -\frac{\widehat{\Theta }}{3}%
D_{\langle k}\widehat{K}_{j\rangle }+\frac{1}{2}\widehat{\sigma }_{\langle
j}^{\,\,\,\,a}D_{k\rangle }\widehat{K}_{a}+\widehat{K}_{\langle k}D^{b}%
\widehat{\sigma }_{j\rangle b}  \notag \\
&&+\frac{1}{2}\widehat{K}_{a}D_{\langle k}\widehat{\sigma }_{j\rangle
}^{\,\,\,\,a}+\frac{\widehat{\Theta }}{3}\left( \widehat{K}+\frac{\widehat{%
\Theta }}{3}\right) \sigma _{kj}-\frac{7\Theta }{6}\widehat{K}_{\langle k}%
\widehat{K}_{j\rangle }+\left( \widehat{K}-\frac{\widehat{\Theta }}{3}%
\right) \widehat{K}_{\langle k}A_{j\rangle }-\frac{\Theta }{6}\widehat{%
\sigma }_{\langle j}^{\,\,\,\,a}\widehat{\sigma }_{k\rangle a}  \notag \\
&&-\frac{1}{2}\left( \widehat{K}-\frac{\widehat{\Theta }}{3}\right) \widehat{%
\sigma }_{\langle k}^{\,\,\,\,\,\,a}\left( \omega _{j\rangle a}+\sigma
_{j\rangle a}\right) -\frac{\Theta }{6}\left( \widehat{K}-\frac{\widehat{%
\Theta }}{3}\right) \widehat{\sigma }_{kj}-\frac{1}{2}\widehat{K}_{\langle
k}\sigma _{j\rangle b}\widehat{K}^{b}-\frac{1}{2}\widehat{\sigma }_{\langle
j}^{\,\,\,\,a}\omega _{k\rangle c}\widehat{\sigma }_{a}^{\,\,\,c}  \notag \\
&&-\frac{1}{2}\sigma _{jk}\widehat{K}^{a}\widehat{K}_{a}+\frac{3}{2}\widehat{%
K}_{\langle k}\omega _{j\rangle a}\widehat{K}^{a}+\widehat{K}_{\langle k}%
\widehat{\sigma }_{j\rangle a}A^{a}+\widehat{\sigma }_{a\langle
j}A_{k\rangle }\widehat{K}^{a}-\frac{1}{2}\widehat{\sigma }_{c}^{\,\,\,\,a}%
\widehat{\sigma }_{\langle k}^{\,\,\,\,\,\,c}\sigma _{j\rangle a}-\frac{%
\widetilde{\kappa }^{2}}{3}\dot{\widetilde{\pi }}_{\langle kj\rangle } 
\notag \\
&&-\frac{\widetilde{\kappa }^{2}}{3}D_{\langle k}\widetilde{q}_{j\rangle }+%
\frac{\widetilde{\kappa }^{2}}{3}\widetilde{\pi }_{\langle k}\widehat{K}%
_{j\rangle }-\frac{4\widetilde{\kappa }^{2}}{3}\widetilde{q}_{\langle
k}A_{j\rangle }-\frac{\widetilde{\kappa }^{2}}{3}\left( \widetilde{\rho }+%
\widetilde{p}\right) \sigma _{jk}-\frac{\widetilde{\kappa }^{2}}{9}\Theta 
\widetilde{\pi }_{jk}-\frac{\widetilde{\kappa }^{2}}{3}\widetilde{\pi }%
_{\langle j}^{\,\,\,\,a}\left( \omega _{k\rangle a}+\sigma _{k\rangle
a}\right) \ , \\
0 &=&\dot{H}_{\langle kj\rangle }+\varepsilon _{ab\langle k}D^{a}E_{j\rangle
}^{\,\,\,\,b}+\frac{1}{2}\varepsilon _{ab\langle k}D^{a}\widehat{\mathcal{E}}%
_{j\rangle }^{\,\,\,b}+\Theta H_{kj}-3\sigma _{a\langle k}H_{j\rangle
}^{\,\,\,a}-\omega _{a\langle k}H_{j\rangle }^{\,\,\,a}  \notag \\
&&-2\varepsilon _{\langle k}^{\ \ \ ab}E_{j\rangle a}A_{b}-\frac{1}{2}%
\varepsilon _{\langle k}^{\ \ \ ab}\sigma _{j\rangle a}\widehat{\mathcal{E}}%
_{b}-\frac{3}{2}\widehat{\mathcal{E}}_{\langle j}\omega _{k\rangle }-\frac{1%
}{2}\varepsilon _{\langle k}^{\ \ \ \ cd}\widehat{\sigma }_{j\rangle
c}D_{d}\left( \widehat{K}-\frac{\widehat{\Theta }}{3}\right)  \notag \\
&&+\frac{1}{2}\left( \widehat{K}-\frac{\widehat{\Theta }}{3}\right)
\varepsilon _{ab\langle k}D^{a}\widehat{\sigma }_{j\rangle }^{\,\,\,\,b}+%
\frac{1}{2}\varepsilon _{ab\langle k}D^{a}\widehat{\sigma }_{j\rangle c}%
\widehat{\sigma }^{cb}+\frac{1}{2}\varepsilon _{ab\langle k}\widehat{\sigma }%
_{j\rangle c}D^{a}\widehat{\sigma }^{cb}  \notag \\
&&+\widehat{\Theta }\omega _{\langle k}\widehat{K}_{j\rangle }+\frac{%
\widehat{\Theta }}{3}\varepsilon _{\langle k}^{\ \ \ ab}\sigma _{j\rangle a}%
\widehat{K}_{b}-\frac{1}{2}\varepsilon _{ab\langle k}\widehat{K}_{j\rangle
}D^{a}\widehat{K}^{b}-\frac{1}{2}\varepsilon _{ab\langle k}\widehat{K}%
^{b}D^{a}\widehat{K}_{j\rangle }  \notag \\
&&+\frac{1}{2}\varepsilon _{\langle k}^{\ \ \ ab}\sigma _{j\rangle b}%
\widehat{\sigma }_{a}^{\,\,\,\,c}\widehat{K}_{c}-\frac{3}{2}\widehat{\sigma }%
_{\langle j}^{\,\,\,\,a}\omega _{k\rangle }\widehat{K}_{a}-\frac{\widetilde{%
\kappa }^{2}}{3}\varepsilon _{ab\langle k}D^{a}\widetilde{\pi }_{j\rangle
}^{b}-\widetilde{\kappa }^{2}\widetilde{q}_{\langle j}\omega _{k\rangle }-%
\frac{\widetilde{\kappa }^{2}}{3}\varepsilon _{\langle k}^{\ \ \ \ ab}\sigma
_{j\rangle a}\widetilde{q}_{b}\ ,
\end{eqnarray}%
\begin{eqnarray}
0 &=&D^{a}E_{ak}-\frac{1}{2}D^{a}\widehat{\mathcal{E}}_{ak}+\frac{1}{3}D_{k}%
\mathcal{E}-3H_{ka}\omega ^{a}+\varepsilon _{k}^{\ \ ab}H_{ac}\sigma
_{b}^{c}+\frac{\Theta }{3}\widehat{\mathcal{E}}_{k}-\frac{1}{2}\left(
3\omega _{ka}+\sigma _{ka}\right) \widehat{\mathcal{E}}^{a}-\frac{2\widehat{%
\Theta }}{9}D_{k}\widehat{\Theta }  \notag \\
&&-\frac{1}{2}\widehat{\sigma }_{ak}D^{a}\left( \widehat{K}-\frac{\widehat{%
\Theta }}{3}\right) -\frac{1}{2}\left( \widehat{K}-\frac{\widehat{\Theta }}{3%
}\right) D^{a}\widehat{\sigma }_{ak}-\frac{1}{2}\widehat{\sigma }%
_{a}^{\,\,\,\,c}D^{a}\widehat{\sigma }_{ck}+\frac{2}{3}\widehat{\sigma }%
^{ab}D_{k}\widehat{\sigma }_{ab}-\frac{1}{2}\widehat{\sigma }_{kb}D^{a}%
\widehat{\sigma }_{a}^{\,\,\,\,b}  \notag \\
&&-\frac{1}{3}\widehat{K}^{a}D_{k}\widehat{K}_{a}+\frac{1}{2}\widehat{K}%
_{a}D^{a}\widehat{K}_{k}+\frac{1}{2}\widehat{K}_{k}D^{a}\widehat{K}_{a}+%
\frac{3}{2}\varepsilon _{k}^{\ \ ad}\widehat{K}_{c}\omega _{a}\widehat{%
\sigma }_{d}^{\,\,\,\,c}-\frac{1}{2}\widehat{\sigma }_{a}^{\,\,\,\,b}%
\widehat{K}^{a}\sigma _{kb}-\frac{2}{9}\widehat{\Theta }\Theta \widehat{K}%
_{k}+\frac{\widehat{\Theta }}{3}\sigma _{kb}\widehat{K}^{b}  \notag \\
&&+\widehat{\Theta }\varepsilon _{k}^{\ \ cd}\widehat{K}_{c}\omega _{d}+%
\frac{\Theta }{3}\widehat{\sigma }_{k}^{\,\,\,\,a}\widehat{K}_{a}+\frac{%
\widetilde{\kappa }^{2}}{3}D^{a}\widetilde{\pi }_{ak}-\frac{\widetilde{%
\kappa }^{2}}{6}D_{k}\left( \widetilde{\rho }-\widetilde{\pi }+\widetilde{p}%
\right) +\frac{2\widetilde{\kappa }^{2}}{9}\Theta \widetilde{q}_{k}-\frac{%
\widetilde{\kappa }^{2}}{3}\widetilde{q}^{a}\left( 3\omega _{ka}+\sigma
_{ka}\right) \ ,
\end{eqnarray}%
\begin{eqnarray}
0 &=&D^{a}H_{ak}+\frac{1}{2}\varepsilon _{k}^{\ \ ab}D_{a}\widehat{\mathcal{E%
}}_{b}-\frac{4\mathcal{E}}{3}\omega _{k}+3E_{ka}\omega ^{a}-\varepsilon
_{k}^{\ \ ab}E_{ac}\sigma _{b}^{\,\,\,\,c}+\frac{1}{2}\widehat{\mathcal{E}}%
_{ak}\omega ^{a}+\frac{1}{2}\varepsilon _{k}^{\ \ ab}\widehat{\mathcal{E}}%
_{ac}\sigma _{b}^{\,\,\,\,c}-\frac{1}{2}\widehat{K}^{c}\omega _{c}\widehat{K}%
_{k}  \notag \\
&&-\frac{1}{3}\varepsilon _{k}^{\ \ ab}\widehat{K}_{b}D_{a}\widehat{\Theta }-%
\frac{\widehat{\Theta }}{3}\varepsilon _{k}^{\ \ ab}D_{a}\widehat{K}_{b}-%
\frac{1}{2}\varepsilon _{abk}\widehat{K}^{c}D^{b}\widehat{\sigma }%
_{c}^{\,\,\,\,a}-\frac{1}{2}\varepsilon _{k}^{\ \ \ ab}\widehat{\sigma }%
_{a}^{\,\,\,\,c}D_{b}\widehat{K}_{c}+\frac{2\widehat{\Theta }}{3}\left( 
\widehat{K}+\frac{\widehat{\Theta }}{3}\right) \omega _{k}-\frac{1}{2}%
\widehat{K}_{a}\widehat{K}^{a}\omega _{k}  \notag \\
&&+\frac{1}{2}\left( \widehat{K}-\frac{\widehat{\Theta }}{3}\right) \widehat{%
\sigma }_{k}^{\,\,\,\,c}\omega _{c}+\frac{1}{2}\varepsilon _{k}^{\ \
ac}\sigma _{ab}\widehat{K}^{b}\widehat{K}_{c}-\frac{1}{2}\widehat{\sigma }%
_{ab}\widehat{\sigma }^{ab}\omega _{k}+\frac{1}{2}\widehat{\sigma }_{ca}%
\widehat{\sigma }_{k}^{\,\,\,\,c}\omega ^{a}+\frac{1}{2}\left( \widehat{K}-%
\frac{\widehat{\Theta }}{3}\right) \varepsilon _{k}^{\ \ ab}\widehat{\sigma }%
_{ac}\sigma _{b}^{\,\,\,\,c}  \notag \\
&&+\frac{1}{2}\varepsilon _{k}^{\ \ ab}\widehat{\sigma }_{da}\widehat{\sigma 
}_{c}^{\,\,\,\,d}\sigma _{b}^{\,\,\,\,c}-\frac{\widetilde{\kappa }^{2}}{3}%
\widetilde{\pi }_{ka}\omega ^{a}+\frac{\widetilde{\kappa }^{2}}{3}%
\varepsilon _{k}^{\ \ ab}D_{a}\widetilde{q}_{b}+\frac{2\widetilde{\kappa }%
^{2}}{3}\left( \widetilde{\rho }+\widetilde{p}\right) \omega _{k}-\frac{%
\widetilde{\kappa }^{2}}{3}\varepsilon _{k}^{\ \ ab}\widetilde{\pi }%
_{a}^{\,\,\,\,c}\sigma _{bc}\ .  \label{DivHab_brane_dot}
\end{eqnarray}%
We note that this system of equations is valid on both sides of the brane.

\section{Kinematical, gravito-electro-magnetic and matter variables for
Bianchi I brane-worlds \label{BianchiBraneWorld}}

In this Appendix we rewrite the Bianchi I brane-world, presented in Ref. 
\cite{sopuerta} in terms of our variables. The brane-world solution contains
an unspecified function $V\left( y\right) $ of a Gauss normal coordinate $y$
(which is however not related to the brane normal).

The kinematical quantities appearing on and outside the brane are presented
in Tables \ref{Table2} and \ref{Table3}, respectively, while the
gravito-electro-magnetic quantities on and outside the brane are given in
Tables \ref{Table4} and \ref{Table5}. Tables \ref{Table3} and \ref{Table5}
contain quantities, which were not computed in Ref. \cite{sopuerta}. 
\begin{table}[h]
\caption{Brane kinematical quantities (first column) for the brane-world 
\protect\cite{sopuerta} (second column; notations and Eq. numbers are from
this reference). }
\label{Table2}%
\begin{equation*}
\begin{tabular}{|c||c|}
\hline
kinematic quantity & for the metric in Ref. \cite{sopuerta} \\ \hline\hline
$A_{a}$ & $0$ \\ \hline
$\Theta \;$ & $\Theta $, as given by Eq. (63) \\ \hline
$\sigma _{ab}$ & $\sigma _{AB}$, as given by Eqs. (64)-(65) \\ \hline
$\omega _{a}\;$ & $0$ \\ \hline
\end{tabular}%
\end{equation*}%
\end{table}
\begin{table}[h]
\caption{Off-brane kinematical quantities (first column) for the brane-world 
\protect\cite{sopuerta} (second column; notations are from this reference). }
\label{Table3}%
\begin{equation*}
\begin{tabular}{|c||c|}
\hline
kinematic quantity & for the metric in Ref. \cite{sopuerta} \\ \hline\hline
$\widehat{K}$ & $\frac{\varepsilon }{\sqrt{1-V^{2}}}\left( \frac{u^{\prime }%
}{4u}+\frac{C_{0}}{u}+\frac{VV^{\prime }}{1-V^{2}}\right) $ \\ \hline
$K$ & $-\frac{1}{\sqrt{1-V^{2}}}\left( \frac{Vu^{\prime }}{4u}+\frac{VC_{0}}{%
u}+\frac{V^{\prime }}{1-V^{2}}\right) $ \\ \hline
$\widehat{\Theta }$ & $\frac{\varepsilon }{\sqrt{1-V^{2}}}\left( \frac{%
3u^{\prime }}{4u}-\frac{C_{0}}{u}\right) $ \\ \hline
$\widehat{A}_{a}$ & $0$ \\ \hline
$\widehat{K}_{a}\;$ & $0$ \\ \hline
$K_{a}$ & $0$ \\ \hline
$L_{a}\;$ & $0$ \\ \hline
$\widehat{\omega }_{a}\;$ & $0$ \\ \hline
$\widehat{\sigma }_{ab}\;$ & $\sum\limits_{i=1}^{3}\widehat{\sigma }%
_{i}e_{ia}e_{ib}\ \ ,\ \ \widehat{\sigma }_{i}=\frac{\varepsilon }{3u\sqrt{%
1-V^{2}}}\left( C_{0}+3C_{i}\right) $ \\ \hline
\end{tabular}%
\end{equation*}%
\end{table}
\begin{table}[h]
\caption{Brane gravito-electro-magnetic quantities (first column) for the
brane-world \protect\cite{sopuerta} (second column; notations and Eq.
numbers are from this reference). }
\label{Table4}%
\begin{equation*}
\begin{tabular}{|c||c|}
\hline
Weyl quantity & for the metric in Ref. \cite{sopuerta} \\ \hline\hline
$\mathcal{E}$ & $-\kappa ^{2}\mathcal{U}$ given by Eq. (53) \\ \hline
$\widehat{\mathcal{E}}_{a}\;$ & $0$ \\ \hline
$\widehat{\mathcal{E}}_{ab}$ & $-\kappa ^{2}\mathcal{P}_{AB}$ given by Eqs.
(54)-(55) \\ \hline
\end{tabular}%
\end{equation*}%
\end{table}
\begin{table}[h]
\caption{Off-brane gravito-electro-magnetic quantities (first column) for
the brane-world \protect\cite{sopuerta} (second column; notations are from
this reference). }
\label{Table5}%
\begin{equation*}
\begin{tabular}{|c||c|}
\hline
Weyl quantity & for the metric in Ref. \cite{sopuerta} \\ \hline\hline
$\mathcal{E}_{a}$ & $0$ \\ \hline
$\mathcal{H}_{a}$ & $0$ \\ \hline
$\mathcal{E}_{ab}$ & $\sum\limits_{i=1}^{3}E_{i}e_{ia}e_{ib}\ \ ,\ E_{i}=-%
\frac{1}{6u^{2}}\left[ \left( C_{0}+3C_{i}\right) u^{\prime }-\frac{3}{2}%
C-2\left( C_{0}^{2}+3C_{i}^{2}\right) \right] $ \\ 
& $\ \ \ \ \ \ \ \ \ \ \ \ \ \ \ \ \ \ \ \ \ \ \ \ +\frac{1}{4u^{2}\left(
1-V^{2}\right) }\left[ \left( C_{0}+3C_{i}\right) u^{\prime }-C+4C_{i}\left(
C_{0}-C_{i}\right) \right] $ \\ \hline
$\mathcal{F}_{ab}$ & $\sum\limits_{i=1}^{3}F_{i}e_{ia}e_{ib}\ \ ,\ \ F_{i}=%
\frac{\varepsilon V}{\left( 1-V^{2}\right) u^{2}}\left[ \frac{C_{0}+3C_{i}}{4%
}u^{\prime }+C_{i}\left( C_{0}-C_{i}\right) -\frac{C}{4}\right] $ \\ \hline
$\mathcal{H}_{ab}$ & $0$ \\ \hline
$\widehat{\mathcal{H}}_{ab}$ & $0$ \\ \hline
\end{tabular}%
\end{equation*}%
\end{table}

The notations for the brane matter variables (after the straightforward
change in the indices from lower case to upper case letters) are identical
in this paper and in Ref. \cite{sopuerta}, with $q_{a}=0$. There are no
off-brane matter variables.

\end{document}